\documentclass[11pt,a4paper]{article}
\pdfoutput=1 
\usepackage{jheppub} 
\def\to{\rightarrow} 
 
\newcommand\as{\alpha_{\rm S}} 

\def\nn{\nonumber}
\def\ep{\epsilon}
\def\eps{\epsilon}
\def\mb{\mathbf}
\def\mc{\mathcal}
\def\dumunu{d^{\mu\nu}(p;n)}
\def\ddmunu{d_{\mu\nu}(p;n)}

\def\Gam{{\rm\Gamma}}
\def\bqt{\mb{q_T}}
\def\qt{$q_T$}
\def\bb{\mb{b}}
\def\bcp{{\boldsymbol{\cal P}}}
\def\bcf{{\boldsymbol{\cal F}}}

\def\bcb{{\boldsymbol{\cal B}}}
\def\cp{{\cal P}}
\def\cf{{\cal F}}
\def\cb{{\cal B}}
\def\bun{{\boldsymbol 1}}
\def\ltap{\raisebox{-.6ex}{\rlap{$\,\sim\,$}} \raisebox{.4ex}{$\,<\,$}} 
\def\gtap{\raisebox{-.6ex}{\rlap{$\,\sim\,$}} \raisebox{.4ex}{$\,>\,$}} 
\title{Collinear functions for QCD resummations}
\author[a]{Stefano Catani} 
\author[b]{and Prasanna K. Dhani}
\affiliation[a]{INFN, Sezione di Firenze and Dipartimento di Fisica e Astronomia,
Universit\`a di Firenze,\\ I-50019 Sesto Fiorentino, Florence, Italy}
\affiliation[b]{INFN, Sezione di Genova,\\ Via Dodecaneso 33, I-16146 Genova, Italy}
\emailAdd{catani@fi.infn.it}
\emailAdd{dhani@ific.uv.es}
\abstract{The singular behaviour of QCD squared amplitudes in the collinear limit is factorized and controlled by splitting kernels with a process-independent structure. We use these kernels
to define collinear functions that can be used in QCD resummation formulae of hard-scattering
observables. Different collinear functions are obtained by integrating the splitting kernels
over different phase-space regions that depend on the hard-scattering observables of interest. The collinear functions depend on an auxiliary vector $n^\mu$ that can be either
light-like $(n^2=0)$ or time-like $(n^2 > 0)$. In the case of transverse-momentum dependent
(TMD) collinear functions, we show that the use of a time-like auxiliary vector avoids the rapidity divergences, which are instead present if $n^2=0$.
The perturbative computation of the collinear functions lead to infrared (IR) divergences
that can be properly factorized with respect to IR finite functions that embody
the logarithmically-enhanced collinear contributions to hard-scattering cross sections.
We evaluate various collinear functions and their $n^\mu$ dependence at ${\cal O}(\as)$.
We compute the azimuthal-correlation component of the TMD collinear functions at
${\cal O}(\as^2)$, and we present the results of the ${\cal O}(\as^2)$ contribution
of linearly-polarized gluons to transverse-momentum resummation formulae.
Beyond ${\cal O}(\as^2)$ the collinear functions of initial-state colliding partons
are process dependent, as a consequence of the violation of strict collinear factorization of QCD squared amplitudes.}
\arxivnumber{2208.05840}
\begin{document}
\maketitle
\section{Introduction}
\label{sec:intro}
The perturbative QCD calculations of a wide class of hard-scattering observables lead to logarithmically-enhanced contributions that are due to the radiation of soft and collinear partons (see, e.g., refs.~\cite{Luisoni:2015xha, Becher:2014oda} and references therein).
These large contributions have to be computed at high perturbative orders in the QCD  coupling $\as$, and possibly resummed to all orders in perturbation theory.
For instance, in the case of hadron collisions two topical observables that can be treated through resummation are the transverse momentum of produced high-mass systems
\cite{{Collins:1984kg}}
and the $N$-jettiness shape variable
\cite{Stewart:2010tn}.

In this paper we consider the logarithmically-enhanced contributions produced by collinear radiation. These contributions originate from the singular behaviour of QCD scattering amplitudes in the multiparton collinear limit. At the squared amplitude level, the singular behaviour is factorized and controlled by perturbative splitting kernels that have a process-independent structure
\cite{Bern:1993qk, Bern:1994zx, Bern:1995ix, Kosower:1999xi, Bern:1999ry, Catani:1999ss, Catani:2003vu, Catani:2011st}.

We exploit collinear factorization of the squared amplitudes to introduce collinear functions that contribute to QCD resummation formulae for hard-scattering cross sections.
The collinear functions have a process-independent structure and are obtained by integration of the splitting kernels over the observable-dependent phase space. Different phase-space constraints lead to corresponding collinear functions for different hard-scattering observables. Specifically, we consider differential collinear functions that upon integration lead to transverse-momentum dependent (TMD) collinear functions and beam functions, which can be used for transverse-momentum and $N$-jettiness resummations, respectively.

Our collinear functions depend on an auxiliary vector $n^\mu$, which is future-like
(i.e., $n^2 \geq 0$). Such auxiliary vector directly appears in the expressions of the splitting kernels for collinear factorization of the squared amplitudes. Applications of collinear factorization typically use a {\it light-like} ($n^2 = 0$) auxiliary vector.
We also use a {\it time-like} ($n^2 > 0$) auxiliary vector. In the case of TMD collinear functions we show that the time-like vector avoids the presence of rapidity divergences, which instead occur if $n^2=0$ 
\cite{Collins:2008ht, Collins:2011zzd, Becher:2010tm, Echevarria:2011epo, Chiu:2012ir}. 

 Our collinear functions are, in general, {\it process dependent for radiation from initial-state colliding partons}. Their process dependence originates from the violation of strict collinear factorization \cite{Catani:2011st}, namely, from the corresponding process dependence of the splitting kernels for amplitude factorization in the space-like (SL)
collinear region. The SL collinear functions are actually process independent up to
${\cal O}(\as^2)$, and their process dependence only occurs at higher perturbative orders.
In the case of radiation from final-state fragmenting partons, the collinear functions are process independent as a consequence of the validity of strict collinear factorization 
in the time-like (TL) collinear region.

In the TL region our TMD and beam functions with $n^2 = 0$ are equivalent to the corresponding parton level functions defined by using 
Soft Collinear Effective Theory (SCET) \cite{Bauer:2000ew, Bauer:2000yr, Bauer:2001yt, Bauer:2002nz}.
This perturbative equivalence directly follows from the relation \cite{Ritzmann:2014mka}
between SCET collinear functions and collinear factorization of squared amplitudes.
In the SL region, setting $n^2 = 0$ a similar equivalence applies up to  ${\cal O}(\as^2)$.

The collinear functions can be computed order-by-order in $\as$ through the phase-space integration of the corresponding perturbative expansion of the splitting kernels. The perturbative computation of the collinear functions lead to infrared (IR) 
divergences that can be properly factorized with respect to the IR finite contributions to the hard-scattering cross sections. In this paper we discuss these perturbative features, and we present the explicit calculation at ${\cal O}(\as^2)$ of the azimuthal-correlation components of the collinear functions. In particular, from our calculation of the TMD functions we derive the ${\cal O}(\as^2)$ contributions of linearly-polarized gluons to transverse-momentum resummation, and we find agreement with the results of independent calculations for both the SL \cite{Luo:2019bmw, Gutierrez-Reyes:2019rug}
and TL \cite{Luo:2019bmw} regions.

The outline of the paper is as follows. In section~\ref{sec:qcdres} we recall the known structure of the collinear contributions to the formalism of transverse-momentum and $N$-jettiness resummations.
Section~\ref{sec:colfun} is devoted to introduce the collinear functions. We first recall
the collinear factorization formula of QCD squared amplitudes, and then we define the differential collinear functions, the TMD functions and the beam functions.
The perturbative features of the SL collinear functions are illustrated in 
section~\ref{sec:sltmd}. We perform the explicit calculation of the collinear functions at 
${\cal O}(\as)$, we present a detailed discussion of their dependence on the auxiliary vector $n^\mu$, and we explain their IR factorization structure. 
In section~\ref{sec:azcorr} we carry out the calculation at ${\cal O}(\as^2)$ of 
the azimuthal-correlation component of the SL collinear functions, and we present the ensuing results for the contribution of linearly-polarized gluons to transverse-momentum resummation. Section~\ref{sec:tlfunct} is devoted to the perturbative features of the TL collinear functions, including the ${\cal O}(\as^2)$ calculation of the their azimuthal-correlation components. A brief summary of the paper is presented in section~\ref{sec:sum}.

\section{Transverse-momentum and \boldmath $N$-jettiness resummations}
\label{sec:qcdres}

In this section we briefly recall some main features of perturbative QCD resummations for two `classical' hard-scattering observables.
We consider transverse-momentum \cite{Collins:1984kg}
and $N$-jettiness \cite{Stewart:2010tn} resummations and, in particular,
the structure of the corresponding contributions due to partonic collinear radiation.

QCD transverse-momentum resummation is fully developed for the inclusive-production
processes of high-mass colourless systems (e.g., vector and Higgs bosons) in
hadron-hadron collisions. In the kinematical region where the transverse momentum
\qt\ of the produced system is much smaller than its invariant mass $M$, the perturbative QCD computation of the \qt-differential cross section leads to large logarithmic contributions of the type $\ln^n (M^2/q_T^2)$. The resummation procedure organizes and systematically sums these large contributions to all perturbative orders in the coupling $\as$.

In the following we specifically refer to the transverse-momentum resummation formalism of refs.~\cite{Collins:1984kg, Catani:2000vq, Catani:2010pd, Catani:2013tia}. Other equivalent formulations of transverse-momentum resummation, based either on 
%transverse-momentum dependent (TMD) 
TMD factorization 
\cite{Collins:2011zzd}
or on SCET methods 
\cite{Bauer:2000ew, Bauer:2000yr, Bauer:2001yt, Bauer:2002nz},
are presented and discussed in refs.~\cite{Becher:2010tm, Collins:2011zzd, Echevarria:2011epo, Chiu:2012ir}.

Transverse-momentum resummation \cite{Collins:1984kg, Catani:2000vq, Catani:2010pd, Catani:2013tia}
is conveniently carried out in impact parameter space, where the impact parameter
vector $\bb$ is the Fourier conjugate variable of the 
transverse-momentum vector $\bqt$. The differential cross section $d\sigma/d^2 \bqt$
is then obtained by inverse Fourier transformation of the result in $\bb$ space.

We directly consider and refer to the notation in ref.~\cite{Catani:2013tia}
(see, in particular, eqs.~(6)--(16) therein). The $\bb$ space cross section at 
$bM \gg 1$ is expressed in terms of the parton distribution functions (PDFs)
of the colliding hadrons and of perturbatively calculable factors. In this paper
we are mainly interested in the process-independent partonic factors 
$C_{ca}$ (see eqs.~(11) and (14) in ref.~\cite{Catani:2013tia}). Here the subscript $a$ ($a=q,{\bar q}, g$) denotes the type of initial-state colliding parton,
while the subscript $c$  ($c=q,{\bar q}, g$) refers to the parton that produces the high-mass system through hard scattering. The factors $C_{ca}$ have a definite dynamical origin \cite{Catani:2000vq}: they are due to the initial-state partonic transition $a \to c + X$ produced by final-state partonic radiation ($X$) that
is collinear to the parton $a$. In the context of formulations of transverse-momentum resummation that are based on SCET methods, the factors $C_{ca}$
are directly related to the so-called `matching coefficients' between TMD parton distributions and customary PDFs.

The quark collinear function $C_{qa}$ depends on the longitudinal-momentum fraction $z$ that is transferred in the collinear-radiation process, and it is computable
as a power series expansion in $\as$.
We write its perturbative expansion as follows
\begin{align}
\label{eq:colq}
    C_{qa}\left(z;\as\right) = \delta_{qa}\delta(1-z) + \frac{\as}{\pi } \, C_{qa}^{(1)}\left(z\right)+ \sum_{m=2}^{\infty} \left( \frac{\as}{\pi } \right)^m C_{qa}^{(m)}\left(z\right).
\end{align}
The antiquark collinear function $C_{{\bar q}a}$ is directly related to the quark collinear function through the relation $C_{{\bar q}a}= C_{q{\bar a}}$, which follows from charge conjugation invariance.

The gluon collinear function $C_{ga}^{\mu\nu}$ has a richer structure since it also depends on the Lorentz indices $\mu$ and $\nu$ of the gluon that produces the high-mass system ($\mu$ and $\nu$ are the Lorentz indices of the gluon in 
the hard-scattering amplitude and its complex-conjugated amplitude, respectively).
The structure of the partonic tensor is \cite{Catani:2010pd}
\begin{align}
\label{eq:colg}
C_{ga}^{\mu\nu}\left(z;p_1,p_2,\mb{b};\as\right) = d^{\mu\nu}(p_1,p_2) \,C_{ga}(z;\as)+D^{\mu\nu}(p_1,p_2;\mathbf{b}) \,G_{ga}(z;\as),
\end{align}
where
\begin{align}
\label{d12tens}
d^{\mu\nu}(p_1,p_2) &= -g^{\mu\nu}+\frac{p_1^{\mu}p_2^{\nu}+p_2^{\mu}p_1^{\nu}}{p_1p_2},
\\
\label{D12tens}
D^{\mu\nu}(p_1,p_2;\mathbf{b}) &=d^{\mu\nu}(p_1,p_2)- 2 \frac{b^{\mu}b^{\nu}}{\mb{b}^2}.
\end{align}
The light-like vectors $p_1^\mu$ and $p_2^\mu$ ($p_i^2=0, \;i=1,2$) 
in eq.~(\ref{eq:colg}) denote the momenta of the initial-state colliding partons or, equivalently, the directions of the momenta of the two colliding hadrons as treated in the massless approximation. In a reference frame in which the colliding hadrons are back-to-back, we can consider light-cone coordinates and we have
$p_1^\mu = (p_1^{+},\mathbf{0_T},0)$ and $p_2^\mu = (0,\mathbf{0_T},p_2^{-})$.
The momentum $b^\mu =  (0, \bb, 0)$ in eq.~(\ref{D12tens}) is the impact parameter vector in the four-dimensional notation
($b^\mu b_\mu= - \bb^2, p_1b = p_2b=0$).

The gluon collinear functions $C_{ga}$ and $G_{ga}$ in eq.~(\ref{eq:colg}) have the following perturbative expansions:
 \begin{align}
\label{azavgcolg}
   C_{ga}\left(z;\as\right) &= \delta_{ga}\delta(1-z) + \frac{\as}{\pi} \,C_{ga}^{(1)}\left(z\right)+ \sum_{m=2}^{\infty} 
\left(\frac{\as}{\pi}\right)^m \,
 C_{ga}^{(m)}\left(z\right),
\\
\label{azcorcolg}
   G_{ga}\left(z;\as\right) &=  \frac{\as}{\pi} \,G_{ga}^{(1)}\left(z\right)+ \left(\frac{\as}{\pi}\right)^2 \,G_{ga}^{(2)}\left(z\right) +\sum_{m=3}^{\infty} \left(\frac{\as}{\pi}\right)^m \,G_{ga}^{(m)}\left(z\right).
\end{align}
Charge conjugation invariance implies the relations $C_{ga}=C_{g{\bar a}}$ and $G_{ga}=G_{g{\bar a}}$.
We note that the expansion of $C_{ga}$ in eq.~(\ref{azavgcolg})  is completely analogous to that of the quark function
$C_{qa}$ in eq.~(\ref{eq:colq}). At variance, the perturbative expansion of $G_{ga}$ in eq.~(\ref{azcorcolg}) starts at $\cal{O}(\as)$.

The structure of eq.~(\ref{eq:colg}) is the consequence of collinear correlations \cite{Catani:2010pd}
that are produced by the evolution of the colliding hadrons into gluon partonic states. In particular,
the contribution of the tensor factor $D^{\mu \nu}$ in eq.~(\ref{eq:colg}) leads to spin and azimuthal correlations 
\cite{Catani:2010pd, Catani:2017tuc}
 in the hard-scattering production of the observed high-mass 
system at small values of \qt. This contribution is sometimes denoted as the contribution of linearly-polarized gluons
\cite{Mulders:2000sh} to TMD factorization and transverse-momentum resummation.
The size of the azimuthal correlations of collinear origin is controlled by the perturbative function $G_{ga}$ in 
eq.~(\ref{azcorcolg}). The quark collinear function $C_{qa}$ in eq.~(\ref{eq:colq}) and the gluon collinear function 
$C_{ga}$ in eq.~(\ref{azavgcolg}) do not lead to spin- and azimuthal-correlation effects.

We recall \cite{Catani:2000vq,Catani:2013tia} that the quark and gluon collinear functions in eqs.~(\ref{eq:colq}) and (\ref{eq:colg})
are precisely defined (and computable) modulo the following resummation-scheme transformations:
\begin{align}
\label{eq:qscheme}
C_{qa}(z;\as) &\to h_q(\as) \, C_{qa}(z;\as), \\
\label{eq:gscheme}
C_{ga}^{\mu \nu}(z;\as) &\to h_g(\as) \, C_{ga}^{\mu \nu}(z;\as),
\end{align}
where $h_a(\as)= 1+ \sum_{m=1}^{\infty} 
%h_a^{(m)} 
(\as/\pi)^m \,h_a^{(m)}$ ($a=q,g$) is an arbitrary perturbative function that does not depend on the
momentum fraction $z$.
As a consequence of the arbitrariness in eqs.~(\ref{eq:qscheme}) and (\ref{eq:gscheme}), the explicit results
for $C_{qa}$ and $C_{ga}^{\mu \nu}$ have to be accompanied by the specification of a resummation scheme.
In the computation of the \qt\ cross section, the resummation scheme dependence of the collinear functions cancels a 
corresponding dependence of the other factors that contribute to the transverse-momentum resummation formula
\cite{Catani:2000vq}.

The azimuthally-uncorrelated quark and gluon collinear functions $C_{qa}$ and $C_{ga}$ in eqs.~(\ref{eq:colq}) and 
(\ref{azavgcolg}) are known up to next-to-next-to-next-to-leading order (N$^3$LO) in QCD perturbation theory,
namely up to ${\cal O}(\as^3)$. The first process-independent computation of the next-to-leading order (NLO) terms
$C_{qa}^{(1)}(z)$ and $C_{ga}^{(1)}(z)$ was carried out in ref.~\cite{deFlorian:2001zd}.
The next-to-next-to-leading order (NNLO) terms $C_{qa}^{(2)}(z)$ and $C_{ga}^{(2)}(z)$ were first computed
in refs.~\cite{Catani:2011kr, Catani:2012qa, Gehrmann:2012ze, Gehrmann:2014yya}. Subsequent independent computations
of these NNLO terms were presented in refs.~\cite{Echevarria:2016scs, Luo:2019hmp, Luo:2019bmw}.
The N$^3$LO terms $C_{qa}^{(3)}(z)$ and $C_{ga}^{(3)}(z)$ have been computed very recently and independently by two research groups
\cite{Luo:2019szz, Ebert:2020yqt, Luo:2020epw}.

The azimuthally-correlated gluon collinear functions $G_{ga}(z;\as)$ in eqs.~(\ref{eq:colg}) and (\ref{azcorcolg})
are known up to ${\cal O}(\as^2)$. The first-order coefficients $G_{ga}^{(1)}$ are resummation-scheme independent, 
and they read \cite{Catani:2010pd}
\begin{equation}
\label{eq:g1ga}
G_{ga}^{(1)}(z)= C_a \;\frac{1-z}{z}, \quad \quad a=q,g,
\end{equation}
where $C_a$ is the Casimir colour coefficient of the parton $a$, with $C_q=C_F=(N_c^2-1)/(2N_c)$ and $C_g=C_A=N_c$ in
SU$(N_c)$ QCD with $N_c$ colours. The second-order terms $G_{ga}^{(2)}(z)$ have been obtained more recently
by the independent computations of refs.~\cite{Luo:2019bmw} and 
\cite{Gutierrez-Reyes:2019rug}.

The transverse-momentum resummation formalism can be extended to processes that are related by `kinematical crossing'
to the hadroproduction processes of high-mass colourless systems. Crossing related processes are double-inclusive hadron production
in $e^+e^-$ annihilation \cite{Collins:1981uk, Collins:2011zzd}  and single-inclusive hadron production 
in deep-inelastic lepton--hadron scattering \cite{Nadolsky:1999kb, Ji:2004wu, Collins:2011zzd}.
The corresponding transverse-momentum resummation formulae \cite{Collins:1981uk, Nadolsky:1999kb, Ji:2004wu, Collins:2011zzd}
are analogous to that for hadron--hadron collisions and they involve a main difference through the replacement of the PDFs
of the colliding hadrons with the parton fragmentation functions (PFFs) of the triggered hadrons in the final state.
In the resummation formulae the PFFs are convoluted with computable perturbative functions that embody the effect of QCD radiation
collinear to the final-state partons that are produced by the hard scattering and that fragment in the triggered hadrons.
These perturbative functions for time-like (TL) collinear evolution have the same structure as the initial-state collinear functions 
$C_{qa}$, $C_{ga}$ and $G_{ga}$ in eqs.~(\ref{eq:colq}), (\ref{eq:colg}), (\ref{azavgcolg}) and (\ref{azcorcolg}), and they are denoted
by $C_{qa}^{\rm TL}$, $C_{ga}^{\rm TL}$ and $G_{ga}^{\rm TL}$ in this paper.

The azimuthally-uncorrelated quark and gluon collinear functions $C_{qa}^{\rm TL}$ and 
$C_{ga}^{\rm TL}$ are known up to the N$^3$LO.
The NLO terms $C_{qa}^{{\rm TL} (1)}(z)$ and $C_{ga}^{{\rm TL} (1)}(z)$ were obtained in refs.~\cite{Nadolsky:1999kb} and \cite{Echevarria:2016scs}. The NNLO terms 
$C_{qa}^{{\rm TL} (2)}(z)$ and $C_{ga}^{{\rm TL} (2)}(z)$ were computed in refs.~\cite{Echevarria:2016scs} 
and refs.~\cite{Luo:2019hmp, Luo:2019bmw}. The N$^3$LO terms 
$C_{qa}^{{\rm TL} (3)}(z)$ and $C_{ga}^{{\rm TL} (3)}(z)$ have been obtained very recently
through the independent computations of refs.~\cite{Luo:2020epw, Ebert:2020qef}.
The azimuthally-correlated gluon collinear functions $G_{ga}^{\rm TL}$ have been evaluated up to ${\cal O}(\as^2)$.
The first-order coefficients $G_{ga}^{{\rm TL} (1)}$ are resummation-scheme independent, and they are
\begin{align}
\label{eq:gg1tl}
G_{gg}^{{\rm TL} (1)}(z) &= \;\,  C_A \; z (1-z),   \\
\label{eq:gq1tl}
G_{gq}^{{\rm TL} (1)}(z) &= - T_R  \,z (1-z), 
\end{align}
where $T_R=1/2$. 
%We note that the initial-state, or space-like (SL), expressions  $G_{ga}^{(1)}$ 
%in Eq.~(\ref{eq:g1ga})
%and the TL expressions in Eqs.~(\ref{eq:gg1tl}) and (\ref{eq:gq1tl}) are directly 
%related through SL-TL crossing symmetry
%\cite{Luo:2019bmw} (see also Sect.~\ref{})\footnote{*** to be specified}. 
The second-order coefficients $G_{ga}^{{\rm TL} (2)}$
were computed in ref.~\cite{Luo:2019bmw}. Our independent computation of 
$G_{ga}^{{\rm TL} (2)}$ 
(see section~\ref{sec:tlazcorr}) confirms the results of ref.~\cite{Luo:2019bmw}.

In the case of hadron--hadron collisions, the formalism of transverse-momentum resummation can be extended to production
processes of high-mass colourful systems (i.e., systems that contain particles with QCD colour charge). Example of such 
extension are those for the associated production of a (vector or Higgs) boson and a jet 
\cite{Sun:2018icb, Sun:2016kkh, Chien:2019gyf, Hatta:2021jcd, Buonocore:2021akg, Chien:2022wiq}
and for the production of heavy quarks 
\cite{Li:2013mia, Catani:2014qha, Catani:2021cbl}. In particular, in the case of the hadroproduction of a heavy-quark pair, the
transverse-momentum resummation formalism is fully developed up to next-to-next-to-leading logarithmic accuracy
and through the explicit computation of all the resummation factors up to NNLO 
\cite{Li:2013mia, Catani:2014qha, Angeles-Martinez:2018mqh, Catani:2019iny}.

The extension from colourless to colourful systems produces significant differences within the formalism of transverse-momentum resummation.
These differences are basically due to QCD radiation from the colourful particles of the observed high-mass system and, hence,
they are mostly related to soft radiation at wide angles with respect to the directions of the initial-state colliding partons.
Beyond the NNLO level of perturbative accuracy, non-abelian soft wide-angle interactions of absorptive origin lead to
violation of strict (i.e., process-independent) factorization of collinear radiation from the initial-state partons 
\cite{Catani:2011st}.  Therefore, the quark and gluon collinear functions $C_{qa}$ and $C_{ga}^{\mu \nu}$ in eqs.~(\ref{eq:colq}) 
and (\ref{eq:colg}) are expected not to be process-independent contributions to transverse-momentum resummation for the
production of high-mass colourful systems. They are certainly process independent up to ${\cal O}(\as^2)$ 
(see the accompanying comments to eqs.~(\ref{sltmdg}) and (\ref{sltmdq})
and section~\ref{sec:tmdfun}), but they can acquire process-dependent structures starting from some higher 
perturbative orders \cite{Catani:2011st, Forshaw:2012bi, Dixon:2019lnw}
and consistently with studies on the violation of generalized TMD factorization \cite{Collins:2007nk, Rogers:2010dm}.

The $N$-jettiness $\tau_N$ \cite{Stewart:2010tn} is a shape variable that measures the amount of radiation that accompanies the 
hard-scattering production of $N$ distinct hadronic jets in hadron and lepton collisions. The limit $\tau_N \to 0$ corresponds to an almost exclusive configuration of the $N$ jets. In this limit, or generically in the region where $\tau_N \ll Q$ 
($Q$ is the typical hard scale of the process), the perturbative computation of the $N$-jettiness cross section produces large logarithmic contributions of the type $\ln(Q/\tau_N)$. These large contributions can be organized and treated by the $N$-jettiness 
resummation formalism.

The $N$-jettiness resummation formula \cite{Stewart:2010tn} has a process-independent structure and it includes various factors 
that embody the effect of the radiation of soft and collinear partons in the final state. In the case of hadron collisions one of the factors
in the $N$-jettiness resummation formula is the beam function of the colliding hadron \cite{Stewart:2009yx}.
The beam function is due to QCD radiation that is collinear to the direction of the initial-state colliding hadron (parton), and it depends on the `transverse virtuality' of the parton that enters the hard scattering after collinear evolution.

At small values of transverse virtuality $t$ (which correspond to small values of the $N$-jettiness $\tau_N$)
the beam function ${\cal B}_c(z;t)$ of the parton $c$ is related to the customary PDF $f_a$ of the parton $a$ through the following convolution structure \cite{Stewart:2009yx}:
\begin{equation}
\label{eq:beam}
{\cal B}_c(z;t)=\sum_a
\int_z^1 \frac{dx}{x} \;{\widehat I}_{ca}(x,t;\mu_F, \as(\mu_F^2)) \;f_a(z/x;\mu_F^2),
\end{equation}
where $z$ is the fraction of the hadron longitudinal momentum carried by the parton $c$, and $\mu_F$ is the evolution scale of 
the PDF $f_a$. The convolution kernels  ${\widehat I}_{ca}$ ($c,a=q,{\bar q},g$) are known as `matching coefficients' 
of the beam function and they are perturbatively computable as power series in 
the QCD coupling $\as$.

The matching coefficients ${\widehat I}_{ca}$ for $N$-jettiness resummation and the collinear functions $C_{ca}$ in 
eqs.~(\ref{eq:colq}) and (\ref{azavgcolg}) for transverse-momentum resummation originate from the same underlying dynamics,
namely from QCD radiation that is collinear to the colliding parton $a$. Note that the radiation that contributes to the beam function
is integrated over the entire azimuthal region and, consequently, in the case of $N$-jettiness resummation 
there is no analogue of the azimuthal-correlation function $G_{ga}$ in eq.~(\ref{azcorcolg}).

In view of the renormalization properties of the beam function \cite{Stewart:2009yx} it is convenient to consider its Laplace transformation
with respect to the transverse virtuality.
 Correspondingly, we introduce the Laplace transformation
$I_{ca}$ of the matching coefficient ${\widehat I}_{ca}$ in eq.~(\ref{eq:beam}), and we define
\begin{equation}
\label{lsmc}
I_{ca}(z,\sigma;\mu_F, \as(\mu_F^2))  \equiv \int_0^{+\infty} dt \; e^{-\sigma t}\; {\widehat I}_{ca}(z,t;\mu_F, \as(\mu_F^2)),
\end{equation}
where $\sigma$ is the Laplace space variable that is conjugated to the transverse virtuality $t$.

The matching coefficients $I_{ca}$
 have the following perturbative expansion:
\begin{align}
\label{matchexp}
   I_{ca}\left(z,\sigma;\mu_F,\as(\mu_F^2)\right) =& \delta_{ca}\delta(1-z) + 
\frac{\as(\mu_F^2)}{\pi}I_{ca}^{(1)}\left(z,\sigma;\mu_F\right)
\nn\\
&+ \sum_{m=2}^{\infty} 
\left(\frac{\as(\mu_F^2)}{\pi}\right)^m I_{ca}^{(m)}\left(z,\sigma;\mu_F\right),
\end{align}
and they fulfil the relation $I_{ca} = I_{{\bar c} {\bar a}}$, which follows from charge-conjugation invariance.
The first-order coefficients $I_{ca}^{(1)}$ were first computed in refs.~\cite{Stewart:2010qs, Berger:2010xi}.
The evaluation of the second-order coefficients $I_{ca}^{(2)}$ was performed in refs.~\cite{Gaunt:2014xga, Gaunt:2014cfa}.
Partial results for the N$^3$LO coefficients $I_{ca}^{(3)}$ were obtained in refs.~\cite{Melnikov:2018jxb, Melnikov:2019pdm, Behring:2019quf,
Baranowski:2020xlp}.
The complete N$^3$LO results are presented in ref.~\cite{Ebert:2020unb}.

The beam function for $N$-jettiness resummation has a corresponding TL function, known as fragmenting jet function \cite{Procura:2009vm}.
We do not consider this TL function in this paper.

In our previous discussion on transverse-momentum resummation we have mentioned the possible occurrence of high-order factorization breaking effects of collinear radiation. We note that similar factorization breaking effects can affect $N$-jettiness resummation for multijet
production in hadron--hadron collisions.

\section{Collinear functions}
\label{sec:colfun}

In this paper we compute the collinear functions of section~\ref{sec:qcdres}
by starting from the evaluation of QCD scattering amplitudes.
At the bare level the computation exhibits ultraviolet (UV) and IR divergences. We regularize both divergences by working in
$d=4 - 2\ep$ space-time dimensions. In particular, we use the scheme of conventional dimensional regularization (CDR) 
\cite{Bollini:1972ui, Ashmore:1972uj, Cicuta:1972jf, Gastmans:1973uv},
in which on-shell gluons have $d-2$ physical states of spin polarizations and on-shell massless quarks (or antiquarks) have 2 spin polarization states. The dimensional regularization scale is denoted by $\mu_0$.

\subsection{Collinear factorization of scattering amplitudes}
\label{sec:colfactamp}

QCD scattering amplitudes are singular in the kinematical configurations in which two or more momenta of their external massless
partons become collinear. The singular behaviour in the collinear limit is described by a factorization formula 
\cite{Bern:1993qk, Bern:1994zx, Bern:1995ix, Kosower:1999xi, Bern:1999ry, Catani:1999ss, Catani:2003vu, Catani:2011st}
that has a universal (i.e., process-independent) structure.

We write the collinear factorization formula in its most general form as follows (see also ref.~\cite{Catani:2011st})
\begin{equation}
\label{colfact}
|  {\cal M}\left( \{ q_i \}; k_1, \dots, k_N \right) |^2 =
\langle  {\cal M}\bigl( \{ q_i \}; {\tilde k} \bigr) | 
\; \bcp\bigl( \{ q_i \};  k_1, \dots, k_N ; n\bigr) \;
|  {\cal M}\bigl( \{ q_i \}; {\tilde k} \bigr) \rangle + \dots,
\end{equation}
where the dots on the right-hand side denote non-singular terms in the collinear limit.
Here, $\cal M$ denotes the on-shell scattering amplitude of a generic hard-scattering process, and
$|{\cal M}|^2$ is the corresponding squared amplitude summed over the spins and colours of its external particles.
In eq.~(\ref{colfact}) we are considering the limit in which the momenta $k_1, \dots, k_N$ of $N$ external massless QCD partons
(gluons, quarks and antiquarks) of $\cal M$ become collinear. The momenta of the other external particles of $\cal M$
are $q_1,q_2,\dots$ and so forth. The dependence on the momenta and quantum numbers of the non-collinear particles
is generically denoted as dependence on $\{ q_i \}$. The singular behaviour in the collinear limit is embodied by the factor
$\bcp$, while ${\cal M}\bigl( \{ q_i \}; {\tilde k} \bigr)$ denotes the scattering amplitude that is obtained
from ${\cal M}\left( \{ q_i \}; k_1, \dots, k_N \right)$ by replacing the $N$ collinear partons with a single parent parton with momentum
${\tilde k} $ (${\tilde k} $ is the collinear limit of $\sum_{i=1}^N k_i$).  In the right-hand side of eq.~(\ref{colfact}) we are using the colour+spin
space notation of ref.~\cite{Catani:1996vz}, so that $|  {\cal M} \rangle $ and $\langle {\cal M} | $ are vectors in the colour+spin space 
of the external particles of $\cal M$ and, correspondingly, the collinear splitting kernel $\bcp$ is an operator acting onto this vector space
(i.e., $\bcp$ is a matrix in the colour and spin indices of the external particles of ${\cal M}\bigl( \{ q_i \}; {\tilde k}\bigr)$).

The scattering amplitude $\cal M$ can be computed in QCD perturbation theory as a power series (loop) expansion in $\as$. We note that
the factorization formula (\ref{colfact}) is valid to all perturbative orders and, consequently, the collinear splitting kernel $\bcp$ has a corresponding loop expansion in powers of $\as$. The dependences on $\as$ and on the CDR parameters $\ep$ and $\mu_0$ are not explicitly
denoted in the arguments of $\cal M$ and $\bcp$.  

In eq.~(\ref{colfact}), $\tilde k$ and $k_1, \dots, k_N$ are the {\it outgoing} 
momenta of the corresponding external partons.
The scattering amplitude $\cal M$ (and the kernel $\bcp$) is evaluated in different physical kinematical regions depending on the sign
of the `energies' (i.e., time components) of the outgoing momenta. If the energies of $k_1, \dots, k_N$ are all positive, we are dealing
with the TL collinear region, in which all the collinear partons are produced in the physical final state of the hard-scattering process.
If one (or more) of the collinear partons has negative energy, we are considering the SL collinear region. The parton with negative energy corresponds to its antiparton in the physical initial state of the hard-scattering process. The distinction between TL and SL collinear regions is,
in general, very relevant. Indeed, in the case of the TL collinear region the splitting kernel $\bcp$ has the relevant property of being completely
process independent: it does not depend on the momenta and quantum numbers of the non-collinear partons in $\cal M$.
%\cite{}
This property of strict (process-independent) collinear factorization is instead violated in the SL collinear regions \cite{Catani:2011st},
where the collinear splitting kernel $\bcp$ can depend on the non-collinear particles of $\cal M$ and, hence, on the specific hard-scattering process.
In both the TL and SL collinear regions, the splitting kernel $\bcp$ depends on an auxiliary vector $n$, as discussed in section~\ref{sec:diffcollfun}.

The splitting kernels $\bcp$ for the various partonic collinear configurations at ${\cal O}(\as)$ are well known (see, e.g., section~4.3
in ref.~\cite{Catani:1996vz}), and they are directly proportional to the real-emission contributions to the Altarelli-Parisi kernels for the
leading order (LO) evolution of the PDFs. The collinear splitting kernels $\bcp$ at ${\cal O}(\as^2)$ are fully known 
\cite{Campbell:1997hg, Catani:1998nv, Catani:1999ss, Bern:1998sc, Bern:1999ry, Kosower:1999rx, Sborlini:2013jba}
for both the TL and SL collinear regions \cite{Catani:2011st}. Various contributions to the splitting kernels $\bcp$ at ${\cal O}(\as^3)$
have been computed in 
refs.~\cite{DelDuca:1999iql, Birthwright:2005ak, Birthwright:2005vi, DelDuca:2019ggv, DelDuca:2020vst, 
Catani:2003vu, Sborlini:2014mpa, Sborlini:2014kla, Badger:2015cxa, Czakon:2022fqi, Bern:2004cz, Badger:2004uk, Duhr:2014nda}, 
and some results on the SL collinear regions are presented in refs.~\cite{Catani:2011st, Dixon:2019lnw}.

\subsection{Differential (unintegrated) collinear functions and the auxiliary
vector $n^\mu$}
\label{sec:diffcollfun}

The splitting kernels $\bcp$ of the factorization formula (\ref{colfact})
describe collinear emission at the fully exclusive level. We use these splitting kernels to introduce collinear-radiation functions at a more inclusive level.

We first consider the TL collinear region. In this case the splitting kernels $\bcp$ are process independent, and they have a non-trivial dependence only on the flavours, spins and momenta of the collinear partons. We introduce the subscript 
$c \to a_1 \dots a_N$ in $\bcp_{c \to a_1 \dots a_N}$ to denote the dependence on the flavours: $a_i$ $(i=1,\dots,N)$ is the flavour of the collinear parton with momentum $k_i$, and $c$ is the flavour of the parent collinear parton (i.e., the flavour of the parton with momentum $\tilde k$ in 
${\cal M}\bigl( \{ q_i \}; {\tilde k}\bigr)$). The TL splitting kernel $\bcp$ is proportional to the unit matrix
%, $\bun$, 
in color space and it is also proportional to the unit matrix
in the spin indices of the non-collinear partons. The dependence of 
$\bcp_{c \to a_1 \dots a_N}$ on the spin of the parent collinear parton $c$ can be instead non-trivial \cite{Catani:1999ss}
and it is different for the cases $c=q,{\bar q}$ and $c=g$.

In the case of the TL collinear splitting of a quark or antiquark, spin correlations are completely absent \cite{Catani:1999ss}. Projecting the kernel 
$\bcp$
onto basis vectors in colour + spin space, we have
\begin{align}
\label{qsplit}
&\langle s; r_i, \cdots | 
\, \bcp_{c \to a_1 \cdots a_N}\bigl(k_1, \dots, k_N ; n\bigr) \,
| s^\prime; r_i^\prime, \cdots \rangle  \\ \nn
& \;\;\; =
\cp_{c \to a_1 \cdots a_N}\bigl(k_1, \dots, k_N ; n\bigr) \;
\delta^{s s^\prime} \;\langle r_i, \cdots | 
\,\bun \,| r_i^\prime, \cdots \rangle\;, \quad \;\; c=q,{\bar q},
\end{align}
%\begin{align}
%\label{qsplit}
%\langle s; r_i, \cdots | 
%\, \bcp_{c \to a_1 \cdots a_N}\bigl(k_1, \dots, k_N ; n\bigr) \,
%| s^\prime; r_i^\prime, \cdots \rangle  
% =
%\cp_{c \to a_1 \cdots a_N}\bigl(k_1, \dots, k_N ; n\bigr) \;
%\delta^{s s^\prime} &\langle r_i, \cdots | 
%\,\bun \,| r_i^\prime, \cdots \rangle\;, \nn \\  
%&\quad \;\; c=q,{\bar q} \;,
%\end{align}
where $\bun$ is the unit matrix in colour+spin space.
%Here
In eq.~(\ref{qsplit})  
$s$ and $s^\prime$ are the spin indices of the parent collinear parton $c$ in 
$\langle \cal M |$ and $| \cal M \rangle$, respectively, while the indices 
$r_i, \cdots$ and $r_i^\prime, \cdots$ denote the other spin and colour indices of the external particles in ${\cal M}\bigl( \{ q_i \}; {\tilde k}\bigr)$.

Spin correlations are instead present in the case of the collinear splitting of a gluon \cite{Catani:1999ss}, and we write
%\begin{align}
%\label{gsplit}
%&\langle \mu; r_i, \cdots |
%\, \bcp_{g \to a_1 \cdots a_N}\bigl(k_1, \dots, k_N ; n\bigr) \,
%| \nu; r_i^\prime, \cdots \rangle  \\ \nn
%& \;\;\; =
%\cp^{\mu \nu}_{g \to a_1 \cdots a_N}\bigl(k_1, \dots, k_N ; n\bigr) \;
%\langle r_i, \cdots | 
%\,\bun \,| r_i^\prime, \cdots \rangle\;, 
%\end{align}
%%%%%%%%%%%%%%%
\begin{equation}
\label{gsplit}
\langle \mu; r_i, \cdots |\bcp_{g \to a_1 \cdots a_N}\bigl(k_1, \dots, k_N ; n\bigr)| \nu; r_i^\prime, \cdots \rangle  
  =
\cp^{\mu \nu}_{g \to a_1 \cdots a_N}\bigl(k_1, \dots, k_N ; n\bigr)\langle r_i, \cdots |\bun| r_i^\prime, \cdots \rangle, 
\end{equation}
where $\mu$ and $\nu$ are the Lorentz indices of the parent collinear gluon in 
$\langle \cal M |$ and $| \cal M \rangle$, respectively.

The scalar kernel $\cp_{c \to a_1 \cdots a_N}$ ($c=q,\bar q$)
in eq.~(\ref{qsplit})
and the tensor kernel $\cp^{\mu \nu}_{g \to a_1 \cdots a_N}$ in eq.~(\ref{gsplit}) are $c$-number functions (i.e., they are not matrices in colour and spin indices).
The tensor dependence of $\cp^{\mu \nu}_{g \to a_1 \cdots a_N}$ \cite{Catani:1999ss}
is due to terms that are proportional to either the metric tensor $g^{\mu \nu}$
or to quadratic terms of the type $k_{i \,T}^\mu k_{j \,T}^\nu$ $(i,j=1,\dots,N)$, where
$k_{i \,T}^\mu$ is the transverse momentum of the $i$-th collinear parton (the parton with momentum $k_i$) with respect to the collinear direction.
The remaining dependence of $\cp_{c \to a_1 \cdots a_N}$ ($c=q,\bar q,g$)
is due to scalar functions of the collinear momenta $k_1, \dots, k_N$. These functions are the sub-energies $s_{ij}= 2 k_i k_j$ and the ratios $x_i/x_j$
of the longitudinal-momentum fractions $x_i$ and $x_j$ of the momenta  $k_i$ and
$k_j$ with respect to the collinear direction.

The most general definition \cite{Catani:1999ss} of the longitudinal-momentum fractions of the collinear partons is obtained by introducing an auxiliary reference
vector $n^\mu$ that is far away from the collinear direction. Then the collinear splitting kernels depend on the ratios $x_i/x_j$ that are defined as
\begin{equation}
\label{xratio}
\frac{x_i}{x_j} = \frac{nk_i}{nk_j}.
\end{equation}
In the literature the reference vector $n^\mu$ is usually chosen to be a light-like
vector (i.e., $n^2=0$). Indeed this choice is very convenient for direct specific computations of the collinear kernels \cite{Catani:1999ss, Catani:2003vu, Sborlini:2013jba, DelDuca:2019ggv, DelDuca:2020vst}
and for many applications of the collinear factorization formula (\ref{colfact}).
However, we emphasize that we can also set $n^2\neq 0$.

In this paper we introduce and use a {\it time-like} auxiliary vector $n^\mu$
($n^2 > 0$), in addition to using also the customary light-like choice. Note that we do not modify any formal expression of the splitting kernels $\bcp$ in the literature. These kernels depend on $n^\mu$ through the ratios in 
eq.~(\ref{xratio}),
and we use the freedom of arbitrarily choosing $n^2 \geq 0$ in the collinear limit.
This arbitrariness follows from the fact that changing the value of $n^2$ produces
ratios $x_i/x_j$ that differ between themselves by terms of order 
%$\mb{k_{i\,T}}$ or  $\mb{k_{j\,T}}$, 
$k_{i\,T}$ or  $k_{j\,T}$,
which therefore vanish in the collinear limit. In other words, varying $n^2$ from
$n^2=0$ to $n^2 \geq 0$ in the kernel $\bcp$ of eq.~(\ref{colfact}) only produces differences in terms
that are non-singular in the collinear limit (these terms can be regarded as
`power corrections' in the context of squared amplitude computations  in the collinear limit). Using $n^2 \geq 0$ we also note that the quantity $n k_i$
($i=1,\dots,N$) can vanish\footnote{Considering a space-like auxiliary vector $n^\mu$
($n^2 < 0$), the scalar product  $n k_i$ vanishes also at a finite value of $k_i$ inside the physical region (though far from the collinear limit). We do not introduce and use space-like auxiliary vectors  in our splitting kernels and collinear functions (see also some related comments at the end of section~\ref{sec:tmdfun}).}
only if $k_i \to 0$ (using $n^2 = 0$, $n k_i$
vanishes also if $k_i$ is collinear to $n$).  
We observe that, independently of the value of $n^2$, the collinear kernels
$\bcp$ (and the ratios in eq.~(\ref{xratio})) are invariant under the rescaling 
$n^\mu \to \xi \,n^\mu$, where $\xi$ is an arbitrary parameter.

After our discussion of the structure of eq.~(\ref{colfact}) in the TL collinear region, we define differential TL collinear functions ${\cf}^{\rm TL}_{ca}$
($c,a=g, q, \bar q$) as follows. We consider the production of a parton of flavour $a$ and momentum $p^\mu$ in the physical final state and we fully integrate over the accompanying collinear radiation by keeping its total ($d$-dimensional) momentum $k$ fixed. 

If the parent collinear parton $c$ is a gluon, we have to take into account
the spin correlations in eq.~(\ref{gsplit}), and the precise definition of the collinear function $\cf^{\rm{TL} \, \mu \nu}_{ga}$ is
\begin{align}
\label{gcolfun}
    \cf^{\rm{TL} \,\mu \nu}_{ga}(p,k;n) =& \sum_{N=2}^{+\infty}
\left[\prod_{m=1}^{N-1}\int \frac{d^d k_m}{(2\pi)^{d-1}} \delta_{+}\left(k_m^2\right)\right] \;\delta^{(d)}\!\!\left(k-\sum_{i=1}^{N-1} k_i\right) \nn \\
& \times \sum_{a_1,\dots,a_{N-1}}
\frac{{\widetilde \cp}_{g\rightarrow a_1 \dots a_N}^{\mu \nu}\left(k_1,\dots,k_N;n\right)}{{\rm SF}\left(a_1,\dots,a_{N-1}\right)}{\Bigg |}_{\substack{k_N = \,p \\
a_N = \,a}},
\end{align}
%\substack{i,j \,\in H \\ i \,\neq\, j}
where ${\rm SF}\left(a_1,\dots,a_{N-1}\right)$ is the Bose symmetry factor for the identical particles in the set $\{ a_1,\dots,a_{N-1} \}$ (e.g., 
${\rm SF}\left(a_1,\dots,a_{N-1}\right)= (N-1)!$ if all these partons $a_i$ are gluons).

The parton momentum $p^\mu$ precisely specifies the collinear direction, and we can use a light-cone reference frame where $p^\mu= (p^+,\mathbf{0_T},0)$ with
$p^+ > 0$. In this frame we have $k_i^\mu = (k_i^+,\mathbf{k_{i\, T}},k_i^-)$
and $k^\mu = (k^+,\mathbf{k_{T}},k^-)$. The auxiliary time-like vector $n$ has coordinates $n^\mu=( n^+, \mathbf{0_T}, n^-)$, with $n^2=2n^+n^- > 0$. The case of a light-like vector $n^\mu$ is obtained by setting $n^+ = 0$.

The gluonic kernel ${\widetilde \cp}^{\mu \nu}$ in eq.~(\ref{gcolfun}) is related to the collinear kernel in eq.~(\ref{gsplit}) as follows
\begin{equation}
\label{gtker}
{\widetilde \cp}_{g\rightarrow a_1 \dots a_N}^{\mu \nu}\left(k_1,\dots,k_N;n\right) =
d^\mu_{\;\; \mu^\prime}(p;n) \;{\cp}_{g\rightarrow a_1 \dots a_N}^{\mu^\prime \nu^\prime}\left(k_1,\dots,k_N;n\right) \;d_{\nu^\prime}^{\;\;\; \nu}(p;n),
\end{equation}
where the spin polarization tensor $\dumunu$ is 
\begin{equation}
\label{dnten}
\dumunu = -g^{\mu\nu}+\frac{p^{\mu}n^{\nu}+n^{\mu}p^{\nu}}{np}-\frac{n^2p^{\mu}p^{\nu}}{(np)^2}.
\end{equation}
The use of ${\widetilde \cp}^{\mu \nu}$ in eq.~(\ref{gcolfun}) removes purely longitudinal terms, proportional to $p^\mu$ or $p^\nu$, from 
$\cf^{\rm{TL} \,\mu \nu}_{ga}$ (such terms are physically harmless, since they do not contribute to eq.~(\ref{colfact}) as a consequence of the gauge invariance relation $p_\mu {\cal M}^\mu\bigl( \{ q_i \}; {\tilde k}\bigr) =0$).
The function $\cf^{\rm{TL} \,\mu \nu}$ depends on the vectors $p,k,n$ and it is orthogonal to both $p$ and $n$. Therefore it has the following decomposition in tensor structures:
\begin{equation}
\label{fincor}
\cf^{\rm{TL} \,\mu \nu}_{ga}(p,k;n) = 
d^{\mu \nu}(p;n) \;\cf^{\rm{TL}}_{ga,\,{\rm az.in.}}(p,k;n)
+ D^{\mu\nu}(p,n;\mb{k_T},\ep) \;\cf^{\rm{TL}}_{ga,\,{\rm corr.}}(p,k;n),
\end{equation}
where
\begin{equation}
\label{Dtenep}
D^{\mu\nu}(p,n;\mb{k_T},\ep) = d^{\mu\nu}(p;n)-(d-2)\,
\frac{k_T^{\mu}k_T^{\nu}}{\mb{k_T}^2}.
\end{equation}
The tensor $D^{\mu\nu}$ in eq.~(\ref{fincor}) leads to correlations with respect to the azimuthal angle of the transverse-momentum vector $\mb{k_T}$. The scalar function $\cf^{\rm{TL}}_{ga,\,{\rm corr.}}$ controls the size of the azimuthal correlations of $\cf^{\rm{TL} \,\mu \nu}_{ga}$. The azimuthal-independent component
of $\cf^{\rm{TL} \,\mu \nu}_{ga}$ is proportional to the scalar function 
$\cf^{\rm{TL}}_{ga,\,{\rm az.in.}}$.

The tensors in eqs.~(\ref{dnten}) and (\ref{Dtenep}) are the $d$-dimensional generalization of those in eqs.~(\ref{d12tens}) and (\ref{D12tens}). They fulfil the following relations:
\begin{align}
\ddmunu D^{\mu\nu}(p,n;\mb{k_T},\ep)
&=0\;,
\nn\\
\ddmunu \,\dumunu&= d-2\;,
%\nn
\\
D_{\mu\nu}(p,n;\mb{k_T},\ep)
D^{\mu\nu}(p,n;\mb{k_T},\ep)
&=(d-2)(d-3)\;, \nn
\end{align}
with $d-2= 2-2\eps$ and $d-3=1-2\eps$.

The scalar functions $\cf^{\rm{TL}}_{ga,\,{\rm az.in.}}$ and
$\cf^{\rm{TL}}_{ga,\,{\rm corr.}}$ can be directly expressed in terms of the collinear splitting kernels ${\cp}_{g\rightarrow a_1 \dots a_N}^{\mu\nu}$.
These expressions are obtained from eq.~(\ref{gcolfun}) through the replacements
$\cf^{\rm{TL} \,\mu \nu}_{ga} \to 
\{ \cf^{\rm{TL}}_{ga,\,{\rm az.in.}}, \cf^{\rm{TL}}_{ga,\,{\rm corr.}} \}$
and ${\widetilde \cp}^{\mu\nu} \to \{ 
\cp^{\rm az.in.}, \cp^{\rm corr.} \}$ in the left-hand and right-hand sides, respectively. The corresponding collinear splitting kernels 
$\cp^{\rm az.in.}$ and $\cp^{\rm corr.}$ are obtained from eq.~(\ref{gtker}),
and they are given by the following relations:
\begin{align}
\label{avker}
\cp_{g\rightarrow a_1 \dots a_N}^{\rm az.in.} &= \frac{\ddmunu}{d-2} \,
{\widetilde \cp}_{g\rightarrow a_1 \dots a_N}^{\mu\nu}
= \frac{\ddmunu}{d-2} \,
{\cp}_{g\rightarrow a_1 \dots a_N}^{\mu\nu} \;,
\\
\label{corrker}
\cp_{g\rightarrow a_1 \dots a_N}^{\rm corr.}
&= \frac{D_{\mu\nu}(p,n;\mb{k_T},\ep)}{(d-2)(d-3)} \,
{\widetilde \cp}_{g\rightarrow a_1 \dots a_N}^{\mu\nu}
= \frac{D_{\mu\nu}(p,n;\mb{k_T},\ep)}{(d-2)(d-3)} \,
{\cp}_{g\rightarrow a_1 \dots a_N}^{\mu\nu} \;.
\end{align}

The TL collinear function $\cf^{\rm{TL}}_{ca}(p,k;n)$ $(c=q,\bar q)$ of a parent collinear fermion $c$ (quark or antiquark) is defined analogously to the gluon collinear function $\cf^{\rm{TL}}_{ga}(p,k;n)$, taking into account the relevant simplification of the absence of spin correlations in the collinear splitting kernels ${\cp}_{c\rightarrow a_1 \dots a_N}$ (see eq.~(\ref{qsplit})).
Therefore, we simply perform the replacements 
$\cf^{\rm{TL} \,\mu \nu}_{ga} \to \cf^{\rm{TL}}_{ca}$ and
${\widetilde \cp}_{g\rightarrow a_1 \dots a_N} \to 
{\cp}_{c\rightarrow a_1 \dots a_N}$ in eq.~(\ref{gcolfun}), and we define
\begin{align}
\label{qcolfun}
    \cf^{\rm{TL}}_{ca}(p,k;n) =& \sum_{N=2}^{+\infty}
\left[\prod_{m=1}^{N-1}\int \frac{d^d k_m}{(2\pi)^{d-1}} \delta_{+}\left(k_m^2\right)\right] \;\delta^{(d)}\!\!\left(k-\sum_{i=1}^{N-1} k_i\right) \nn \\
& \times \sum_{a_1,\dots,a_{N-1}}
\frac{{\cp}_{c\rightarrow a_1 \dots a_N}\left(k_1,\dots,k_N;n\right)}{{\rm SF}\left(a_1,\dots,a_{N-1}\right)}{\Bigg |}_{\substack{k_N = \, p \\
a_N = \, a}}, \quad \;\;\; c=q,{\bar q},
\end{align}
where ${\cp}_{c\rightarrow a_1 \dots a_N}$ is the collinear splitting kernel in the right-hand side of eq.~(\ref{qsplit}).

We can briefly and straightforwardly illustrate our definition of differential collinear functions for the SL collinear regions. Our main point is that the SL collinear splitting kernels $\bcp$ in the factorization formula (\ref{colfact})
are, in general, process dependent and, in particular, they can depend on the colour indices and momenta of the non-collinear partons and on the colour indices of the parent collinear parton $c$
in the hard-scattering process.

The general extension $\bcf_{ca}$ of the TL collinear functions in 
eqs.~(\ref{gcolfun}) and (\ref{qcolfun}) to the SL collinear regions 
is as follows
\begin{align}
\label{slcolfun}
    \bcf_{ca}(\{q_i\};p,k;n) =& \sum_{N=2}^{+\infty}
\left[\prod_{m=1}^{N-1}\int \frac{d^d k_m}{(2\pi)^{d-1}} \delta_{+}\left(k_m^2\right)\right] \;\delta^{(d)}\!\!\left(k-\sum_{i=1}^{N-1} k_i\right) \\
& \times \sum_{a_1,\dots,a_{N-1}}
\frac{{\widetilde \bcp}_{{\bar c}\rightarrow a_1 \dots a_N}\left(\{q_i\};k_1,\dots,k_N;n\right)}{{\rm SF}\left(a_1,\dots,a_{N-1}\right)}{\Bigg |}_{\substack{k_N=\,- p \\
a_N = \, {\bar a}}} \;\; \frac{\mathcal{N}_c(\eps)}{\mathcal{N}_a(\eps)}, \quad c=g,q,{\bar q}.\nn
\end{align}
The SL collinear functions $\bcf_{ca}(\{q_i\};p,k;n)$ depend on the flavour $a$ of the parton with momentum $p^\mu= (p^+,\mathbf{0_T},0)$ (with
$p^+ > 0$) that collides in the physical initial state of the scattering process.
The flavour $c$ refers to the incoming parton of the hard-scattering process
after the radiation of the final-state collinear partons with total momentum 
$k^\mu$. Similarly to the TL collinear functions, the auxiliary vector 
$n^\mu=( n^+, \mathbf{0_T}, n^-)$ is time-like ($n^2=2n^+n^- > 0$), in general
(the light-like case can be obtained by setting $n^+ = 0$).

The function ${\mathcal{N}_a(\eps)}$ in the right-hand side of 
eq.~(\ref{slcolfun})
depends on the number of space-time dimensions, and it is $\mathcal{N}_a(\eps) = (-1)^{2S_a}n_s(a,\eps)n_c(a)$,
where $S_a$ denotes the spin of the parton with flavour $a$, 
$n_s(a,\eps)$ is number of spin polarization states of that parton, and
$n_c(a)$ is its number of colours. Therefore, we have 
$\mathcal{N}_{q}(\eps) = \mathcal{N}_{{\bar q}}(\eps)= -2 N_c$ 
%for $b=q, \bar q$, 
and $\mathcal{N}_g(\eps) = 2(1-\eps)(N_c^2-1)$.

The SL function $\bcf_{ca}$ is the contribution of initial-state colliner radiation at the cross section level. The cross section is proportional to the square of the scattering amplitude $\cal M$, averaged over the spins and colours of the initial-state partons. This average procedure introduces the factor 
$\mathcal{N}_c(\eps)/\mathcal{N}_a(\eps)$ in the right-hand side of 
eq.~(\ref{slcolfun}) (the factor $(-1)^{2S_a}$ in $\mathcal{N}_a(\eps)$ is due to the crossing of the parton $a$ from the final state to the initial state of the scattering process).

The spin dependence of $\bcp$ is similar in the TL and SL collinear regions. The SL
collinear kernel $\bcp_{c \rightarrow a_1 \dots a_N}$ is proportional to the unit matrix in the spin indices of the non-collinear partons and it depends on the spin of the parent collinear parton $c$ as in eqs.~(\ref{qsplit}) and (\ref{gsplit}).
Analogously to the TL collinear functions in 
eqs.~(\ref{qcolfun}) and (\ref{gcolfun}), the kernel 
${\widetilde \bcp}_{{c}\rightarrow a_1 \dots a_N}$ in 
eq.~(\ref{slcolfun}) is exactly equal to ${\bcp}_{{c}\rightarrow a_1 \dots a_N}$
in the quark or antiquark cases $c=q, \bar q$, and it is given as in 
eq.~(\ref{gtker}) in the gluon case $c=g$.

As we have already stated, the kernels $\bcp$ in the SL collinear region are, in
general, process dependent and, correspondingly, the SL collinear functions 
$\bcf_{ca}$ in eq.~(\ref{slcolfun}) are also process dependent. We consider the perturbative expansion of $\bcp$ in terms of the unrenormalized QCD coupling 
%$\alpha^u_{\rm S}$ 
$\as^u$, and we write
\begin{equation}
\label{cpexp}
\bcp = \sum_{L=0}^{+\infty} \bcp^{(L)},
\end{equation}
where $\bcp^{(0)}$ is the tree-level contribution to $\bcp$, $\bcp^{(1)}$
is its one-loop contribution, and so forth. The term $\bcp_{{c}\rightarrow a_1 \dots a_N}^{(L)}$ is proportional to $(\as^u)^{N-1+L}$. The process dependence of $\bcp$ and, hence, of $\bcf_{ca}$ 
first occurs at ${\cal O}(\as^3)$ and it is due to the one-loop and two-loop contributions $\bcp_{{c}\rightarrow a_1 a_2 a_3}^{(1)}$ and 
$\bcp_{{c}\rightarrow a_1 a_2}^{(2)}$ \cite{Catani:2011st, Forshaw:2012bi, Dixon:2019lnw}, while $\bcp_{{c}\rightarrow a_1 a_2 a_3 a_4}^{(0)}$ is process
independent\footnote{The tree-level kernels 
$\bcp_{{c}\rightarrow a_1 \dots a_N}^{(0)}$ ($N \geq 2$) are process independent
for both TL and SL collinear radiation.}. In particular, 
such process dependence
at ${\cal O}(\as^3)$ leads to non-abelian colour-matrix structures in 
$\bcp$ and $\bcf_{ca}$ in the case of 
collinear radiation from 
scattering amplitudes for the production of two or more QCD hard partons (jets, heavy quarks or hadrons) in parton--parton (hadron--hadron) collisions.

Considering perturbative contributions at ${\cal O}(\as)$ and ${\cal O}(\as^2)$,
the SL collinear kernels $\bcp$ and functions $\bcf$ are process independent 
and proportional to the unit matrix in the colour space of the hard-scattering partons.
Therefore, we can factorize such overall (and trivial) colour space dependence
in both sides of eq.~(\ref{slcolfun}), and we can simply deal with $c$-number 
SL collinear functions, analogously to the TL collinear case. Such SL collinear
functions are denoted as  
\begin{equation}
\label{slcfun}
\cf^{\, \mu \nu}_{ga}(p,k;n) \;, \;\; \cf_{ga,\,{\rm az.in.}}(p,k;n) \;,
\;\;\cf_{ga,\,{\rm corr.}}(p,k;n) \;\;{\rm and} \;\;\cf_{ca}(p,k;n) \;(c=q,{\bar q}) 
\;,
\end{equation}
and they are analogous to the TL collinear functions in 
eqs.~(\ref{gcolfun}), (\ref{fincor}), and (\ref{qcolfun}).

The SL collinear kernel $\bcp$ in eq.~(\ref{colfact}) and the collinear function $\bcf_{ca}$
in eq.~(\ref{slcolfun}) are proportional to the unit matrix in colour space also in the case of scattering amplitudes 
${\cal M}\bigl( \{ q_i \}; {\tilde k} \bigr)$ with a single external non-collinear QCD parton in addition to the parent collinear parton with momentum ${\tilde k}$.
This feature is valid to arbitrary orders in $\as$ since it simply follows from the fact that the colour space of such scattering amplitudes is one dimensional. Considering this class of processes, 
the trivial colour space dependence factorizes and
we can directly deal with $c$-number collinear functions $\cf_{ca}$ also in the SL regions. The production of high-mass colourless systems, which is considered in section~\ref{sec:qcdres},
is included in this class of processes.

We note that our TL and SL collinear functions, $\cf_{ca}^{\rm TL}$ and 
$\bcf_{ca}$, in eqs.~(\ref{gcolfun}), (\ref{qcolfun}), and (\ref{slcolfun}) are 
related to corresponding SCET differential functions, namely,
to the differential fragmenting jet functions and the differential beam functions of refs.~\cite{Jain:2011iu, Mantry:2009qz, Mantry:2010mk}. Their relation follows from the fact that all these functions are differential with respect to the 
($d$-dimensional) momentum $k^\mu$ of the final-state collinear radiation.
The differential functions of refs.~\cite{Jain:2011iu, Mantry:2009qz, Mantry:2010mk}
are defined as matrix elements of appropriate SCET operators and, considering
parton matrix elements, they lead to quantities that can directly be compared with
$\cf_{ca}^{\rm TL}$ and $\bcf_{ca}$ at small values of $k^-$ and 
$\mathbf{k_{T}}$ (i.e., in the collinear region where 
$k^+ = {\cal O}(p^+), k^- \ll p^+$ and $k_T^2 \ll (p^+)^2$). In the following
we can briefly comment on some general aspects of this comparison.

We first note that the SCET differential functions
%differential jet and beam functions 
%of Refs.~\cite{Jain:2011iu, Mantry:2009qz, Mantry:2010mk} 
depend on a {\it light-like}
vector $n^\mu$, which specifies the direction of Wilson line operators that enter
the definition of the SCET operators 
\cite{Bauer:2000ew, Bauer:2000yr, Bauer:2001yt, Bauer:2002nz}.
The differential fragmenting jet functions \cite{Jain:2011iu} at partonic level directly correspond to our TL collinear functions  $\cf_{ca}^{\rm TL}(p,k;n)$ with the choice of a light-like auxiliary vector $n^\mu$. Such direct correspondence
is also true between the differential beam functions 
\cite{Jain:2011iu, Mantry:2009qz, Mantry:2010mk} and our SL collinear functions 
$\bcf_{ca}$ with $n^2=0$, but the correspondence is limited up to ${\cal O}(\as^2)$.
Indeed the SCET differential functions 
%differential beam functions of Refs.~\cite{Jain:2011iu, Mantry:2009qz, Mantry:2010mk} 
are process independent, while our SL collinear functions $\bcf_{ca}$ 
become process dependent at ${\cal O}(\as^3)$ and higher orders.
We also note that the TL and SL collinear functions, 
$\cf_{ca}^{\rm TL}$ and $\bcf_{ca}$, with a time-like auxiliary vector  $n^\mu$
do not directly correspond to 
SCET differential functions.
%the partonic differential functions of 
%Refs.~\cite{Jain:2011iu, Mantry:2009qz, Mantry:2010mk}.
Perturbative computations of SCET differential jet and beam functions 
%(with $n^2=0$) 
are presented in refs.~\cite{Jain:2011iu, Mantry:2009qz, Mantry:2010mk,Gaunt:2014xxa, Gaunt:2020xlc}.
The perturbative calculation of the SL and TL collinear functions, $\bcf_{ca}$ and $\cf_{ca}^{\rm TL}$, with $n^2 \geq 0$ 
%(with both $n^2=0$ and $n^2 > 0$) 
is presented in sections.~\ref{sec:sldiff}, \ref{sec:azcorrdiff}, \ref{sec:tldiff}
and \ref{sec:tlazcorr}.

The differential collinear functions in eqs.~(\ref{gcolfun}), (\ref{qcolfun}), 
and (\ref{slcolfun}) are defined by integrating the splitting kernels $\bcp$
with the constraint of fixing the total momentum $k$ of the final-state collinear radiation. 
Applications of the collinear factorization formula (\ref{colfact}) to different types of
hard-scattering observables can require different or additional phase-space constraints
(e.g., constraints related to jet definitions or to angular/rapidity limitations)
on the momenta of the produced collinear partons. These constraints can be implemented on the phase-space integrations in the right-hand side of 
eqs.~(\ref{gcolfun}), (\ref{qcolfun}) and (\ref{slcolfun}), thus leading to the definition of corresponding collinear functions. Some main features of these functions
(e.g., their $n^\mu$ dependence and their process dependence in the SL collinear
region) are equal
to those of the collinear functions that we explicitly consider in this paper.

\subsection{TMD collinear functions and beam functions}
\label{sec:intcolfun}

The TL and SL differential collinear functions that we have introduced in section~\ref{sec:diffcollfun}
can be used to define inclusive functions that are directly related to the perturbative computation and resummation
of large logarithmic contributions to hard-scattering observables. In the following we define TMD collinear functions and
beam functions that lead to the resummation coefficients which we have discussed in section~\ref{sec:qcdres}.

We \,first\, consider \,the\, TL\, collinear\, region. We\, use \,the\, gluon\, collinear\, function\, $\cf^{\rm{TL} \,\mu \nu}_{ga}(p,k;n)$ 
of eq.~(\ref{gcolfun}) and we define the gluon TMD function  $F^{\rm{TL} \,\mu \nu}_{ga}$ by integrating over
the radiated collinear momentum $k$ 
as follows 
\begin{align}
%\label{eq:masterqtgg}
\label{tlgtmd}
 F^{\rm{TL} \,\mu \nu}_{ga}(z;p/z,\mb{q_T};n) =& \delta(1-z) \;\delta^{(d-2)}(\mb{q_T})
\;\, \delta_{ga} \;\, d^{\mu \nu}(p;n) \nn \\
&+
\int d^dk \;\delta^{(d-2)}(\mb{k_T}+\mb{q_T}) 
\;\,\delta\!\left(\frac{k^{+}}{p^{+}}-\frac{1-z}{z}\right)
\;\cf^{\rm{TL} \,\mu \nu}_{ga}(p,k;n).
\end{align}
The TMD function  $F^{\rm{TL} \, \mu \nu}_{ga}$ describes the inclusive perturbative fragmentation
of a gluon into a parton $a$. The collinear fragmentation process transfers 
the transverse momentum $\mb{q_T}$ and the longitudinal-momentum fraction $z$, with $0 \leq z \leq 1$.
The gluon function $F^{\rm{TL}\,\mu \nu}_{ga}$ depends on the Lorentz
indices (and, hence, on the spin polarizations) of the fragmenting gluon.
As a consequence of eq.~(\ref{fincor}), the tensor function $F^{\rm{TL} \,\mu \nu}_{ga}$ fulfils the following decomposition:
%in scalar components
\begin{align}
\label{tmdfincor}
F^{\rm{TL} \,\mu \nu}_{ga}(z;p/z,\mb{q_T};n)  =& 
d^{\mu \nu}(p;n)\;\;F^{\rm{TL}}_{ga,\,{\rm az.in.}}\!\!\left(z;\mb{q_T}^2,\frac{n^2 \mb{q_T}^2}{(2np/z)^2}\right) \nn \\
&+ D^{\mu\nu}(p,n;\mb{q_T},\ep)\;\;F^{\rm{TL}}_{ga,\,{\rm corr.}}\!\!\left(z;\mb{q_T}^2,\frac{n^2 \mb{q_T}^2}{(2np/z)^2}\right),
\end{align}
where $F^{\rm{TL}}_{ga,\,{\rm az.in.}}$ and $F^{\rm{TL}}_{ga,\,{\rm corr.}}$ are the azimuthally-independent
and azimuthally-correlated components, respectively. These components are straightforwardly related to the collinear functions
$\cf^{\rm{TL}}_{ga,\,{\rm az.in.}}$ and $\cf^{\rm{TL}}_{ga,\,{\rm corr.}}$ of eq.~(\ref{fincor}). We have
\begin{align}
\label{tmdazin}
 F^{\rm{TL} }_{ga,\,{\rm az.in.}}\!\!\left(z;\mb{q_T}^2,\frac{n^2 \mb{q_T}^2}{(2np/z)^2}\right) =& \delta(1-z) \;\delta^{(d-2)}(\mb{q_T}) \;\, \delta_{ga}
 \nn \\
&+ 
\int d^dk \;\delta^{(d-2)}(\mb{k_T}+\mb{q_T}) 
\,\delta\!\left(\frac{k^{+}}{p^{+}}-\frac{1-z}{z}\right)
\cf^{\rm{TL}}_{ga,\,{\rm az.in.}}(p,k;n),
\\
%\end{align}
%\begin{align}
\label{tmdcorr}
 F^{\rm{TL} }_{ga,\,{\rm corr.}}\!\!\left(z;\mb{q_T}^2,\frac{n^2 \mb{q_T}^2}{(2np/z)^2}\right) 
=& 
\int d^dk \;\delta^{(d-2)}(\mb{k_T}+\mb{q_T}) 
\;\,\delta\!\left(\frac{k^{+}}{p^{+}}-\frac{1-z}{z}\right)
\;\cf^{\rm{TL}}_{ga,\,{\rm corr.}}(p,k;n).
\end{align}
The quark (or antiquark) TMD function $F^{\rm{TL}}_{ca}$ ($c=q, {\bar q}$) is defined analogously to the gluon TMD function 
in eq.~(\ref{tlgtmd}), and we have
\begin{align}
%\label{eq:masterqtqq}
\label{tlqtmd}
 F^{\rm{TL}}_{ca}\!\left(z;\mb{q_T}^2,\frac{n^2 \mb{q_T}^2}{(2np/z)^2}\right) 
 =& \delta(1-z) \;\delta^{(d-2)}(\mb{q_T}) \;\, \delta_{ca} \nn \\
&+ 
\int d^dk \;\delta^{(d-2)}(\mb{k_T}+\mb{q_T}) 
\;\,\delta\!\left(\frac{k^{+}}{p^{+}}-\frac{1-z}{z}\right)
\;\cf^{\rm{TL}}_{ca}(p,k;n),
\end{align}
where $\cf^{\rm{TL}}_{ca}(p,k;n)$ is the collinear function in eq.~(\ref{qcolfun}) .

We note that $F^{\rm{TL} }_{ga,\,{\rm az.in.}}, F^{\rm{TL} }_{ga,\,{\rm corr.}}$, and
$F^{\rm{TL}}_{ca} \;(c=q,{\bar q})$ are scalar functions that depend on $z$ and the vectors $p^\mu, n^\mu, \mb{q_T}$. Therefore they depend on the scalar quantities
$z, \mb{q_T}^2$ and $\frac{n^2 \mb{q_T}^2}{(2np/z)^2}$, as explicitly denoted by their
argument in eqs.~(\ref{tmdfincor})-(\ref{tlqtmd}). In particular, the dependence on
$\frac{n^2 \mb{q_T}^2}{(np)^2}$ is a consequence of the invariance under the arbitrary rescaling $n^\mu \to \xi n^\mu$.

As discussed in the following sections, the perturbative computation of the TL TMD functions $F^{\rm TL}_{ca}$ ($c=g,q,{\bar q}$) leads to IR divergences that can be factorized in terms of IR singular and IR finite contributions. The IR finite contributions are directly related to the collinear functions $C^{\rm TL}_{ca}$ of 
section~\ref{sec:qcdres}.

The general SL TMD function ${\bf F}_{ca}$ is obtained by analogy with the TL functions in eqs.~(\ref{tlgtmd}) and (\ref{tlqtmd}) and by taking into account that the SL collinear function $\bcf_{ca}$ in eq.~(\ref{slcolfun}) is, in general, process
dependent. The explicit definition of ${\bf F}_{ca}$ is
\begin{align}
\label{sltmdgen}
 {\bf F}_{ca}(\{q_i\};z;zp,\mb{q_T};n) =& \bun \;
\delta(1-z) \;\delta^{(d-2)}(\mb{q_T}) \;\, \delta_{ca}
%\;\bun 
\nn \\
&+ z
\int d^dk \;\,\delta^{(d-2)}(\mb{k_T}+\mb{q_T}) 
\;\,\delta\!\left(\frac{k^{+}}{p^{+}}-1+ z\right)
\;\bcf_{ca}(\{q_i\};p,k;n),
\end{align}
where the symbol $\bun$ in the right-hand side denotes the unit matrix in colour+spin space. 
The TMD function ${\bf F}_{ca}$ is a process-dependent operator that acts onto the colour and spin indices of the external QCD partons of the scattering amplitude vectors
$|  {\cal M}\bigl( \{ q_i \}; {\tilde k} \bigr) \rangle$ and 
$\langle  {\cal M}\bigl( \{ q_i \}; {\tilde k} \bigr) |$ (see eq.~(\ref{colfact})).
The parton $c$ in ${\bf F}_{ca}(\{q_i\};z;zp,\mb{q_T};n)$ carries the 
transverse momentum $\mb{q_T}$ and the fraction $z$ (with $0 \leq z \leq 1$) of the momentum $p$ of the initial-state parton $a$.

Using the SL collinear function $\bcf_{ca}$ we also introduce the partonic beam function ${\bcb}_{ca}$ as follows
\begin{align}
\label{slbeamgen}
 {\bcb}_{ca}(\{q_i\};z;zp,t;n) =& \bun \;
\delta(1-z) \;\delta(t) \;\, \delta_{ca}
%\;\bun 
\nn \\
&+ z
\int d^dk \;\delta(t - 2zpk) 
\;\,\delta\!\left(\frac{k^{+}}{p^{+}}-1+ z\right)
\;\bcf_{ca}(\{q_i\};p,k;n),
\end{align}
where the kinematical variable $t$ denotes the transverse virtuality and $z$
(with $0 \leq z \leq 1$) is the longitudinal-momentum fraction that is transferred by the initial-state parton $a$ to the colliding parton $c$. Similarly to
the SL TMD function in eq.~(\ref{sltmdgen}), the beam function ${\bcb}_{ca}$ is, in general, a process-dependent operator in colour and spin space.

The SL functions ${\bf F}_{ca}$ and $\bcb_{ca}$ are relevant in the context of transverse-momentum resummation and $N$-jettiness resummation for general hard-scattering processes, respectively. The process-dependent features of 
${\bf F}_{ca}$ and $\bcb_{ca}$ directly follow from the corresponding features of the SL collinear function $\bcf_{ca}$, which have been discussed in 
section~\ref{sec:diffcollfun}.

In particular, considering either computations up to ${\cal O}(\as^2)$
or computations (at arbitrary orders in $\as$) for processes with two hard-scattering partons,
${\bf F}_{ca}$ and $\bcb_{ca}$ are proportional to the unit matrix in colour space.
In these cases the trivial colour space dependence can be factorized with respect to 
$c$-number TMD functions $F_{ca}$ and beam functions $\cb_{ca}$ for the SL collinear region.
In the case of TMD functions we can deal with the gluon functions
\begin{equation}
\label{sltmdg}
F^{\, \mu \nu}_{ga}(z;zp,\mb{q_T};n) \;, 
\;\; 
F_{ga,\,{\rm az.in.}}\!\!\left(z;\mb{q_T}^2,\frac{n^2 \mb{q_T}^2}{(2zpn)^2}\right) 
\;,
\;\;
F_{ga,\,{\rm corr.}}\!\!\left(z;\mb{q_T}^2,\frac{n^2 \mb{q_T}^2}{(2zpn)^2}\right) 
\;,
\end{equation}
and the quark (or antiquark) functions 
\begin{equation}
\label{sltmdq}
F_{ca}\!\left(z;\mb{q_T}^2,\frac{n^2 \mb{q_T}^2}{(2zpn)^2}\right), \quad (c=q,{\bar q}),
\end{equation}
which are the SL analogue of the TL TMD functions in 
eqs.~(\ref{tlgtmd})--(\ref{tlqtmd}).
In the case of the gluon beam functions, the integration over $\mb{k_T}$ in 
eq.~(\ref{slbeamgen}) cancels the contribution of the azimuthal-correlation component of 
$\cf^{\, \mu \nu}_{ga}(p,k;n)$ (i.e., $\cb_{ga}^{\mu \nu} = d^{\mu \nu} \cb_{ga}$),
and we can simply deal with $c$-number scalar functions
\begin{equation}
\label{beamgqqbar}
\cb_{ca}\!\left(z;t,\frac{n^2 \,t}{(2zpn)^2}\right), \quad (c=g,q,{\bar q}),
\end{equation}
for both the gluon and quark (or antiquark) partonic channels.

In the following sections we discuss the perturbative computation of the functions in eqs.~(\ref{sltmdg}), (\ref{sltmdq}), and (\ref{beamgqqbar}). After appropriate factorization of IR divergences, these TMD functions lead to the SL collinear coefficients $C_{qa}$ and 
$C^{\mu \nu}_{ga}$ in eqs.~(\ref{eq:colq}) and (\ref{eq:colg}),
while the partonic beam functions lead to the matching coefficients in 
eqs.~(\ref{eq:beam}) and (\ref{matchexp}).

Setting $n^2=0$, our TMD functions (see eqs.~(\ref{tlgtmd}), (\ref{tlqtmd}) and 
(\ref{sltmdgen})) and beam functions (see eq.~(\ref{slbeamgen}))
are related to the partonic matrix elements of corresponding SCET function operators. 
Such relation between collinear splitting kernels 
%with $n^2=0$
and SCET functions was already observed
in ref.~\cite{Ritzmann:2014mka}, and it is also used in recent high-order perturbative computations \cite{Melnikov:2018jxb, Melnikov:2019pdm, Behring:2019quf, Baranowski:2020xlp,
Ebert:2020lxs}. In particular (see the related comments on differential collinear functions
at the end of section~\ref{sec:diffcollfun}), our collinear functions and the partonic SCET functions are perturbatively equivalent in the TL region.
% while the equivalence in the SL region is limited up to ${\cal O}(\as^2)$.
In the SL region the equivalence is, in general, limited up to ${\cal O}(\as^2)$.
Such limitation follows from the general process dependence of the collinear splitting kernels $\bcp$ beyond ${\cal O}(\as^2)$. Our SL collinear functions directly acquire
the general process dependence of the collinear splitting kernels. The SL partonic SCET functions instead have a process-independent form, and they are directly related to the collinear limit and corresponding splitting kernels of the QCD scattering amplitudes 
for a specific class of processes, namely, the production processes of high-mass colourless systems (see section~\ref{sec:qcdres}). Therefore, the all-order equivalence between our collinear functions and the partonic SCET functions in the SL region is specific for this class of processes.

\section{SL collinear functions: IR factorization and perturbative results}
\label{sec:sltmd}

In this section and in the following section we discuss the perturbative calculation of the SL collinear functions introduced in sections~\ref{sec:diffcollfun} and \ref{sec:intcolfun}.
We consider explicit computations up to ${\cal O}(\as^2)$ and, hence, we simply refer to the process-independent $c$-number functions of eq.~(\ref{slcfun}) and 
eqs.~(\ref{sltmdg})--(\ref{beamgqqbar}).

\subsection{Differential collinear functions at ${\cal O}(\as)$}
\label{sec:sldiff}

The perturbative expansion of the collinear functions in eq.~(\ref{slcfun}) can be written as follows
\begin{equation}
\label{fexp12}
\cf(p,k;n) = \cf^{(1R)}(p,k;n) + 
\left[ \; \cf^{(2R)}(p,k;n) + \cf^{(1R1V)}(p,k;n) \; \right] 
+ {\cal O}(\as^3).
\end{equation}
The notation in eq.~(\ref{fexp12}) applies to any specific collinear function and, therefore, we have not explicitly denoted the corresponding subscripts and superscripts in $\cf$. The terms in the right-hand side of eq.~(\ref{fexp12})
are directly related to the loop expansion of the collinear splitting kernels
$\bcp$ (see eq.~(\ref{cpexp})). The contribution to $\cf$ at ${\cal O}(\as)$ is due to
$\cf^{(1R)}$, which corresponds to single real emission in the final state at the tree-level, and it is obtained by using the tree-level kernel $\bcp^{(0)}_{c \to a_1a_2}$
in the right-hand side of eq.~(\ref{slcolfun}). The contributions to $\cf$ at
${\cal O}(\as^2)$ are $\cf^{(2R)}$ (double real emission at the tree level)
and $\cf^{(1R1V)}$ (single real emission with one-loop virtual corrections).
The terms $\cf^{(2R)}$ and $\cf^{(1R1V)}$ are obtained from eq.~(\ref{slcolfun})
by using the tree-level kernel $\bcp^{(0)}_{c \to a_1a_2a_3}$ and the one-loop
kernel $\bcp^{(1)}_{c \to a_1a_2}$, respectively.

We express the perturbative contributions in eq.~(\ref{fexp12}) in terms of the unrenormalized (bare) QCD coupling $\alpha^u_{\rm S}$, which is related as follows to 
$\as(\mu_R^2)$, the renormalized coupling at the scale $\mu_R$ in the
$\overline{{\rm MS}}$ renormalization scheme:
\begin{align}
\label{eq:alphasren}
    \alpha^u_{\rm S} \,\mu_0^{2\eps} \,S_{\eps} = \as(\mu_R^2) \,\mu_R^{2\eps}\left[1-\frac{\as(\mu_R^2)}{\pi}\frac{\beta_0}{\eps} +\mc{O}\left(\as^2(\mu_R^2)\right)   \right],
\end{align}
where $\beta_0= (11 C_A - 2 N_f)/12$ and $N_f$ is the number of massless quark flavours. 
% $\beta_0 $ is the first perturbative coefficient of the QCD $\beta$ function.
The $d$-dimensional spherical factor $S_{\eps}$ is
$S_{\eps}=\left(4 \pi \,e^{-\gamma_E} \right)^{\eps}$ and $\gamma_E$ is the Euler number
($\gamma_E=0.5772\dots$).

The definition in eq.~(\ref{slcolfun}) leads to a direct relation between 
$\cf^{(1R)}$ and the well-known expressions of the collinear kernels 
$\bcp^{(0)}_{c \to a_1a_2}$ (see, e.g., refs.~\cite{Catani:1999ss,Catani:1996vz}).
In the case of the azimuthally-independent functions 
$\cf^{(1R)}_{ca, \,\text{az.in.}}$, the collinear kernels are proportional to ${\widehat P}_{ca}(x; \ep)$, which are the $d$-dimensional real emission contributions to the Altarelli--Parisi splitting functions for the LO evolution of the PDFs. We have
\begin{equation}
\label{f1R}
\cf^{(1R)}_{ca, \,\text{az.in.}}(p,k;n) = \frac{\alpha^u_{\rm S} \,\mu_0^{2\eps}
\,S_{\eps}}{\pi} \;\frac{e^{\eps\gamma_E}}{\pi^{1-\eps}} 
\;\frac{\delta_{+}(k^2)}{pk} \;\frac{1}{z_n} \;{\widehat P}_{ca}(z_n; \ep),
\;\;\; \quad c=g,q,{\bar q},
\end{equation}
where we have introduced the notation 
$\cf^{(1R)}_{ca} \equiv \cf^{(1R)}_{ca, \,\text{az.in.}}$ ($c=q,{\bar q}$) for the 
collinear functions in the quark and antiquark partonic channels.
The azimuthal-correlation contribution in the gluon channel 
is
%\begin{align}
%\label{fggcor}
%&\mathcal{F}^{(1R)}_{gg, \,\text{corr.}}(p,k;n) =- \;\frac{\alpha^u_{\rm S} \,\mu_0^{2\eps}
%\,S_{\eps}}{\pi} \;\frac{e^{\eps\gamma_E}}{\pi^{1-\eps}}
%\;\frac{\delta_{+}(k^2) \,C_A}{pk} \;\frac{1-z_n}{z_n^2}\,,
%\\
%\label{fgqcor}
%&\mathcal{F}^{(1R)}_{ga, \,\text{corr.}}(p,k;n) =- \;\frac{\alpha^u_{\rm S} \,\mu_0^{2\eps} %\,S_{\eps}}{\pi} \;\frac{e^{\eps\gamma_E}}{\pi^{1-\eps}} 
%\;\frac{\delta_{+}(k^2) \,C_F}{pk} \;\frac{1-z_n}{z_n^2}\,, \quad\quad a=q,{\bar q}
%\;\;.
%\end{align}
\begin{equation}
\label{fgacor}
\mathcal{F}^{(1R)}_{ga, \,\text{corr.}}(p,k;n) =- \;\frac{\alpha^u_{\rm S} \,\mu_0^{2\eps} \,S_{\eps}}{\pi} \;\frac{e^{\eps\gamma_E}}{\pi^{1-\eps}} 
\;\frac{\delta_{+}(k^2)}{pk} \;C_a \;\frac{1-z_n}{z_n^2} 
%\quad\quad a=q,{\bar q},g
,
\end{equation}
where $C_a$ is the Casimir colour coefficient of the parton $a=q,{\bar q},g$, as in eq.~(\ref{eq:g1ga}). 
The expressions of $\mathcal{F}^{(1R)}$ in eqs.~(\ref{f1R}) and (\ref{fgacor})
depend on the auxiliary vector $n^\mu$ through the variable $z_n$,
\begin{equation}
\label{znvar}
z_n = \frac{n (p-k) }{np}.
\end{equation}
In the exact collinear limit (i.e., $k^- = 0$) the parent hard-scattering parton $c$ in $\mathcal{F}^{(1R)}_{ca}$ carries the momentum $z_n p^\mu$, independently of the value of $n^2$.

The explicit expressions of the real emission kernels ${\widehat P}_{ca}(x; \ep)$ are
\begin{align}
&{\widehat P}_{qq}(x; \ep) = \frac{1}{2} \;C_F 
\left[ \frac{1+x^2}{1-x} - \ep (1-x)\right],
\label{rqq}\\
&{\widehat P}_{qg}(x; \ep) =\frac{1}{2} \;T_R 
\left[ 1- \frac{2x(1-x)}{1-\ep}\right],
\label{rqg}\\
&{\widehat P}_{gg}(x; \ep) = C_A 
\left[ \frac{x}{1-x} + \frac{1- x}{x} + x(1-x) \right],
\label{rgg}\\
&{\widehat P}_{gq}(x; \ep) = \frac{1}{2} \;C_F 
\left[ \frac{1+ (1-x)^2}{x} - \ep x\right],
\label{rgq}\\
&{\widehat P}_{q{\bar q}}(x; \ep) = {\widehat P}_{q{q}^\prime}(x; \ep)
= {\widehat P}_{q{\bar q}^\prime}(x; \ep) = 0,
\label{rqqprime}
\end{align}
where $q$ and $q^\prime$ denote quarks with different flavour.
The expressions of ${\widehat P}_{ca}$ for the remaining partonic channels are obtained by using the relation ${\widehat P}_{ca}(x; \ep)= 
{\widehat P}_{{\bar c}{\bar a}}(x; \ep)$, which follows from charge conjugation invariance. For subsequent use we also introduce the decomposition of 
${\widehat P}_{ca}$ in their singular (${\widehat P}^{\rm \,sing}_{ca}$) and regular
(${\widehat P}^{\rm \,reg}_{ca}$) parts around the point $x=1$. We write
\begin{align}
\label{psr}
{\widehat P}_{ca}(x; \ep) =& {\widehat P}^{\rm \,sing}_{ca}(x)
+ {\widehat P}^{\rm \,reg}_{ca}(x; \ep),
\\
\label{psin}
{\widehat P}^{\rm \,sing}_{ca}(x) =& \delta_{ca} \;\frac{A_c^{(1)}}{1-x},
%\quad {\rm with} \;\; A_q^{(1)}=A_{\bar q}^{(1)}= C_F \;, \;A_g^{(1)} = C_A \;\;,
\\
\label{preg}
{\widehat P}^{\rm \,reg}_{ca}(x; \ep) =& {\widehat P}^{\rm \,reg}_{ca}(x; \ep=0)
+ \ep \;{\widehat P}^{\prime}_{ca}(x; \ep).
\end{align}
The coefficients $A_c^{(1)}$ ($c=q, {\bar q}, g$) in eq.~(\ref{psin}) measure the intensity of soft and collinear gluon radiation from the parton $c$, and we have
\begin{equation}
\label{a1coef}
A_q^{(1)}=A_{\bar q}^{(1)}= C_F, \quad \quad \;A_g^{(1)} = C_A.
\end{equation}
The explicit expressions of ${\widehat P}^{\rm \,reg}_{ca}(x; \ep=0)$ and
${\widehat P}^{\prime}_{ca}(x; \ep)$ are readily obtained by direct comparison of 
eqs.~(\ref{rqq})--(\ref{rqqprime}) and eqs.~(\ref{psr})--(\ref{preg}).

As discussed at the end of section~\ref{sec:diffcollfun},
we can relate the collinear functions $\cf_{ca}(p,k;n)$ to the SCET differential beam functions   
\cite{Jain:2011iu, Mantry:2009qz, Mantry:2010mk}. More precisely, setting $n^2=0$
and introducing the variables $\mb{q_T}= \mb{k_T}$ and $t=2zpk$, the first-order collinear function $\mathcal{F}^{(1R)}_{ca}(p,k;n)$ is equal to the first-order contribution to the partonic differential beam function ${\cal B}_{ca}(z,t,\mb{q_T})$ of ref.~\cite{Jain:2011iu}.
The expressions in eqs.~(\ref{f1R}) and (\ref{fgacor})
agree with the first-order results presented in ref.~\cite{Jain:2011iu}.

\subsection{TMD functions}
\label{sec:tmdfun}

The SL TMD functions $F_{ca}$ are obtained from the differential collinear functions
$\cf_{ca}$ by using eq.~(\ref{sltmdgen}). The contribution of ${\cal O}(\as)$ to 
$F_{ca}$ is denoted by $F_{ca}^{(1R)}$, and it is obtained from the corresponding contribution
$\cf_{ca}^{(1R)}$ to the differential collinear functions. At ${\cal O}(\as^2)$, the terms
$\cf_{ca}^{(2R)}$ and $\cf_{ca}^{(1R1V)}$ in eq.~(\ref{fexp12}) produce corresponding contributions to the TMD functions that are denoted as
$F_{ca}^{(2R)}$ and $F_{ca}^{(1R1V)}$, respectively.

The explicit expressions of $F_{ca}^{(1R)}$ are obtained through eq.~(\ref{sltmdgen})
by using the corresponding expressions of $\cf_{ca}^{(1R)}(p,k;n)$ in 
eqs.~(\ref{f1R}) and (\ref{fgacor}). At this perturbative order the integration over the momentum $k$ in 
eq.~(\ref{sltmdgen}) is trivial, and $F_{ca}^{(1R)}$ turns out to be proportional to the overall factor $1/\mathbf{q_T}^2$ times a function of $z$ and $\mathbf{q_T}^2$. This functional dependence on $z$ and $\mathbf{q_T}^2$ is due to the corresponding dependence on $z_n$ in eqs.~(\ref{f1R}) and (\ref{fgacor}). The overall factor $1/\mathbf{q_T}^2$ is singular in the limit
$\mb{q_T} \to 0$, and it corresponds to the singular behaviour of QCD cross sections that we can consistently compute by exploiting the collinear factorization formula 
(\ref{colfact}). Therefore, considering the residual dependence of $F_{ca}^{(1R)}$
on $z_n$ we can set $\mb{q_T} = 0$, provided the limit $\mb{q_T} \to 0$ is smooth and non-singular. 

We immediately discuss the dependence on the auxiliary vector $n^\mu$, which affects 
$F_{ca}^{(1R)}$ through the variable $z_n$ (see eq.~(\ref{znvar})). The key point regards the effect of the singular contribution ${\widehat P}^{\rm \,sing}_{ca}(z_n)$
(see eqs.~(\ref{psr})--(\ref{preg})) to $\cf_{ca}^{(1R)}$ and, hence, to $F_{ca}^{(1R)}$.
Such contribution is proportional to the following factor:
\begin{equation}
\label{znsing}
\frac{1}{1-z_n} = \frac{np}{nk} = \frac{p^+}{k^+ + \frac{n^2}{2 np}  \frac{k^-}{n^-} p^+}
= \frac{1}{1-z +\frac{n^2 \mathbf{q_T}^2}{(1-z) (2 np)^2}},
\end{equation}
where in the last equality we have implemented the kinematics of the TMD collinear function at ${\cal O}(\as)$ (i.e., $k^2=0, k^+=(1-z)p^+, \mb{k_T}= - \mb{q_T}$).
Setting $n^2=0$, the factor in eq.~(\ref{znsing}) becomes $(1-z)^{-1}$ and, therefore, it is divergent (and not integrable over $z$) at $z=1$. Correspondingly, the first-order contributions $F_{ca}^{(1R)}$ to the TMD collinear functions are divergent.
Such divergences, which are known as 
rapidity divergences 
\cite{Collins:2008ht, Collins:2011zzd, Becher:2010tm, Echevarria:2011epo, Chiu:2012ir} in the literature, are a general feature of SCET formulations of TMD functions, and they can be treated by introducing appropriate regularization 
procedures \cite{Collins:2011zzd, Chiu:2011qc, Becher:2011dz, Echevarria:2015usa, 
Li:2016axz, Ebert:2018gsn}. In the following computations
of the TMD functions we use $n^2 > 0$, thus
avoiding rapidity divergences. Further comments
on the origin of the rapidity divergences are postponed 
to the final part of this subsection.

Using a time-like auxiliary vector $n^\mu$, the factor in eq.~(\ref{znsing}) is not divergent, and it can be further approximated in the limit $\mb{q_T} \to 0$. Indeed,
setting $\lambda = n^2 \mb{q_T}^2/{(2pn)^2}$ in eq.~(\ref{znsing}), we can use the following relation:
\begin{align}
\label{tmdplus}
\frac{1}{1-z +\frac{\lambda}{1-z}} &= \left( \frac{1-z}{(1-z)^2 +\lambda} \right)_+
+\delta(1-z) \int_0^1 dz^\prime \frac{1-z^\prime}{(1-z^\prime)^2 +\lambda} \nn \\
&= \frac{1}{2} \;\ln\left(\frac{1}{\lambda}\right) \;\delta(1-z) 
+ \left( \frac{1}{1-z} \right)_+ + {\cal O}({\sqrt \lambda}),
\end{align}
where the symbol $\bigl( f(z) \bigr)_+$ denotes the customary `plus-distribution'
of the function $f(z)$ with respect to the variable $z$.
The term of ${\cal O}({\sqrt \lambda}) \sim {\cal O}(\mb{q_T})$ in eq.~(\ref{tmdplus})
smoothly vanishes in the limit $\mb{q_T} \to 0$ and, therefore, it can be neglected in the computation of $F_{ca}^{(1R)}$. We can similarly neglect other smooth terms 
in the limit $\mb{q_T} \to 0$ by using $z_n= z + {\cal O}(\mb{q_T}^2)$ in the remaining
$z_n$ dependence of $F_{ca}^{(1R)}$.

In summary, we obtain the following first-order results for the azimuthal-independent component of the TMD collinear functions:
\begin{align}
\label{tmdqt}
&F^{(1R)}_{ca,\,{\rm az.in.}}\!\!\left(z;\mb{q_T}^2,\frac{n^2 \mb{q_T}^2}{(2zpn)^2}\right) 
= \frac{\alpha^u_{\rm S} \,\mu_0^{2\eps}
\,S_{\eps}}{\pi} \;\frac{e^{\eps\gamma_E}}{\pi^{1-\eps} \,\mathbf{q_T}^2}  \\
& \quad \times \left\{{\widehat P}^{\rm \,reg}_{ca}(z; \ep) + \delta_{ca} \;A_c^{(1)}
\left[ \left( \frac{1}{1-z} \right)_+ 
- \frac{1}{2} \;\ln\left( \frac{n^2 \mb{q_T}^2}{(2zpn)^2} \right) \,\delta(1-z) 
\right] \right\},  \;\;\;\;\;\; c=g,q,{\bar q},\nn
\end{align}
where, similar to eq.~(\ref{f1R}), we have introduced the notation 
$F^{(1R)}_{ca} \equiv F^{(1R)}_{ca, \,\text{az.in.}}$ ($c=q,{\bar q}$) for the TMD
collinear functions in the quark and antiquark partonic channels.
The first-order result for the azimuthal-correlation components in the gluon channel is
%\begin{align}
%\label{tmdgcor}
%&F^{(1R)}_{gg,\,{\rm corr.}}\!\!\left(z;\mb{q_T}^2,\frac{n^2 \mb{q_T}^2}{(2zpn)^2}\right)
% =- \;\frac{\alpha^u_{\rm S} \,\mu_0^{2\eps}
%\,S_{\eps}}{\pi} \;\frac{e^{\eps\gamma_E}}{\pi^{1-\eps} \,\mathbf{q_T}^2}
%\;C_A \frac{1-z}{z}\,,
%\\
%\label{tmdqcor}
%&F^{(1R)}_{ga,\,{\rm corr.}}\!\!\left(z;\mb{q_T}^2,\frac{n^2 \mb{q_T}^2}{(2zpn)^2}\right)
% =- \;\frac{\alpha^u_{\rm S} \,\mu_0^{2\eps} \,S_{\eps}}{\pi} \;\frac{e^{\eps\gamma_E}}{\pi^%{1-\eps} \,\mathbf{q_T}^2} 
%\; C_F\;\frac{1-z}{z}\,, \quad\quad a=q,{\bar q}
%\;\;.
%\end{align}
\begin{equation}
\label{tmdgacor}
F^{(1R)}_{ga,\,{\rm corr.}}\!\!\left(z;\mb{q_T}^2,\frac{n^2 \mb{q_T}^2}{(2zpn)^2}\right)
 =- \;\frac{\alpha^u_{\rm S} \,\mu_0^{2\eps} \,S_{\eps}}{\pi} \;\frac{e^{\eps\gamma_E}}{\pi^{1-\eps} \,\mathbf{q_T}^2} 
\; C_a\;\frac{1-z}{z}. 
%\quad\quad a=q,{\bar q},g \;\;.
\end{equation}
We note that the result in eq.~(\ref{tmdgacor}) is independent of the auxiliary vector 
$n^\mu$.
We recall that we have neglected terms of relative order $\mb{q_T}$ in the right-hand sides
of eqs.~(\ref{tmdqt}) and (\ref{tmdgacor}). Such terms are actually controlled by the parameter
$n^2 \mb{q_T}^2/{(2pn)^2}$. However, we remark on the fact that we are not performing approximations in the limit $n^2 \to 0$. In our framework the value of $n^2$ is arbitrary,
and it is not regarded as a small expansion parameter.

We introduce the Fourier transformation of the TMD collinear function $F_{ca}$ to the purpose of having a more direct relation with the discussion in section~\ref{sec:qcdres}
on the transverse-momentum resummation formalism in impact parameter space.
The Fourier transformation ${\widetilde F}_{ca}$ in $\mb{b}$ space of the TMD collinear function $F_{ca}$ for the quark and antiquark partonic channels is
\begin{equation}
\label{tmdqb}
{\widetilde F}_{ca}\!\left( z;\frac{\mb{b}^2}{b_0^2},\frac{n^2 b_0^2}{(2zpn)^2 \,\mb{b}^2}\right)
\equiv \int d^{d-2}\mb{q_T}\;e^{-i\mb{b}.\mb{q_T}} \;{F}_{ca}\!\left(z;\mb{q_T}^2,\frac{n^2 \mb{q_T}^2}{(2zpn)^2}\right), \quad c=q,{\bar q},
\end{equation}
where the impact parameter $\mb{b}$ is a $(d-2)$-dimensional vector. The numerical coefficient $b_0=2 e^{-\gamma_E}$ is conveniently and customarily introduced in $\mb{b}$ space resummation formulae (see, e.g., ref.~\cite{Catani:2013tia}).
We note that the $n^\mu$ dependence of ${\widetilde F}_{ca}$ ($c=q,{\bar q}$) occurs through
the variable  $n^2 (b_0^2/\mb{b}^2)/(2zpn)^2$. In the gluon partonic channel we introduce
the Fourier transformation ${\widetilde F}^{\mu \nu}_{ga}$ of the TMD tensor function 
${F}^{\mu \nu}_{ga}$, and we have
\begin{equation}
\label{tmdgb}
{\widetilde F}^{\mu \nu}_{ga}(z; zp, \mb{b};n)
\equiv \int d^{d-2}\mb{q_T}\;e^{-i\mb{b}.\mb{q_T}} 
\;{F}_{ga}^{\mu \nu}(z; zp, \mb{q_T};n).
\end{equation}
Analogously to the TMD gluon function in eq.~(\ref{tmdfincor}),
the $\mb{b}$ space tensor function ${\widetilde F}^{\mu \nu}_{ga}$ can be decomposed in its azimuthal-independent and azimuthal-correlation components, 
${\widetilde F}_{ga,\,{\rm az.in.}}$ and ${\widetilde F}_{ga,\,{\rm corr.}}$:
\begin{align}
\label{tmdgbaz}
{\widetilde F}^{\mu \nu}_{ga}(z; zp, \mb{b};n)  =& 
d^{\mu \nu}(p;n) 
\;\;{\widetilde F}_{ga,\,{\rm az.in.}}\!\!\left(z;\frac{\mb{b}^2}{b_0^2},
\frac{n^2 b_0^2}{(2zpn)^2 \,\mb{b}^2}\right) \nn \\
&+ D^{\mu\nu}(p,n;\mb{b},\ep) 
\;\;{\widetilde F}_{ga,\,{\rm corr.}}\!\!\left(z;\frac{\mb{b}^2}{b_0^2},
\frac{n^2 b_0^2}{(2zpn)^2 \,\mb{b}^2}\right).
\end{align}
Similarly to ${\widetilde F}_{ca}$ in eq.~(\ref{tmdqb}), the scalar functions
${\widetilde F}_{ga,\,{\rm az.in.}}$ and ${\widetilde F}_{ga,\,{\rm corr.}}$
depend on the auxiliary vector $n^\mu$ through
the variable  $n^2 (b_0^2/\mb{b}^2)/(2zpn)^2$.

The fixed-order perturbative contributions to the TMD functions ${F}_{ca}$
depend on $\mb{q_T}$ through powers and logarithms. To move from ${F}_{ca}$
to ${\widetilde F}_{ca}$ we have to perform the Fourier transformation of this type of functional dependence on $\mb{q_T}$. The required most general Fourier transformations are as follows
\begin{align}
\label{ftav}
\int d^{d-2}\mb{q_T}\,e^{-i\mb{b}.\mb{q_T}}\frac{\ln^m(\mb{q_T}^2)}{(\mb{q_T}^2)^{1+\delta}}
=&\pi^{1-\ep} \; \frac{d^m}{d\rho^m}\bigg{|}_{\rho=0} 
\left[ \left(\frac{\mb{b}^2}{4}\right)^{\ep+\delta-\rho}\frac{\Gam(\rho-\ep-\delta)}{\Gam(1+\delta-\rho)} \right],
\\
%\end{align}
%\begin{align}
\label{ftcor}
\int d^{d-2}\mb{q_T}\,e^{- i\mb{b}.\mb{q_T}}\frac{\ln^m(\mb{q_T}^2)}{(\mb{q_T}^2)^{1+\delta}} D^{\mu\nu}(p,n;\mb{q_T},\ep) 
=& - D^{\mu\nu}(p,n;\mb{b},\ep) \;\pi^{1-\ep} \nn 
\\
&\times \;\frac{d^m}{d\rho^m}\bigg{|}_{\rho=0}
\left[ \left(\frac{\mb{b}^2}{4}\right)^{\ep+\delta-\rho}\frac{\Gam(1+\rho-\ep-\delta)}{\Gam(2+\delta-\rho)} \right],
\end{align}
where eqs.~(\ref{ftav}) and (\ref{ftcor}) refer to azimuthal-independent and  
azimuthal-correlation contributions, respectively. The function $\Gam(x)$ denotes the Euler
$\Gamma$-function of the variable $x$. These basic Fourier transformations are sufficient to
go from $\mb{q_T}$ space to $\mb{b}$ space in the computation of the TMD functions at
arbitrary perturbative orders.

The perturbative expansion of the azimuthal-independent component of the TMD collinear functions in $\mb{b}$ space is as follows
\begin{align}
\label{bspaceexp}
{\widetilde F}_{ca, \,{\rm az.in.}}\!\left( z;\frac{\mb{b}^2}{b_0^2},\frac{n^2 b_0^2}{(2zpn)^2 \,\mb{b}^2}\right) = \delta_{ca} \;\delta(1-z) 
+ {\widetilde F}^{(1R)}_{ca, \,{\rm az.in.}}\!\left( z;\frac{\mb{b}^2}{b_0^2},\frac{n^2 b_0^2}{(2zpn)^2 \,\mb{b}^2}\right)  + {\cal O}(\as^2),
\nn\\
c=g,q,{\bar q},
\end{align}
where, analogously with the notation in eq.~(\ref{tmdqt}) we have defined 
${\widetilde F}_{ca} \equiv {\widetilde F}_{ca, \,{\rm az.in.}}$ for $c=q,{\bar q}$.
The first-order term ${\widetilde F}^{(1R)}_{ca, \,{\rm az.in.}}$ is obtained from 
eq.~(\ref{tmdqt}) by using the Fourier transformations in eq.~(\ref{ftav}) with
$\delta=0$ and $m=0,1$. We obtain the following result:
\begin{align}
&{\widetilde F}^{(1R)}_{ca,\,{\rm az.in.}}\!\left(z;\frac{\mb{b}^2}{b_0^2},
\frac{n^2 b_0^2}{(2zpn)^2 \,\mb{b}^2}\right) = \frac{\alpha^u_{\rm S}\,S_{\eps}}{\pi} 
\;\left( \frac{\mu_0^2 \,\mb{b}^2}{b_0^2} \right)^{+\ep}
\;\frac{e^{- \eps\gamma_E} \, \Gamma(1-\ep)}{(-\eps)} 
\Bigg\{ {P}^{(1)}_{ca}(z) + \ep \;{\widehat P}^{\prime}_{ca}(z; \ep)
\nn\\
\label{azin1b}
&- \delta_{ca}\,\delta(1-z) \left[ 
\frac{A_c^{(1)}}{2} \left( \frac{1}{\ep} + \psi(1-\ep) - \psi(1) 
+ \ln \frac{n^2 b_0^2}{(2zpn)^2 \,\mb{b}^2}  \right) + \frac{\gamma_c}{2}\right]
\Bigg\}, \quad c=g,q,{\bar q},
\end{align}
where $\psi(x)= d \ln \Gamma(x)/dx$ is
the Euler $\psi$-function, and we have introduced the lowest-order Altarelli--Parisi kernel ${P}^{(1)}_{ca}(x)$ for the evolution of the PDFs:
\begin{equation}
\label{AP1}
{P}^{(1)}_{ca}(x) = {\widehat P}^{\rm \,reg}_{ca}(x; \ep=0) + \delta_{ca}
\left[ A_c^{(1)} \left(\frac{1}{1-x}\right)_+ + \frac{1}{2} \;\gamma_c \;\delta(1-x) \right].
\end{equation}
The functions ${\widehat P}^{\rm \,reg}_{ca}$ and 
${\widehat P}^{\prime}_{ca}$ are given in eqs.~(\ref{psr})--(\ref{preg})
and the coefficients $\gamma_c \;(c=q,{\bar q},g)$ are
\begin{equation}
\label{gacoef}
\gamma_q =\gamma_{\bar q}= \frac{3}{2} C_F, \quad \quad \;\gamma_g= \frac{1}{6}
\left( 11 C_A - 2 N_f \right).
\end{equation}
The azimuthal-correlation component ${\widetilde F}_{ga,\,{\rm corr.}}$ of the gluon 
TMD function ${\widetilde F}^{\mu \nu}_{ga}$ in eq.~(\ref{tmdgbaz}) has the following perturbative expansion:
\begin{align}
&{\widetilde F}_{ga,\,{\rm corr.}}\!\!\left(z;\frac{\mb{b}^2}{b_0^2},
\frac{n^2 b_0^2}{(2zpn)^2 \,\mb{b}^2}\right) = 
{\widetilde F}^{(1R)}_{ga,\,{\rm corr.}}\!\!\left(z;\frac{\mb{b}^2}{b_0^2},
\frac{n^2 b_0^2}{(2zpn)^2 \,\mb{b}^2}\right) \nn \\
& \quad \quad + 
\left[ {\widetilde F}^{(2R)}_{ga,\,{\rm corr.}}\!\!\left(z;\frac{\mb{b}^2}{b_0^2},
\frac{n^2 b_0^2}{(2zpn)^2 \,\mb{b}^2}\right) + 
{\widetilde F}^{(1R1V)}_{ga,\,{\rm corr.}}\!\!\left(z;\frac{\mb{b}^2}{b_0^2},
\frac{n^2 b_0^2}{(2zpn)^2 \,\mb{b}^2}\right) \right]
+ {\cal O}(\as^3).
\label{bspacecorr}
\end{align}
The first-order term ${\widetilde F}^{(1R)}_{ga,\,{\rm corr.}}$ is obtained from 
eq.~(\ref{tmdgacor}) by using the Fourier transformation in 
eq.~(\ref{ftcor}) with $\delta=0$ and $m=0$. The result is
\begin{equation}
\label{1corb}
{\widetilde F}^{(1R)}_{ga,\,{\rm corr.}}\!\!\left(z;\frac{\mb{b}^2}{b_0^2},
\frac{n^2 b_0^2}{(2zpn)^2 \,\mb{b}^2}\right) =
\frac{\alpha^u_{\rm S}\,S_{\eps}}{\pi} 
\;\left( \frac{\mu_0^2 \,\mb{b}^2}{b_0^2} \right)^{\!+\ep}
\;e^{- \eps\gamma_E} \, \Gamma(1-\ep) \;C_a \;\frac{1-z}{z}.
\end{equation}
The results in eqs.~(\ref{azin1b}) and (\ref{1corb}) are valid in arbitrary $d=4-2\ep$ space-time dimensions. Considering the physical four-dimensional limit $\ep \to 0$, we see
that the azimuthal-correlation components are finite while the azimuthal-independent terms are divergent. The divergences are due to double and single poles, $1/\ep^2$ and $1/\ep$,
and they have an IR (soft and collinear) origin.

More generally, the perturbative computation at ${\cal O}(\as^n)$
of the TMD functions in $\mb{b}$ space leads to divergent pole terms $1/\ep^m$ with
$1 \leq m \leq 2n$. These divergences are of UV and IR origin. The UV divergences are removed by using eq.~(\ref{eq:alphasren}) which is used
to express the bare coupling $\alpha^u_{\rm S}$ in terms of the renormalized coupling
$\as(\mu_R^2)$. The IR divergences are then factorizable.

The TMD functions in $\mb{b}$ space fulfil the following IR factorization formulae:
\begin{align}
\label{IRfactav}
{\widetilde F}_{ca,\,{\rm az.in.}}\!\!\left(\!z;\frac{\mb{b}^2}{b_0^2},
\frac{n^2 b_0^2}{(2zpn)^2 \,\mb{b}^2}\!\right) &= 
Z_c\!\left(\!\as(b_0^2/\mb{b}^2), \frac{n^2 b_0^2}{(2zpn)^2 \,\mb{b}^2}\!\right) \\
&\times \sum_b \int_z^1 \frac{dx}{x}
{\widetilde C}_{cb}\left(\!x;\as(b_0^2/\mb{b}^2), \ep, \frac{n^2 b_0^2}{(2zpn)^2 \,\mb{b}^2}\!\right){\widetilde \Gamma}_{ba}(z/x;b_0^2/\mb{b}^2),
%\end{align}
%\begin{align}
\nn\\
\label{IRfactcor}
{\widetilde F}_{ga,\,{\rm corr.}}\!\!\left(\!z;\frac{\mb{b}^2}{b_0^2},
\frac{n^2 b_0^2}{(2zpn)^2 \,\mb{b}^2}\!\right) &= 
Z_g\!\left(\!\as(b_0^2/\mb{b}^2), \frac{n^2 b_0^2}{(2zpn)^2 \,\mb{b}^2}\!\right)\\
&\times \sum_b \int_z^1 \frac{dx}{x}
{\widetilde G}_{gb}\!\left(\!x;\as(b_0^2/\mb{b}^2), \ep, \frac{n^2 b_0^2}{(2zpn)^2 \,\mb{b}^2}\!\right){\widetilde \Gamma}_{ba}(z/x;b_0^2/\mb{b}^2).\nn
\end{align}
Note that in the right-hand side of eqs.~(\ref{IRfactav}) and (\ref{IRfactcor}) we use the renormalization scale
$\mu_R^2=b_0^2/\mb{b}^2$. Therefore, the various functions $Z_c, {\widetilde \Gamma}_{ba},
{\widetilde C}_{cb}$ and ${\widetilde G}_{gb}$ depend on $\as(b_0^2/\mb{b}^2)$.
A generic renormalization scale $\mu_R$ can be introduced in a straightforward way by expressing $\as(b_0^2/\mb{b}^2)$ in terms of $\as(\mu_R^2), \ln(b_0^2 \mu_R^2/\mb{b}^2)$
and $\ep$ (see eq.~(\ref{eq:alphasren})).

The factor ${\widetilde \Gamma}_{ba}(x;\mu_F^2)$ in the right-hand side of 
eqs.~(\ref{IRfactav}) and (\ref{IRfactcor}) is the customary collinear-divergent function
that defines the scale-dependent PDF $f_b(z;\mu_F^2)$ in the $\overline{{\rm MS}}$
factorization scheme. We have
\begin{equation}
\label{pdfren}
f_b(z;\mu_F^2) = \sum_a \int_z^1 \frac{dx}{x} \;{\widetilde \Gamma}_{ba}(z/x;\mu_F^2) \;
f^{(0)}_a(x),
\end{equation}
where $f^{(0)}_a(x)$ is the bare (scale-independent) PDF of the parton $a$.
The perturbative expansion of ${\widetilde \Gamma}_{ba}$ is
\begin{equation}
\label{wgamexp}
{\widetilde \Gamma}_{ba}(z;\mu_F^2) = \delta_{ba} \;\delta(1-z) 
- \frac{\as(\mu^2_F)}{\pi} \;\frac{{P}^{(1)}_{ba}(z)}{\ep} + {\cal O}(\as^2),
\end{equation}
where ${P}^{(1)}_{ba}(z)$ is the lowest-order Altarelli--Parisi kernel in 
eq.~(\ref{AP1}).

After factorization of the collinear-divergent $\ep$ poles embodied in 
${\widetilde \Gamma}_{ba}$,
the $\mb{b}$ space TMD functions still contain IR divergences that are factorizable in the
perturbative functions $Z_c$ of 
eqs.~(\ref{IRfactav}) and (\ref{IRfactcor}). The functions ${\widetilde C}_{cb}$ and ${\widetilde G}_{gb}$ are then finite and independent of $n^2$ 
(i.e., $n^2 (b_0^2/\mb{b}^2)/(2zpn)^2$) in the limit $\ep \to 0$, order-by-order in the perturbative expansion in powers of $\as(b_0^2/\mb{b}^2)$.

We note some main features of the IR factor $Z_c$ of ${\widetilde F}_{ca}$ in 
eqs.~(\ref{IRfactav}) and (\ref{IRfactcor}). The factor $Z_c$ depends on the hard-scattering parton $c$ of ${\widetilde F}_{ca}$, and it is independent of the initial-state parton $a$.
The dependence of $Z_c$ on the parton $c$ occurs through the perturbative coefficients
(see eq.~(\ref{z1tmd})) and the momentum $zp$ (i.e., $n^2 (b_0^2/\mb{b}^2)/(2zpn)^2$), which is the longitudinal momentum transferred to the parton $c$ by the collinear emission.
The function $Z_c$ has no additional dependence on the momentum fraction $z$. We remark on the fact that in the gluon channel, the IR factor $Z_g$ is the {\it same} for both the azimuthal-independent and azimuthal-correlation components
(i.e., $Z_g$ is the overall IR factor of the tensor ${\widetilde F}^{\mu \nu}_{ga}$ in 
eq.~(\ref{tmdgbaz})).
The factor $Z_c$ embodies $\ep$ poles and also possible IR finite contributions that depend on $n^2$. Moreover, $Z_c$ can include arbitrary IR finite terms that are independent of 
$n^2$ (see eq.~(\ref{z1tmd})). Such arbitrariness corresponds to the IR factorization scheme dependence of eqs.~(\ref{IRfactav}) and (\ref{IRfactcor}), and it is related to the resummation-scheme dependence \cite{Catani:2000vq,Catani:2013tia} 
of the transverse-momentum resummation formalism of QCD cross sections.

Comparing the structure of eq.~(\ref{IRfactav}) with the results in 
eqs.~(\ref{bspaceexp}) and (\ref{azin1b}), we can immediately derive the explicit expression of $Z_c$ up to ${\cal O}(\!\as\!)$. Setting ${\widetilde \lambda}\!=\!n^2 (b_0^2/\mb{b}^2)/(2zpn)^2$,
we obtain
\begin{align}
Z_c(\as, {\widetilde \lambda} ) =& 1 + \frac{\as}{\pi} \;Z_c^{(1)}({\widetilde \lambda} )+ {\cal O}(\as^2),
\\
\label{z1tmd}
Z_c^{(1)}({\widetilde \lambda} ) =& \frac{1}{2} \left[ A^{(1)}_c \left( \frac{1}{\ep^2}
+ \frac{1}{\ep} \ln {\widetilde \lambda} \right) + \frac{1}{\ep} \,\gamma_c \right]
- \frac{\pi^2}{24} A^{(1)}_c + h_c^{(1)} + {\widetilde h}_c^{(1)}(\ep,{\widetilde \lambda} ),
\end{align}
where $h_c^{(1)}$ and ${\widetilde h}_c^{(1)}(\ep,{\widetilde \lambda} )$ are introduced to specify the scheme
dependence of $Z_c$ (we have $h_c^{(1)}= h_{\bar c}^{(1)}$ and ${\widetilde h}_c^{(1)} = 
{\widetilde h}_{\bar c}^{(1)}$ because of charge conjugation invariance). The function ${\widetilde h}_c^{(1)}(\ep,{\widetilde \lambda} )$
vanishes in the limit $\ep \to 0$. The coefficient $h_c^{(1)}$ is related to the 
resummation-scheme dependence (see eqs.~(\ref{eq:qscheme}) and (\ref{eq:gscheme})).
Setting $h_c^{(1)}=\pi^2  A^{(1)}_c/24$ corresponds to a `minimal' scheme in which 
$Z_c^{(1)}$ contains only $\ep$ pole contributions. The hard scheme 
\cite{Catani:2013tia} is instead obtained by setting $h_c^{(1)}=0$ (see eq.~(\ref{c1coeff})).

The IR finite function ${\widetilde C}_{ca}$ of the azimuthal-independent component
of the TMD function in eq.~(\ref{IRfactav}) has the following perturbative expansion:
\begin{equation}
{\widetilde C}_{ca}(z;\as,\ep,{\widetilde \lambda}) = \delta_{ca} \;\delta(1-z) +
\frac{\as}{\pi} \;{\widetilde C}_{ca}^{(1)}(z;\ep,{\widetilde \lambda})
+ \left(\frac{\as}{\pi}\right)^2 \;{\widetilde C}_{ca}^{(2)}(z;\ep,{\widetilde \lambda})
+ {\cal O}(\as^3),\,\, c=q,{\bar q},g.
\label{ctilde}
\end{equation}
The limit $\ep \to 0$ of this function is finite and independent of ${\widetilde \lambda}$.
The limit gives the collinear functions in eqs.~(\ref{eq:colq}) and (\ref{azavgcolg}),
namely ${\widetilde C}_{ca}^{(m)}(z;\ep=0,{\widetilde \lambda}) = C_{ca}^{(m)}(z)
\;\; (c=q,{\bar q},g)$. We obtain
\begin{equation}
\label{c1coeff}
{C}_{ca}^{(1)}(z) = - {\widehat P}^{\prime}_{ca}(z; \ep=0) - 
\delta_{ca}\;\delta(1-z) \;{h}_c^{(1)}, \quad \quad c=q,{\bar q},g,
\end{equation}
in agreement with the known results in the literature. In particular, as first derived in ref.~\cite{deFlorian:2001zd},
the non-trivial $z$ dependence of ${C}_{ca}^{(1)}(z)$ is due to 
${\widehat P}^{\prime}_{ca}(z; \ep=0)$ (see eqs.~(\ref{psr})--(\ref{preg})), which is the contribution of ${\cal O}(\ep)$ to the $d$-dimensional real emission kernel 
${\widehat P}_{ca}(z; \ep)$
in eqs.~(\ref{rqq})--(\ref{rqqprime}). In the hard scheme \cite{Catani:2013tia} 
${C}_{ca}^{(1)}(z)$ has no contributions proportional to $\delta(1-z)$ and, therefore, 
$h_c^{(1)}=0$. In explicit form and for a general scheme we have
\begin{align}
    C_{qq}^{(1)}(z) =& \frac{1}{2}C_F \;(1-z) - {h}_q^{(1)} \;\delta(1-z),
    \\
    C_{qg}^{(1)}(z) =& T_R \;z(1-z),
    \\
    C_{gg}^{(1)}(z) =& - {h}_g^{(1)} \;\delta(1-z),
    \\
    C_{gq}^{(1)}(z) =& \frac{1}{2}C_F \;z,
    \\
    C_{q\bar{q}}^{(1)}(z)=&C_{qq^\prime}^{(1)}(z)=C_{q\bar{q}^\prime}^{(1)}(z)=0,
\end{align}
and we recall that the expressions in the other partonic channels are obtained by the charge
conjugation relation $C_{ca}^{(1)}(z)= C_{{\bar c}{\bar a}}^{(1)}(z)$.
The complete $\ep$ dependence of ${\widetilde C}_{ca}^{(1)}$ is
\begin{align}
\label{c1all}
{\widetilde C}_{ca}^{(1)}(z;\ep,{\widetilde \lambda})&= 
- \,e^{- \eps\gamma_E} \, \Gamma(1-\ep) \;{\widehat P}^{\prime}_{ca}(z; \ep)
+ \frac{1 -e^{- \eps\gamma_E} \, \Gamma(1-\ep) }{\ep} {P}^{(1)}_{ca}(z)  \\
&
- \delta_{ca}\;\delta(1-z) \; \bigg\{\! 
 \frac{1 -e^{- \eps\gamma_E} \, \Gamma(1-\ep) }{2\ep}
\left[
A_c^{(1)} \left( \frac{1}{\ep} + \ln {\widetilde \lambda}  \right) + \gamma_c \right]
 \nn \\
&- \frac{e^{- \eps\gamma_E} \, \Gamma(1-\ep) }{2\ep} A_c^{(1)} 
\left( \psi(1-\ep) - \psi(1) \right)- \frac{\pi^2}{24} A^{(1)}_c + h_c^{(1)} + {\widetilde h}_c^{(1)}(\ep,{\widetilde \lambda} )
\!\bigg\},c\!=\!q,{\bar q},g.\nn
\end{align}
Such dependence is relevant to compute the contributions at higher orders in $\as$ through the implementation of the IR factorization formula in eq.~(\ref{IRfactav}).

The\, perturbative\, expansion\, of\, the\, azimuthal-correlation\, component\,
${\widetilde F}_{ga,\,{\rm corr.}}$ in eq.~(\ref{bspacecorr}) starts at ${\cal O}(\as)$. Therefore, no IR factorization procedure is required at this perturbative order.
The IR finite function ${\widetilde G}_{ga}$ in eq.~(\ref{IRfactcor}) has the following perturbative expansion:
\begin{equation}
{\widetilde G}_{ga}(z;\as,\ep,{\widetilde \lambda}) = 
\frac{\as}{\pi} \;{\widetilde G}_{ga}^{(1)}(z;\ep,{\widetilde \lambda})
+ \left(\frac{\as}{\pi}\right)^2 \;{\widetilde G}_{ga}^{(2)}(z;\ep,{\widetilde \lambda})
+ {\cal O}(\as^3). 
\label{gtilexp}
\end{equation}
The first-order term ${\widetilde G}_{ga}^{(1)}$ is simply obtained from eq.~(\ref{1corb})
by expressing $\alpha^u_{\rm S}$ in terms of $\as(b_0^2/\mb{b}^2)$, and we obtain
\begin{equation}
\label{gtil1}
{\widetilde G}_{ga}^{(1)}(z;\ep,{\widetilde \lambda})= 
e^{- \eps\gamma_E} \, \Gamma(1-\ep) \;C_a \;\frac{1-z}{z}.
\end{equation}
In the four-dimensional limit $\ep \to 0$, ${\widetilde G}_{ga}$ gives the 
transverse-momentum resummation function $G_{ga}$ in 
eqs.~(\ref{eq:colg}) and (\ref{azcorcolg}),
and specifically we have ${\widetilde G}_{ga}^{(m)}(z;\ep=0,{\widetilde \lambda}) = 
G_{ga}^{(m)}(z)$. By inspection of eq.~(\ref{gtil1}) we see that 
${\widetilde G}_{ga}^{(1)}(z;\ep=0,{\widetilde \lambda}) = C_a (1-z)/z$, in
agreement with the known result \cite{Catani:2010pd} reported in eq.~(\ref{eq:g1ga}).

We briefly comment on the relation between the IR factorization formulae in
eqs.~(\ref{IRfactav}) and (\ref{IRfactcor}) and the transverse-momentum resummation formulae
of QCD cross sections (see the discussion in section~\ref{sec:qcdres}). The TMD functions in
eqs.~(\ref{IRfactav}) and (\ref{IRfactcor}) embody the collinear contributions to the cross section, and they lead to the resummation coefficients ${\widetilde C}_{ca}$ and
${\widetilde G}_{ga}$ (i.e., $C_{ca}$ and $C^{\mu \nu}_{ga}$ of 
eqs.~(\ref{eq:colq}) and (\ref{eq:colg}) in the limit $\ep \to 0$) and to the IR singular factors $Z_c$. In the small-$q_T$ (or large-$\mb{b}^2$) region, the cross section receives
additional relevant contributions from soft, but non-collinear, radiation. Such contributions have to be properly evaluated and combined with the factors $Z_c$ and with the purely-virtual contributions to the cross section.  The combination of all these contributions in transverse-momentum resummation formulae produces the cancellation of the IR divergences, and it leads \cite{Catani:2013tia}
to the large logarithmic terms (i.e., $\ln \mb{b}^2$ terms) of the Sudakov form factor and to the hard (i.e., IR finite) virtual factors.

We remark on the fact that the soft non-collinear contribution to the cross section depends on the auxiliary vector $n^\mu$ that is used in the TMD collinear functions. Indeed, the soft
non-collinear contribution is obtained from the complete soft terms (which are independent of
$n^\mu$) by subtracting the soft limit of the terms (which depend on $n^\mu$) 
included in the TMD collinear functions. This subtraction is necessary to avoid double
counting of soft and collinear terms.
The $n^2$ dependence of the soft non-collinear contribution cancels the $n^2$ dependence of the collinear factors $Z_c$, and the result for the cross section is independent of $n^2$. At ${\cal O} (\as)$ we have explicitly verified this cancellation, and we have reobtained the complete contributions to the transverse-momentum resummation formula \cite{Catani:2013tia}.

We conclude this subsection with some comments on rapidity divergences. The order-by-order
perturbative computation of $q_T$-differential cross sections at any values of $q_T$ can be carried out in exact form without encountering rapidity divergences. From this computation 
one can then (`a posteriori') extract the small-$q_T$ behaviour of the cross section by simply neglecting subdominant terms at small values of $q_T$.
Therefore, the rapidity divergences are an artifact of `a priori' approximations that are introduced to evaluate only the dominant terms of the cross section in the small-$q_T$ region. These `a priori' approximations regard the QCD scattering amplitudes and the
$q_T$-dependent phase space.

The dominant collinear contributions at small $q_T$ can be consistently evaluated by approximating the squared amplitudes  through the collinear factorization formula 
(\ref{colfact})
and by introducing the differential collinear function ${\cf}_{ca}(p,k;n)$
in eq.~(\ref{slcolfun}). In the computation of the cross section the upper value of the
momentum $k^\mu$ is bounded by kinematics and, in particular, $k^- \ltap {\cal O}(p^+)$.
Moreover, the validity of the collinear factorization formula is, strictly speaking,
limited to the collinear region where $k^- \ltap {\cal O}(k^+)$. Introducing the TMD collinear function $F_{ca}$ in eq.~(\ref{sltmdgen}), 
the cross section kinematics is approximated by removing any upper bounds on $k^-$ and integrating over the entire region $0 < k^- < +\infty$. Such approximation is justified only if the collinear factorization formula and the collinear function ${\cf}_{ca}(p,k;n)$
are sufficiently well behaved at large values of $k^-$, so that the large-$k^-$ region
eventually leads only to subdominant terms at small values of $q_T$.

The $n^\mu$ dependence of ${\cf}_{ca}(p,k;n)$ is harmless in the collinear region where
$k^- \ltap {\cal O}(k^+)$, but it is very relevant at large values of $k^-$. If $n^2=0$,
the $n^\mu$ dependence of ${\cf}_{ca}(p,k;n)$ in the limit  $k^- \to +\infty$ produces the rapidity divergences. However, if $n^2 > 0$, ${\cf}_{ca}(p,k;n)$ is integrable over the 
large-$k^-$ region and, actually, it is effectively (dynamically) damped in the region
where $k^- \gtap {\cal O}(2 (n^-)^2\, k^+/n^2)$ (see, e.g., eqs.~(\ref{znsing}) and 
(\ref{norapdiv})),
namely, outside the region of validity of the collinear approximation of the squared amplitudes. Therefore, if $n^2 > 0$, the TMD function $F_{ca}$ in eq.~(\ref{sltmdgen})
can be consistently used to approximate and evaluate the dominant collinear contributions to the $q_T$ cross sections at small values of $q_T$. We note that the dependence of 
${\cf}_{ca}(p,k;n)$ on the auxiliary time-like vector $n^\mu$ affects the soft limit 
$k \to 0$ and, consequently, the small-$q_T$ limit of the TMD function depends on $n^2$.
As previously mentioned, such $n^2$ dependence is properly compensated and cancelled by the corresponding $n^2$ dependence of the soft non-collinear contributions to the cross section.

The rapidity divergences of the TMD function $F_{ca}$ in eq.~(\ref{sltmdgen}) can also be avoided by considering collinear functions ${\cf}_{ca}(p,k;n)$ that are defined
as in eq.~(\ref{slcolfun}) by using a {\it space-like} auxiliary vector $n^\mu$ ($n^2 < 0$).
However, the use of a space-like vector $n^\mu$ can introduce unphysical divergences at finite values of $k^-$. These features can be clearly seen by setting 
$n^\mu=( n^+, \mathbf{0_T}, n^-)$ with $n^2=2n^+n^- < 0$ and by considering the computation
of ${\cf}_{ca}(p,k;n)$ and $F_{ca}$ at ${\cal O}(\as)$. If $n^2 < 0$, the factor $1/(1-z_n)$
of eq.~(\ref{znsing}) leads to a dynamical damping in the large-rapidity region where
$k^- \gg k^+$ (similarly to the case in which $n^2 > 0$), but it also produces a divergence at the value $k^-= - 2 (n^-)^2\, k^+/n^2$.
This divergence has to be regularized (for instance, we can perform the replacement
$1/(1-z_n) \to {\rm PV}[1/(1-z_n)]$, where PV denotes the {\it principal value} prescription)
to carry out the integration over $k^-$ and 
to properly define the TMD function $F_{ca}$. 
In view of the unnecessary complications (with respect to using $n^2 > 0$) of introducing and, consequently, regularizing unphysical divergences, we do not use a space-like auxiliary vector $n^\mu$ in the definition (see section~\ref{sec:colfun})
of the splitting kernels $\bcp$
and of the ensuing collinear and TMD functions.

Auxiliary space-like vectors are used in the formulation of TMD factorization that is worked out in ref.~\cite{Collins:2011zzd}.
However, as briefly discussed below, those auxiliary vectors are not related to the space-like auxiliary vector that can be introduced through the collinear splitting kernels $\bcp$.

%The formalism of Ref.~\cite{Collins:2011zzd} regards TMD factorization for the specific processes of production of high-mass colourless systems and corresponding processes related
%by kinematical crossing (see Sect.~\ref{sec:qcdres}). 
The formalism of ref.~\cite{Collins:2011zzd} regards TMD factorization for the specific class of production processes involving colourless high-mass systems and corresponding processes related
by kinematical crossing (see section~\ref{sec:qcdres}). The TMD functions defined in 
ref.~\cite{Collins:2011zzd} embody both collinear and soft contributions that depend on several Wilson line operators. The dressing of the collinear contributions uses Wilson line operators along the direction $n^\mu$, while the soft contributions depend on Wilson line operators along the directions $n^\mu$ and $n_S^\mu$, where $n_S^\mu$ is a space-like
vector ($n_S^2 < 0$). The auxiliary vector $n^\mu$ in the collinear and soft contributions is light-like 
%$n^2=0$ (although the light-like limit is approached from space-like values). 
(although the limit $n^2=0$ is approached from space-like values). The ensuing TMD functions eventually depend on the space-like vector  $n_S^\mu$ that is introduced through the soft contributions. Such $n_S$ dependence is not related to the auxiliary-vector dependence that is introduced in our collinear functions through the collinear splitting kernels $\bcp$.

\subsection{Beam functions}
\label{sec:beam}

The partonic beam function $\cb_{ca}$ is obtained from the differential collinear function
$\cf_{ca}$ by using eq.~(\ref{slbeamgen}). The contribution of ${\cal O}(\as)$ to $\cb_{ca}$
is denoted by $\cb_{ca}^{(1R)}$, and it is obtained from the corresponding contribution
$\cf_{ca}^{(1R)}$ to the differential collinear function.
The computation of $\cb_{ca}^{(1R)}$ is performed by using the expression
of $\cf_{ca}^{(1R)}(p,k;n)$ in eq.~(\ref{f1R}),
similarly to our computation of the TMD functions in section~\ref{sec:tmdfun}.
The integration over $k$ in eq.~(\ref{slbeamgen}) is trivial at ${\cal O}(\as)$,
and $\cb_{ca}^{(1R)}$ turns out to be proportional to the overall factor
$t^{-1-\ep} \,(1-z)^{-\ep} \,z^\ep$ times a function of $z$ and $t$, which is due to the corresponding dependence of eq.~(\ref{f1R}) on $z_n$ (see eq.~(\ref{znvar})).
Using the collinear factorization formulae, we compute the singular terms of $\cb_{ca}$ in the limit $t \to 0$. Such terms are due to the overall factor $t^{-1-\ep}$
in $\cb_{ca}^{(1R)}$ and, therefore, the residual $z_n$ dependence can be approximated by setting $t=0$, provided the limit $t \to 0$ is smooth and non-singular.
In complete analogy with our discussion of the TMD functions (see section~\ref{sec:tmdfun}),
we can set $z_n= z +{\cal O}(t)$ in all the contributions to $\cb_{ca}^{(1R)}$ with the exception of those that are due to ${\widehat P}^{\rm \,sing}_{ca}(z_n)$ 
(see eqs.~(\ref{psr})--(\ref{preg}))
in eq.~(\ref{f1R}). 

The contribution of ${\widehat P}^{\rm \,sing}_{ca}(z_n)$ to
$\cb_{ca}^{(1R)}$ is proportional to the following factor
\begin{equation}
\label{znsingt}
\frac{1}{1-z_n} = \frac{np}{nk} = \frac{p^+}{k^+ + \frac{n^2}{2 np}  \frac{k^-}{n^-} p^+}
= \frac{1}{1-z +\frac{n^2 t}{z (2 np)^2}},
\end{equation}
where in the last equality we have implemented the kinematics of the beam function  
%at ${\cal O}(\as)$ 
(i.e., $k^+=(1-z)p^+, k^-=t/(2zp^+)$).
Setting $n^2=0$, this factor is divergent (and not integrable over $z$) at $z=1$, 
analogously to the corresponding factor in eq.~(\ref{znsing}) for the TMD functions.
However, as previously noted, $\cb_{ca}^{(1R)}$ contains an overall factor $(1-z)^{-\ep}$,
which is produced by the integration over $\mb{k_T}$ (at ${\cal O}(\as)$ we have
$\mb{k_T}^2=t(1-z)/z$). In the context of dimensional regularization the factor 
$(1-z)^{-\ep}$ regularizes the singular factor $(1-z)^{-1}$ and, therefore, 
$\cb_{ca}^{(1R)}$ is well defined (though it is IR divergent in the limit $\ep \to 0$)
by using a light-like  auxiliary vector $n^\mu$. In other words, if $n^2=0$ the beam function
$\cb_{ca}^{(1R)}$ does not contain rapidity divergences (in contrast to the case of the TMD function $F_{ca}^{(1R)}$). Few other comments on the absence of rapidity divergences are presented at the end of this subsection.

The first-order term $\cb_{ca}^{(1R)}$ is well defined if $n^2=0$ and, also, if $n^2 > 0$.
In the following we consider the explicit computation of the beam function for a generic auxiliary vector with $n^2 \geq 0$.

Independently of the specific value of $n^2$, the contribution of the factor in 
eq.~(\ref{znsingt}) to $\cb_{ca}^{(1R)}$ can be approximated in the relevant limit
$t \to 0$. Considering the effect of the additional factor $(1-z)^{-\ep}$ and setting
$\lambda^{\prime}=n^2 t/(2pn)^2$, we can use the following approximation:
\begin{align}
\label{beamplus}
\frac{(1-z)^{-\ep}}{1-z +\frac{\lambda^{\prime}}{z}} &= \left( \frac{(1-z)^{-\ep}}{1-z +\frac{\lambda^{\prime}}{z}} \right)_+
+\delta(1-z) \int_0^1 dz^\prime \frac{(1-z^\prime)^{-\ep}}{1-z^\prime +\frac{\lambda^{\prime}}{z^\prime}} \nn \\
&= \left[ \,(\lambda^{\prime})^{-\ep} \;\Gamma(1+\ep) \,\Gamma(1-\ep) -1 \right] \frac{1}{\ep} \;\delta(1-z) 
+ \left( \frac{(1-z)^{-\ep}}{1-z} \right)_+ + {\cal O}(\lambda^{\prime}),
\end{align}
which is valid for arbitrary values of $\ep$. Owing to the factor $(\lambda^{\prime})^{-\ep}$ in
eq.~(\ref{beamplus}), we note that the limit $t \to 0$ and the transition from $n^2 >0$ to  $n^2=0$  are not smooth
(i.e., the limits $t\to 0$ or $n^2 \to 0$ do not commute with the limit
$\ep \to 0$). The term
of ${\cal O}(\lambda^{\prime}) \sim {\cal O}({t})$ in the right-hand side of eq.~(\ref{beamplus})
smoothly vanishes in the limit $t \to 0$ (i.e., it vanishes order-by-order in the $\ep$ expansion around $\ep=0$) and, therefore, it can be neglected in the computation of 
$\cb_{ca}^{(1R)}$.

Using eq.~(\ref{beamplus}) and neglecting subdominant terms that smoothly vanish in the limit $t \to 0$, we obtain the following result for the beam function at ${\cal O}(\as)$:
\begin{align}
\label{beam1R}
\cb_{ca}^{(1R)}\!\left(z;t,\frac{n^2 \,t}{(2zpn)^2}\right) \;
=& \frac{\alpha^u_{\rm S} \,\mu_0^{2\eps}
\,S_{\eps}}{\pi} \;\frac{e^{\eps\gamma_E}}{\Gamma(1-\ep)} \;t^{-1-\ep} \;
\Bigl\{ \left( \frac{1-z}{z}\right)^{\!-\ep} {\widehat P}^{\rm \,reg}_{ca}(z; \ep)
\Bigr.
 \\
& + \Bigl. \delta_{ca} \,A_c^{(1)}
\Bigl[ (z^\ep -1) \frac{(1-z)^{-\ep}}{1-z} + \left( \frac{(1-z)^{-\ep}}{1-z} \right)_+
\nn\\
&+ \delta(1-z) 
\frac{
\;\left( \frac{n^2 t}{(2zpn)^2} \right)^{\!-\ep} \;\Gamma(1+\ep) \,\Gamma(1-\ep) -1}{\ep}
\,
\Bigr]  \Bigr\},  \nn
\end{align}
%\begin{align}
%%\label{tmdqt}
%&\cb_{ca}^{(1R)}\!\left(z;t,\frac{n^2 \,t}{(2zpn)^2}\right) \;
%= \frac{\alpha^u_{\rm S} \,\mu_0^{2\eps}
%\,S_{\eps}}{\pi} \;\frac{e^{\eps\gamma_E}}{\Gamma(1-\ep)} \;t^{-1-\ep} \;
%\Bigl\{ \left( \frac{1-z}{z}\right)^{\!-\ep} {\widehat P}^{\rm \,reg}_{ca}(z; \ep)
%\Bigr.
% \\
%& + \Bigl. \delta_{ca} \;A_c^{(1)}
%\left[ (z^\ep -1) \frac{(1-z)^{-\ep}}{1-z} + \left( \frac{(1-z)^{-\ep}}{1-z} \right)_+
%+ \delta(1-z) \, \frac{1}{\ep}
%\left(
%\;\left( \frac{n^2 t}{(2zpn)^2} \right)^{\!-\ep} \;\Gamma(1+\ep) \,\Gamma(1-\ep) -1
%\right)
%\right]  \Bigr\} \;,  \nn
%\end{align}
where the kernels ${\widehat P}^{\rm \,reg}_{ca}(z; \ep)$ are given in 
eqs.~(\ref{rqq})--(\ref{a1coef}). The singular $t$ dependence of eq.~(\ref{beam1R})
is due to the terms $t^{-1-\ep}$ and $t^{-1-2\ep}$, which can be customarily expanded
in the limit $\ep \to 0$ (see, e.g., ref.~\cite{Ritzmann:2014mka}) and lead to
$\ep$ poles and plus-distributions of the variable $t$. Setting $n^2=0$, it is straightforward to check that the $\ep$ expansion of eq.~(\ref{beam1R}) agrees with the expressions of $\cb_{qq}^{(1R)}$ and $\cb_{qg}^{(1R)}$ obtained in 
ref.~\cite{Ritzmann:2014mka}.

In the following we consider the Laplace transformation 
${\widetilde \cb}_{ca}$ of the beam function $\cb_{ca}$ with respect to the transverse virtuality $t$. We define ${\widetilde \cb}_{ca}$ as follows
\begin{equation}
\label{beamlt}
{\widetilde \cb}_{ca}\!\left(z;\sigma,\frac{n^2 \sigma_0}{(2zpn)^2 \sigma}\right)
\equiv \int_0^{+\infty} dt \;e^{- \sigma t} \;\cb_{ca}\!\left(z;t,\frac{n^2 t}{(2zpn)^2}\right)
\;\;,
\end{equation}
where $\sigma$ is the conjugate variable to $t$ in Laplace space, and 
$\sigma_0= e^{-\gamma_E}$ is a customary numerical coefficient 
\cite{Catani:1989ne, Catani:1992ua}
that appears in Mellin or Laplace transformations of plus-distributions.  

In Laplace space, the singular terms of ${\cb}_{ca}$ in the limit
$t \to 0$ become logarithmic contributions of the type $\ln \sigma$, which are large in the limit $\sigma \to + \infty$. 
We use the following result:
\begin{align}
\label{ltres}
\int_{0}^{\infty}dt\, e^{-\sigma t} \;\frac{\ln^m t}{t^{1+\delta}} = \frac{d^m}{d\rho^m}\bigg{|}_{\rho=0} \;\left[
\left(\sigma\right)^{\delta-\rho}\Gam\left(\rho-\delta\right) \right],
\end{align}
which is the most general Laplace transformation that is necessary to go from 
${\cb}_{ca}$ to ${\widetilde \cb}_{ca}$ at arbitrary perturbative orders in $\as$.

The perturbative expansion of ${\widetilde \cb}_{ca}$ is
\begin{equation}
\label{beamexp}
{\widetilde \cb}_{ca}\!\left(z;\sigma,\frac{n^2 \sigma_0}{(2zpn)^2 \sigma}\right)=
\delta_{ca} \,\delta(1-z) +
{\widetilde \cb}_{ca}^{(1R)}\!\left(z;\sigma,\frac{n^2 \sigma_0}{(2zpn)^2 \sigma}\right)
+ {\cal O}(\as^2).
\end{equation}
The first-order contribution ${\widetilde \cb}_{ca}^{(1R)}$ is obtained from the expression of ${\cb}_{ca}^{(1R)}$ in eq.~(\ref{beam1R}) by using the Laplace transformations 
in eq.~(\ref{ltres}) with $\delta=\ep, 2\ep$ and $m=0$. We have
\begin{align}
\label{beamlt1R}
{\widetilde \cb}_{ca}^{(1R)}\!\left(z;\sigma,\frac{n^2 \sigma_0}{(2zpn)^2 \sigma}\right) \;
=& \frac{\alpha^u_{\rm S}
\,S_{\eps}}{\pi} \left(\frac{\mu_0^2 \,\sigma}{\sigma_0}\right)^{\!+\ep} \;
\frac{1}{\ep} \;
\Bigg\{ - \left( \frac{1-z}{z}\right)^{\!-\ep} {\widehat P}^{\rm \,reg}_{ca}(z; \ep)\nn
 \\
& -  \delta_{ca} \,A_c^{(1)}
\Bigg[ (z^\ep -1) \frac{(1-z)^{-\ep}}{1-z} + \left( \frac{(1-z)^{-\ep}}{1-z} \right)_{\! +}
\nn \\
&+\frac{\delta(1-z)}{2\ep} \;\left( 
\left( \frac{n^2 \sigma_0}{(2zpn)^2 \sigma} \right)^{\!-\ep} \; e^{-\eps\gamma_E}
\,\Gamma(1+\ep) \,\Gamma(1-2\ep)
-2 \right) 
\,
\Bigg]  \Bigg\},  
\end{align}
which is the result for arbitrary $d=4-2\ep$ space-time dimensions.
In the limit $\ep \to 0$, ${\widetilde \cb}_{ca}^{(1R)}$ contains double and single poles, 
$1/\ep^2$
and $1/\ep$, of IR origin. At higher perturbative orders, the computation of 
${\widetilde \cb}_{ca}$ leads to $\ep$ poles of UV and IR origins. The UV divergences are removed by renormalizing the bare coupling $\alpha^u_{\rm S}$ 
(see eq.~(\ref{eq:alphasren})), 
while the IR divergences are factorizable.

The IR factorization formula for the beam function ${\widetilde \cb}_{ca}$ in Laplace space
is
\begin{align}
\label{beamfact}
{\widetilde \cb}_{ca}\!\left(z;\sigma,\frac{n^2 \sigma_0}{(2zpn)^2 \sigma}\right)=&
{\cal Z}_c\!\left(\as(\mu_F^2), \sigma, \mu_F; \frac{n^2 \sigma_0}{(2zpn)^2 \sigma}\right)
\nn \\
&\times \sum_b \int_z^1 \frac{dx}{x} \; 
{\widetilde I}_{cb}\!\left(x,\sigma; \mu_F, \as(\mu_F^2), \ep, \frac{n^2 \sigma_0}{(2zpn)^2 \sigma}\right) \; {\widetilde \Gamma}_{ba}(z/x; \mu_F^2)\;,
\end{align}
where the functions ${\cal Z}_c, {\widetilde I}_{cb}$ and ${\widetilde \Gamma}_{ba}$ are expressed in terms of the renormalized coupling $\as(\mu_F^2)$ at the renormalization scale $\mu_R=\mu_F$.
The structure of eq.~(\ref{beamfact}) is closely analogous to that of the corresponding IR
factorization formulae in eqs.~(\ref{IRfactav}) and (\ref{IRfactcor})
for the TMD functions in $\mb{b}$ space.
The factor ${\widetilde \Gamma}_{ba}(x;\mu_F^2)$ in eq.~(\ref{beamfact}) is the 
collinear-divergent function that defines the scale-dependent PDFs $f_b(z;\mu_F^2)$ at the scale $\mu_F$ in the $\overline{{\rm MS}}$ factorization scheme (see eqs.~(\ref{pdfren}) and 
(\ref{wgamexp})). The function ${\cal Z}_c$ embodies $\ep$ poles of IR origin and the function ${\widetilde I}_{cb}$ is finite in the limit $\ep \to 0$. Using a time-like auxiliary vector $n^\mu$, ${\widetilde \cb}_{ca}$ explicitly depends on $n^2$. In the limit
$\ep \to 0$ any IR finite dependence on $n^2$ is absorbed in the factor ${\cal Z}_c$,
so that the function ${\widetilde I}_{cb}$ is independent of $n^2$ and it leads to the matching coefficient ${I}_{cb}$ in eqs.~(\ref{lsmc}) and (\ref{matchexp}).

The explicit expression of ${\cal Z}_c$ up to ${\cal O}(\as)$ is derived by comparing the factorized structure in eq.~(\ref{beamfact}) with the results in 
eqs.~(\ref{beamexp}) and (\ref{beamlt1R}).
Setting
$\widetilde{\lambda}^{\prime}=\frac{n^2 \sigma_0}{(2zpn)^2 \sigma}$, we obtain
\begin{align}
{\cal Z}_c(\as(\mu_F^2),\sigma, \mu_F; \widetilde{\lambda}^{\prime}) =& 1 + 
\frac{\as(\mu_F^2)}{\pi} \;
{\cal Z}_c^{(1)}(\sigma, \mu_F; \widetilde{\lambda}^{\prime}) + {\cal O}(\as^2),\\
%\end{align}
%\begin{align}
\label{z1beam}
{\cal Z}_c^{(1)}(\sigma, \mu_F; \widetilde{\lambda}^{\prime}) =&
 A_c^{(1)} \left[ \frac{1}{\ep^2} + 
 \frac{1}{\ep} \ln \left( \frac{\mu_F^2 \,\sigma}{\sigma_0} \right) 
\right] 
 + \gamma_c \;\frac{1}{2\ep} - A_c^{(1)} \;r^{(1)}(\ep, \widetilde{\lambda}^{\prime}; \mu_F,  \sigma),
\end{align}
where the function $r^{(1)}$ is
\begin{equation}
\label{r1fun}
r^{(1)}\Bigl( \ep, \frac{n^2 \sigma_0}{(2zpn)^2 \sigma}; \mu_F, \sigma \Bigr) = \;\frac{1}{2\ep^2} \;\left(\frac{\mu_F^2 \,\sigma}{\sigma_0}\right)^{\!+\ep} \;  
\left( \frac{n^2 \sigma_0}{(2zpn)^2 \sigma} \right)^{\!-\ep} \; e^{-\eps\gamma_E}
\,\Gamma(1+\ep) \,\Gamma(1-2\ep). 
\end{equation}

Analogously to the IR factor $Z_c$ for the TMD functions (see eq.~(\ref{z1tmd})), 
the beam function factor ${\cal Z}_c$ in eq.~(\ref{beamfact}) has a factorization scheme dependence, which is specified by IR finite contributions. If $n^2=0$, the function
$r^{(1)}$ in eq.~(\ref{r1fun})
vanishes and, consequently, the expression of ${\cal Z}_c^{(1)}$ in 
eq.~(\ref{z1beam}) contains only $\ep$ poles. Therefore, such expression corresponds to a `minimal' scheme, in which ${\cal Z}_c^{(1)}$ has no IR finite contributions.
% if $n^2=0$.
This minimal scheme is customarily used for the SCET definition of the beam functions
\cite{Stewart:2009yx, Stewart:2010qs, Berger:2010xi, Ritzmann:2014mka}.
If $n^2 > 0$, $r^{(1)}$ is not vanishing and, in particular, ${\cal Z}_c^{(1)}$
in eq.~(\ref{z1beam}) embodies the entire dependence (i.e., the dependence at arbitrary orders
in the $\ep$ expansion) of ${\widetilde \cb}_{ca}^{(1R)}$ on $n^2$ 
(see eq.~(\ref{beamlt1R})).
In the limit $\ep \to 0$ the expression of $r^{(1)}$ for $n^2 > 0$ is
\begin{equation}
\label{r1exp}
r^{(1)}\Bigl( \ep, \frac{n^2 \sigma_0}{(2zpn)^2 \sigma}; \mu_F, \sigma \Bigr) = 
\frac{1}{2\ep^2}-\frac{1}{2\ep}
\ln \left(\! \frac{n^2 \sigma_0^2}{(2zpn)^2 \sigma^2 \mu_F^2} \!\right)
+\frac{1}{4}  \ln^2 \left(\! \frac{n^2 \sigma_0^2}{(2zpn)^2 \sigma^2 \mu_F^2} \!\right)
+ \frac{5 \pi^2}{24} + {\cal O}(\ep).
\end{equation}
As we have already noticed (see eq.~(\ref{beamplus}) and accompanying comments),
the limit $n^2 \to 0$ is not smooth (i.e., it does not commute with the limit $\ep \to 0$).
Indeed, the result $r^{(1)}=0$ for $n^2 = 0$ cannot be recovered by setting $n^2 = 0$
in the expression in the right-hand side of eq.~(\ref{r1exp}).

The IR finite contribution ${\widetilde I}_{ca}$ in eq.~(\ref{beamfact}) has the following perturbative expansion:
\begin{equation}
\label{itilexp}
{\widetilde I}_{ca}\!\left(\!\!z,\sigma; \mu_F, \as(\mu_F^2), \ep, \frac{n^2 \sigma_0}{(2zpn)^2 \sigma}\!\!\right) = \delta_{ca} \delta(1-z) + \frac{\as(\mu_F^2)}{\pi}
{\widetilde I}_{ca}^{(1)}\!\left(\!\!z,\sigma; \mu_F, \ep, \frac{n^2 \sigma_0}{(2zpn)^2 \sigma}\!\!\right)+ {\cal O}(\as^2).
\end{equation}
The matching coefficient $I_{ca}$ of eqs.~(\ref{lsmc}) and (\ref{matchexp}) corresponds
to the limit $\ep \to 0$ of ${\widetilde I}_{ca}$. In particular, we have
\begin{equation}
\label{itil1}
{\widetilde I}_{ca}^{(1)}\!\left(z,\sigma; \mu_F, \ep=0, \frac{n^2 \sigma_0}{(2zpn)^2 \sigma}\right) = I_{ca}^{(1)}(z,\sigma; \mu_F). 
\end{equation}
By using the result in eq.~(\ref{beamlt1R}), the IR factorization formula (\ref{beamfact})
and the expression of ${\cal Z}_c^{(1)}$ in eq.~(\ref{z1beam}), the first-order term
${\widetilde I}_{ca}^{(1)}$ is independent of $n^2$ and, setting $\ep=0$, we 
obtain 
\begin{align}
\label{i1eq}
&I_{ca}^{(1)}(z,\sigma; \mu_F) \;
= - {P}^{(1)}_{ca}(z) \,\ln \left(\frac{\mu_F^2 \,\sigma}{\sigma_0}\right) 
 - {\widehat P}^{\prime}_{ca}(z; \ep=0) + 
%\ln \left( \frac{1-z}{z}\right) 
{\widehat P}^{\rm \,reg}_{ca}(z; \ep=0) \,\ln \left( \frac{1-z}{z}\right)
 \\
& + \delta_{ca} \; \left\{ \,A_c^{(1)}
\left[  \left( \frac{ \ln(1-z)}{1-z} \right)_+ - \frac{\ln z}{1-z} \right] 
+ \delta(1-z) \left[\frac{A_c^{(1)}}{2} \ln^2 \left(\frac{\mu_F^2 \,\sigma}{\sigma_0}\right) + \frac{\gamma_c}{2} \ln \left(\frac{\mu_F^2 \,\sigma}{\sigma_0}\right)
\right]  \right\}.  \nn
\end{align}
This result for the matching coefficient $I_{ca}^{(1)}$ agrees with the Laplace transformation
of the known result in the literature \cite{Stewart:2010qs, Berger:2010xi}.
We recall that eq.~(\ref{i1eq}) refers to a minimal scheme in which 
${\cal Z}_c^{(1)}$ has only $\ep$ pole contributions if $n^2=0$.
The generic scheme dependence of the beam functions at ${\cal O}(\as)$ can be explicitly denoted by introducing an $\ep$-independent function in the expressions of ${\cal Z}_c^{(1)}$ and $I_{ca}^{(1)}$,
similarly to the function $h_c^{(1)}$ in 
eqs.~(\ref{z1tmd}) and (\ref{c1coeff}) for the TMD functions. 

We conclude this subsection with some additional comments about the dependence of the beam functions on the auxiliary vector $n^\mu$. As discussed in section~\ref{sec:tmdfun},
in the case of the TMD function $F_{ca}$ the singular dependence on $n^2$ and the rapidity divergences originate in eq.~(\ref{sltmdgen})
from the $k^-$ integration of $\cf_{ca}(p,k;n)$ over the region 
${\cal O}(p^+) \ltap k^- < +\infty$, which lies outside the region of validity of the collinear factorization formula (\ref{colfact})
that is used to define $\cf_{ca}(p,k;n)$.
In the case of the beam function $\cb_{ca}$ the integration of $\cf_{ca}(p,k;n)$
over the momentum $k^\mu$ is specified in eq.~(\ref{slbeamgen}), and it is always bounded
inside the collinear region. Indeed, we have $k^-=t/(2zp^+)$ and the transverse virtuality
$t$ is fixed to be small (i.e., $t \ll {\cal O}((p^+)^2)$) in the computation of the beam function. The $\mb{k_T}$ integration in eq.~(\ref{slbeamgen}) is also bounded to the region
of small values of $\mb{k_T}$ since we have $\mb{k_T}^2 \leq t(1-z)/z$ (which follows
from $k^2 \geq 0$).

In summary, due to its definition in eq.~(\ref{slbeamgen}), the beam function $\cb_{ca}$
can be consistently computed by using both time-like or light-like auxiliary vectors $n^\mu$,
without encountering rapidity divergences. Varying the value of $n^2$ varies the 
behaviour of $\cf_{ca}(p,k;n)$ in soft limit $k \to 0$ and, consequently, the effect of the soft contributions included in the beam function $\cb_{ca}$.
This effect produces the $n^2$ dependence of $\cb_{ca}$, which can be factorized in the function ${\cal Z}_c$ of eqs.~(\ref{beamfact}) and (\ref{z1beam}).
In the computation of cross sections, the $n^2$ dependence of $\cb_{ca}$ is cancelled
by the $n^2$ dependence of the soft non-collinear contributions, through the same mechanism
that leads to the cancellation of the $n^2$ dependence due to the TMD function (see the discussion at the end of section~\ref{sec:tmdfun}).

At ${\cal O}(\as)$ we have explicitly evaluated the beam function by using a time-like auxiliary vector. Similar computations of the beam function $\cb_{ca}$ with $n^2 > 0$
can be carried out at higher orders in $\as$. Admittedly, such computations turn out to be much more cumbersome than those with $n^2=0$ \cite{Gaunt:2014xga, Gaunt:2014cfa, Melnikov:2018jxb, Melnikov:2019pdm, Behring:2019quf, Baranowski:2020xlp, Ebert:2020unb}.

\section{Azimuthal correlations at \boldmath ${\cal O}(\as^2)$ in the SL region}
\label{sec:azcorr}

In this section we consider the perturbative calculation of the azimuthal-correlation terms at ${\cal O}(\as^2)$. We present the results for the differential collinear function
$\cf_{ga,\,{\rm corr.}}(p,k;n)$
and, subsequently, for the TMD function $F_{ga,\,{\rm corr.}}$.

\subsection{Differential collinear functions}
\label{sec:azcorrdiff}

At ${\cal O}(\as^2)$ the differential collinear function $\cf_{ga,\,{\rm corr.}}$ receives the two contributions, 
$\cf_{ga,\,{\rm corr.}}^{(1R1V)}$ and $\cf_{ga,\,{\rm corr.}}^{(2R)}$,
in eq.~(\ref{fexp12}). We discuss them in turn.

The term $\cf_{ca}^{(1R1V)}(p,k;n)$ is due to the one-loop virtual correction
to the final-state emission of a collinear parton with momentum $k$, and it is obtained from eq.~(\ref{slcolfun}) by using the one-loop collinear kernel 
%$\bcp_{{c}\rightarrow a_1 a_2}^{(1)}$.
%$\cp_{{c}\rightarrow a_1 a_2}^{(1)}$.
$\cp_{{c}\rightarrow a_1 a}^{(1)}$.
The one-loop correction to the collinear splitting process 
%$c \rightarrow a_1 a_2$
$c \rightarrow a_1 a$ is known since a long time \cite{Bern:1998sc, Bern:1999ry, Kosower:1999rx, Sborlini:2013jba}.
The original results in the literature regard collinear factorization at the amplitude level
and they refer to the TL collinear region. The extension to the SL region was worked out and discussed in ref.~\cite{Catani:2011st}, by noticing two main features:
one-loop factorization for the collinear splitting 
${c}\rightarrow a_1 a$ is process dependent at the level of the scattering amplitudes, and it becomes process independent at the squared amplitude level. We have computed the SL kernels  
%$\cp_{{c}\rightarrow a_1 a_2}^{(1)}$ 
$\cp_{{c}\rightarrow a_1 a}^{(1)}$
by squaring the amplitude level results of refs.~\cite{Bern:1998sc, Bern:1999ry, Kosower:1999rx, Sborlini:2013jba}
and by applying the prescription of ref.~\cite{Catani:2011st} (see section~7.2 therein)
to derive the results in the SL region.

The expression of $\cf_{ga,\,{\rm corr.}}^{(1R1V)}(p,k;n)$ is directly proportional to
%$\cp_{{g}\rightarrow a_1 a_2}^{(1)}$ 
$\cp_{{g}\rightarrow a_1 a}^{(1)}$ 
(see eqs.~(\ref{corrker}) and (\ref{slcolfun})) and we obtain the following results:
\begin{align}
\label{gg1R1V}
\mc{F}^{(1R1V)}_{gg,\,{\rm corr.}}(p,k;n)
 =&\left(\frac{\alpha^u_s\mu_0^{2\eps}S_{\eps}}{\pi}\right)^2\frac{e^{2\eps\gamma_E}{\rm \Gamma}(1+\eps){\rm \Gamma^2(1-\eps)}}{\pi^{1-\eps}{\rm \Gamma(1-2\eps)}}\frac{C_A\delta_{+}(k^2)}{(2pk)^{1+\eps}}\bigg\{\!\!-\frac{N_f}{6}+\frac{C_A}{6} 
\\
&+C_A\frac{1-z}{z^2}\left[\frac{1}{\eps^2}-\frac{1}{\eps}\Big(\ln(1-z)-2\ln(z)\Big)-\frac{\pi^2}{3}+\frac{1}{2}\ln^2(1-z)\right]\!\!\bigg\} +{\cal O}(\ep),
\nn
\\
\label{gq1R1V}
 \mc{F}^{(1R1V)}_{ga,\,{\rm corr.}}(p,k;n) 
=& -\!\!\left(\!\frac{\alpha^u_s\mu_0^{2\eps}S_{\eps}}{\pi}\!\right)^2\frac{e^{2\eps\gamma_E}{\rm \Gamma}(1+\eps){\rm \Gamma^2(1-\eps)}}{\pi^{1-\eps}{\rm \Gamma(1-2\eps)}}\frac{C_F\delta_{+}(k^2)}{(2pk)^{1+\eps}}\frac{1-z}{z^2}\bigg\{\! C_F\!\left[\!-\frac{2}{\eps^2}-\frac{3}{\eps}-8\right]
 \nn\\
 &+C_A\left[\frac{1}{\eps^2}+\frac{1}{\eps}\left(\frac{11}{3}-2\ln(z)+\ln(1-z)\right)+\frac{76}{9}+\frac{\pi^2}{3}-\frac{1}{2}\ln^2(1-z) \right]
 \nn\\
 &+N_f\left[-\frac{2}{3\eps}-\frac{10}{9}\right]\!\bigg\}+{\cal O}(\ep) \;,  
\quad \quad \quad \quad \quad \quad
\quad \quad (a=q,{\bar q}).
\end{align}
We note that in eqs.~(\ref{gg1R1V}) and (\ref{gq1R1V}) we have neglected contributions of
${\cal O}(\ep)$, since they are not relevant to the four-dimensional limit at 
${\cal O}(\as^2)$. Moreover, we have used the momentum fraction $z$ that is defined as
\begin{equation}
\label{zvar}
z = \frac{\bar{n} (p-k)}{\bar{n}p},
\end{equation}
where the {\it light-like} auxiliary vector ${\bar n}^\mu$ is
\begin{equation}
\label{nbar}
        \bar{n}^{\mu}\equiv n^{\mu}-\frac{n^2p^{\mu}}{2np}.
\end{equation}
In our customary reference frame we have $p^\mu=(p^{+},\mathbf{0_T},0)$,
$n^\mu = (n^{+},\mathbf{0_T},n^{-})$ and, therefore,
${\bar n}^\mu = (0,\mathbf{0_T},n^{-})$ and $z=1-k^+/p^+$.
%We note that the expressions of $\mc{F}^{(1R1V)}_{ga,{\rm corr.}}$ in 
%Eqs.~(\ref{gg1R1V}) and (\ref{gq1R1V}) have a smooth dependence on $n^2$ and they are
%not singular in the limit $n^2 \to 0$, similarly to the corresponding expressions
%of $\mc{F}^{(1R)}_{ga,{\rm corr.}}$ at ${\cal O}(\as)$ (see Eq.~(\ref{fgacor})).

The use of the variable $z = \bar{n} (p-k)/\bar{n}p$ in 
eqs.~(\ref{gg1R1V}) and (\ref{gq1R1V}) deserves some comments, since it is practically equivalent to compute $\mc{F}^{(1R1V)}_{ga,\,{\rm corr.}}$ ($a=g,q,{\bar q}$) by setting 
$n^2=0$. As discussed in sections~\ref{sec:tmdfun} and \ref{sec:beam},
the use of the time-like auxiliary vector $n^\mu$ in the computation of the collinear functions is, in general, essential to avoid unphysical behaviour outside the collinear region. Nonetheless, already at ${\cal O}(\as)$ we noticed that we can correctly compute the collinear functions by setting  $n^2=0$ in most of the $n$-dependent terms of 
$\mc{F}^{(1R)}_{ca,\,{\rm az.in.}}$ in eq.~(\ref{f1R})
and in the entire $n$ dependence of $\mc{F}^{(1R)}_{ga,\,{\rm corr.}}$ in eq.~(\ref{fgacor}).
Therefore, in the expressions of eqs.~(\ref{gg1R1V}) and (\ref{gq1R1V})
we have directly implemented the approximation of setting  $n^2=0$ in terms with a harmless
dependence on $n^2$. The exact dependence of $\mc{F}^{(1R1V)}_{ga,\,{\rm corr.}}$ on
$n^2$ can be recovered from eqs.~(\ref{gg1R1V}) and (\ref{gq1R1V})
by replacing the variable $z$ of eq.~(\ref{zvar}) with the variable $z_n$ of 
eq.~(\ref{znvar}).

The term $\cf_{ga,\,{\rm corr.}}^{(2R)}(p,k;n)$ in eq.~(\ref{fexp12})
is due to the final-state emission of two collinear partons at the tree level. It is obtained starting from the azimuthal-correlation component of the tree-level collinear kernel
$\cp_{{g}\rightarrow a_1 a_2 a}^{(0) \mu \nu}(k_1,k_2,p;n)$
(see eqs.~(\ref{corrker}) and (\ref{slcolfun})). The tree-level kernel is process independent, and the TL and SL regions are straightforwardly related through the replacement $p \leftrightarrow -p$. The explicit expressions of 
$\cp_{{g}\rightarrow a_1 a_2 a}^{(0) \mu \nu}(k_1,k_2,p;n)$ for the various partonic channels
were obtained in refs.~\cite{Catani:1998nv, Catani:1999ss}. We evaluate 
$\cf_{ga,\,{\rm corr.}}^{(2R)}(p,k;n)$ in eq.~(\ref{slcolfun}) by performing the integrations
of the azimuthal-correlation component
of $\cp_{{g}\rightarrow a_1 a_2 a}^{(0)}(k_1,k_2,p;n)$
over $k_1$ and $k_2$ at fixed momentum $k=k_1+k_2$. Such integrations can be carried out to all orders in $\ep$ and for arbitrary values of $n^2$ by using the $d$-dimenional angular integrals of ref.~\cite{Somogyi:2011ir}. In the following we present the results in a simplified form that amounts to neglect contributions at high orders in $\ep$ and to set $n^2=0$ in all the contributions with a harmless (smooth and non-singular) dependence on
$n^2$. In particular, we use the variable $z$ of eq.~(\ref{zvar}) to express part
of the $n$ dependence of $\mathcal{F}^{(2R)}_{ga,\,\text{corr.}}(p,k;n)$.

We write $\mathcal{F}^{(2R)}_{ga,\,\text{corr.}}$ in terms of contributions with different colour factors:
\begin{align}
\label{ggcorr2R}
&\mathcal{F}^{(2R)}_{gg,\,\text{corr.}}(p,k;n) =\left(\frac{\alpha_s^u\mu_0^{2\eps}S_{\eps}}{\pi}\right)^2\frac{e^{2\eps\gamma_E}\,{\rm\Gamma}(1-\epsilon)}{2\pi^{1-\eps}\,{\rm \Gamma}(1-2\epsilon)} \,(k^2)^{-\epsilon}
\\
&\!\times\!\! \left[C_FT_RN_f\mathcal{F}^{(2R),\,C_FT_RN_f}_{gg,\,\text{corr.}}(p,k;n)
+ C_AT_RN_f\mathcal{F}^{(2R),\,C_AT_RN_f}_{gg,\,\text{corr.}}(p,k;n)
+ C_A^2\mathcal{F}^{(2R),\,C_A^2}_{gg,\,\text{corr.}}(p,k;n)\right]\!,
\nn\\
%\end{align}
%\begin{align}
\label{gqcorr2R}
&\mathcal{F}^{(2R)}_{gq,\,\text{corr.}}(p,k;n) = \left(\frac{\alpha^u_s\mu_0^{2\eps}S_{\eps}}{\pi}\right)^2\frac{e^{2\eps\gamma_E}\,{\rm\Gamma}(1-\epsilon)}{2\pi^{1-\eps}\,{\rm \Gamma}(1-2\epsilon)} \,(k^2)^{-\epsilon}\nn\\
&\times \left[ C_F^2\mathcal{F}^{(2R),\,C_F^2}_{gq,\,\text{corr.}}(p,k;n)+ C_AC_F\mathcal{F}^{(2R),\,C_AC_F}_{gq,\,\text{corr.}}(p,k;n)\right],
\end{align}
and\,\, we \,\,recall\,\, that\,\, $\mathcal{F}^{(2R)}_{g{\bar q},\,\text{corr.}}(p,k;n) 
= \mathcal{F}^{(2R)}_{gq,\,\text{corr.}}(p,k;n)$.
The\,\, individual\,\, coefficients \,\,$\mathcal{F}^{(2R),\,C_FT_RN_f}_{gg,\,\text{corr.}}$, $\mathcal{F}^{(2R),\,C_AT_RN_f}_{gg,\,\text{corr.}}$ and $\mathcal{F}^{(2R),\,C_A^2}_{gg,\,\text{corr.}}$ 
of eq.~(\ref{ggcorr2R}) for the gluon-gluon channel are given by
the following expressions:
\begin{align}
\label{ggCF}
\mathcal{F}^{(2R),\,C_FT_RN_f}_{gg,\,\text{corr.}}(p,k;n) =&\frac{(k^2-2pk)^2}{2z^2(pk)^4}\left[\frac{1}{\eps}+4\right]+\frac{(1+z)(k^2-2pk)}{z^2(pk)^3}\left[\frac{1}{\eps}+4\right]
\nn\\
&+\frac{1}{z^2(pk)^2}\left[\frac{1+2z}{\eps}+1+8z\right]+\frac{2}{zpk(k^2-2pk)}\left[\frac{1}{\eps}+1\right],
\\
\label{ggCA}
\mathcal{F}^{(2R),\,C_AT_RN_f}_{gg,\,\text{corr.}}(p,k;n) =&\frac{4(1-z)(3+5\eps)}{9z^2k^2(k^2-2pk)}+\frac{(k^2-2pk)^2}{6z^2(pk)^4}-\frac{k^2-2pk}{6(1-z)(pk)^3}-\frac{1}{3(1-z)(pk)^2}
\nn\\
&+\frac{(1+2z)(k^2-2pk)}{6z^2(pk)^3}+\frac{2+z+z^2}{3z^2(pk)^2}-\frac{2(1-2z)}{3z^2pk(k^2-2pk)},
\\
\label{ggCA2}
\mathcal{F}^{(2R),\,C_A^2}_{gg,\,\text{corr.}}(p,k;n) =&\!-\!\frac{1-z}{z^2k^2(k^2-2pk)}\!\left[\frac{2}{\eps}\!+\!\frac{11}{3}\!+\!\frac{67}{9}\eps\right]\!-\frac{4}{z(k^2-2pk)^2}\frac{1}{\eps}\!+\!\frac{k^2-2pk}{12(1-z)(pk)^3}
\nn\\
&-\frac{(k^2-2pk)^2}{4z^2(pk)^4}\left[\frac{1}{\eps}+\frac{10}{3}\right]-\frac{k^2-2pk}{2z^2(pk)^3}\left[\frac{1+z}{\eps}+\frac{19+20z}{6}\right]
\nn\\
&+\frac{1}{6(1-z)(pk)^2}-\frac{1}{z^2(pk)^2}\left[\frac{2+z}{\eps}+\frac{14+19z+z^2}{6}\right]
\nn\\
&-\frac{1}{z^2pk(k^2-2pk)}\left(\frac{1+4z}{\eps}-\frac{11-28z}{6}\right)+
\mathcal{K}(p,k;n)
\nn\\
&+\mathcal{F}^{n^2-\text{sing.}}(p,k;n),
\end{align}
where the functions $\mathcal{K}$ and $\mathcal{F}^{n^2-\text{sing.}}$
are presented in eqs.~(\ref{fnreg})--(\ref{fn20}).
The individual coefficients $\mathcal{F}^{(2R),\,C_F^2}_{gq,\,\text{corr.}}$ and $\mathcal{F}^{(2R),\,C_AC_F}_{gq,\,\text{corr.}}$ of eq.~(\ref{gqcorr2R}) for the gluon-quark channel
are
\begin{align}
\label{gqCF}
\mathcal{F}^{(2R),\,C_F^2}_{gq,\,\text{corr.}}(p,k;n) = &-\frac{1-z}{z^2k^2(k^2-2pk)}\left[\frac{4}{\eps}+3+7\eps\right]+\frac{k^2-2pk}{4z^2(pk)^3}\left[\frac{1}{\eps}+2 \right]
\\
&-\frac{1}{2z^2(pk)^2}\left[\! \frac{1-z}{\eps}\!+\!3-2z\!\right]\!+\!\frac{1}{2z^2pk(k^2-2pk)}\!\!\left[\!\frac{2(2-z)}{\eps}\!+\!3(1-2z)\right]\!,
\nn
\\
\label{gqCA}
\mathcal{F}^{(2R),\,C_AC_F}_{gq,\,\text{corr.}}(p,k;n)=&\frac{2(1-z)}{z^2k^2(k^2-2pk)}\frac{1}{\eps}-\frac{4}{z(k^2-2pk)^2}\frac{1}{\eps}-\frac{k^2-2pk}{4z^2(pk)^3}\left[\frac{1}{\eps}+2\right]
\nn\\
&-\frac{1}{2z^2(pk)^2}\left[\frac{2+z}{\eps}+1+2z\right]-\frac{1}{z^2pk(k^2-2pk)}\left[\frac{3+2z}{\eps}+z\right]
\nn\\
&+\mathcal{K}(p,k;n)+\mathcal{F}^{n^2-\text{sing.}}(p,k;n).
\end{align}
The function $\mathcal{K}$ in eqs.~(\ref{ggCA2}) and (\ref{gqCA})
has the following expression:
\begin{align}
\label{fnreg}
\mathcal{K}(p,k;n) =& -\frac{1}{z^2(k^2-2pk)}\left[\frac{2(1-z)}{k^2}+\frac{1}{pk}\right]\frac{1}{\eps}-\frac{2}{z^2pk(k^2-2pk)}\left(\frac{\mathbf{k_T}^2}{k^2}\right)^{1+\eps}\left[\frac{1}{\eps}+\frac{\pi^2}{6}\eps\right]
\nn\\
&-\frac{1}{z^2(k^2-2pk)}\left[\frac{6(1-z)}{k^2}-\frac{1}{pk}\right]\ln\left(1+\frac{k^2}{\mathbf{k_T}^2}\right)-\frac{2}{z}\frac{k^2}{pk(k^2-2pk)^2}
\nn\\
&+\ln(z)\left[\frac{4(2-z)pk}{(k^2-2pk)^3}-\frac{2(4-z)}{z(k^2-2pk)^2} \right]+\frac{1}{z^2\mathbf{k_T}^2pk}\ln\left(1-\frac{\mathbf{k_T}^2}{2pk}\right)
\nn\\
&+\left[\frac{1}{z^2pk(k^2-2pk)}-\frac{2(1+z)(1-z)}{z^2k^2(k^2-2pk)}\right]\frac{k^2}{\mathbf{k_T}^2}\ln\left(1+\frac{\mathbf{k_T}^2}{k^2}\right)-\ln\left(1-\frac{k^2}{2pk}\right)
\nn\\
&\times\left[\frac{8(1-z)(2-z)(pk)^2}{k^2(k^2-2pk)^3}-\frac{1}{z^2k^2pk}-\frac{4(6-3z+z^2)pk}{zk^2(k^2-2pk)^2}-\frac{2(2+5z-z^2)}{z^2k^2(k^2-2pk)} \right]
\nn\\
&+\frac{8z(1-z)(2-z)(pk)^2}{\mathbf{k_T}^2(k^2-2pk)^3}\left[\frac{\ln(z)}{1-z}-\frac{2pk}{k^2}\ln\left(1-\frac{k^2}{2pk}\right)\right].
\end{align}
The function $\mathcal{F}^{n^2-\text{sing.}}$ in eqs.~(\ref{ggCA2}) and (\ref{gqCA})
embodies the non-smooth dependence of $\mathcal{F}^{(2R)}_{ga,\,\text{corr.}}$ on 
$n^2$. In the case of a time-like auxiliary vector we find the following result:
\begin{align}
\label{fnsing}
&\mathcal{F}^{n^2-\text{sing.}}(p,k;n) =\frac{4(1-z)}{z^2k^2(k^2-2pk)}\left(\frac{n^2k^2}{4(np)^2(1-z)^2}\right)^{-\eps}\left[\frac{1}{\eps}+\frac{\pi^2}{3}\eps \right]-\frac{2}{z(k^2-2pk)}
\\
&\times\left[\frac{k^2}{pk\left(\mathbf{k_T}^2+\frac{n^2(pk)^2}{(np)^2}\right)}+\frac{2(1-z)}{\mathbf{k_T}^2}\ln\left(\frac{n^2k^2(pk)^2}{(np)^2(k^2+\mathbf{k_T}^2)\left(\mathbf{k_T}^2+\frac{n^2(pk)^2}{(np)^2}\right)}\right)\right], \quad (n^2 > 0).\nn
\end{align}
Performing the calculation of $\mathcal{F}^{(2R)}_{ga,\text{corr.}}$ with a light-like
auxiliary vector we obtain
\begin{align}
\label{fn20}
\mathcal{F}^{n^2-\text{sing.}}(p,k;n) &=-\frac{2}{z(k^2-2pk)}  \\
& \times \left\{-\frac{1}{pk}+\frac{2(1-z)}{\mathbf{k_T}^2}\left[\ln\left(1+\frac{\mathbf{k_T}^2}{k^2}\right)+\left(\frac{k^2}{\mathbf{k_T}^2}\right)^{\eps}\frac{1+\eps}{\eps}\right]\right\}, \quad (n^2 = 0). \nn
\end{align}
We comment on the results in eqs.~(\ref{ggcorr2R})--(\ref{fn20})
by discussing the $\ep$ dependence and the $n^2$ dependence in turn.

The differential collinear function $\mathcal{F}^{(2R)}_{ga,\,\text{corr.}}(p,k;n)$ is singular
in the limit $k^2 \to 0$. The singularities are due to contributions that are proportional
to the factors $(k^2)^{-1-\ep}$ and  $(k^2)^{-1-2\ep}$. Such factors can be expressed in terms of plus-distributions of the variable $k^2$, which are integrable at $k^2 = 0$,
and contact terms of the type $\delta_+(k^2)$, whose coefficients are single poles $1/\ep$.
Owing to these single-pole effective contributions, in eqs.~(\ref{ggCA})--(\ref{fnsing}) the coefficients of the factors that are singular at $k^2 = 0$ are reported up to 
${\cal O}(\ep)$, by neglecting terms at ${\cal O}(\ep^2)$ and higher orders in $\ep$.\footnote{Other comments on the singularities in the limit $k^2 \to 0$ are presented
in section~\ref{sec:azcorrtmd}.}
In all the other contributions to eqs.~(\ref{ggCF})--(\ref{fn20}) that are not singular in the limit $k^2 \to 0$, we limit ourselves to presenting the results up to ${\cal O}(\ep^0)$.

The contributions of the function $\mathcal{F}^{n^2-\text{sing.}}$ to 
$\mathcal{F}^{(2R)}_{ga,\,\text{corr.}}$ have a non-smooth dependence on $n^2$. All the 
other contributions of eqs.~(\ref{ggCF})--(\ref{fnreg}) to 
$\mathcal{F}^{(2R)}_{ga,\,\text{corr.}}$ have a smooth and harmless dependence on $n^2$ and, similarly to
$\mathcal{F}^{(1R1V)}_{ga,\,\text{corr.}}$ in eqs.~(\ref{gg1R1V}) and (\ref{gq1R1V}), they are presented by setting $n^2=0$.

By\, direct\, inspection\, and\, comparison\, of\, eqs.~(\ref{fnsing}) and (\ref{fn20}),\, we\, can\, see\, that\,
$\mathcal{F}^{n^2-\text{sing.}}(p,k;n)$ has a non-smooth and singular dependence 
on $n^2$ in the limit $n^2 \to 0$.
The $n^2$ dependence of $\mathcal{F}^{n^2-\text{sing.}}(p,k;n)$ is also related to the issue
of the rapidity divergences, which we have already considered in 
sections~\ref{sec:tmdfun} and \ref{sec:beam}.
If $n^2=0$, $\mathcal{F}^{n^2-\text{sing.}}$ is given by eq.~(\ref{fn20}) and its contribution to $\mathcal{F}^{(2R)}_{ga,\,\text{corr.}}(p,k;n)$ in the limit $k^- \to +\infty$
is due to the following factor:
\begin{equation}
\label{rapdiv}
\left(k^2 \right)^{-\ep} \;\mathcal{F}^{n^2-\text{sing.}}(p,k;n) \simeq
\frac{2(1-z)}{z^2} \;\frac{\left(\mathbf{k_T}^2 \right)^{-1 -\ep} }{p^+ \;k^-}
\;\frac{1+\eps}{\eps}, \quad \quad k^- \to + \infty \quad (n^2 = 0),
\end{equation}
which is proportional to $1/k^-$. This contribution is not integrable over $k^-$
in the region where $k^- \to +\infty$. This singular behaviour is a manifestation at 
${\cal O}(\as^2)$ of the rapidity divergences that we have discussed in 
section~\ref{sec:tmdfun}. 
If $n^2 > 0$, the expression of $\mathcal{F}^{n^2-\text{sing.}}$ is given by 
eq.~(\ref{fnsing}) and in the limit $k^- \to +\infty$ we have
\begin{align}
\label{norapdiv}
\left(k^2 \right)^{-\ep}\mathcal{F}^{n^2-\text{sing.}}(p,k;n) &\simeq
%- \frac{4(1-z)}{z^2} \frac{\left( 2 k^+ k^- \right)^{-\ep} }{\mathbf{k_T}^2 \,p^+ \;k^-}
 \frac{2(1-z)\left( 2 k^+ k^- \right)^{-\ep} }{z^2 \,\mathbf{k_T}^2 \,p^+ \;k^-}
 \\
 &\times\left[\frac{1}{1+\frac{n^2(pk)^2}{(np)^2 \mathbf{k_T}^2}}-
\ln\left(1+\frac{\mathbf{k_T}^2 (np)^2}{n^2(pk)^2}\right)\right],\; k^- \to + \infty \quad (n^2 > 0), \nn
\end{align}
%where we have neglected $n^2$-independent contributions of ${\cal O}((k^-)^{-2-\ep})$.
where we have neglected contributions of ${\cal O}((k^-)^{-2-2\ep})$ and 
${\cal O}((k^-)^{-2-\ep})$ that are $n^2$-independent at large values 
of $k^-$.
%~\footnote{*** To be rephrased differently?}
The overall factor $(k^-)^{-1-\ep}$ in the right-hand side of eq.~(\ref{norapdiv})
is not integrable\footnote{In our computation we have $\ep < 0$ to regularize the IR divergences. Therefore, the $\ep$-dependent factor $(k^-)^{-\ep}$
does not remove the rapidity divergence due to $1/k^-$ at $k^- \to + \infty$.} as $k^- \to + \infty$.
However, since $n^2(pk)^2/(np)^2= n^2 (k^-)^2/(n^-)^2$, both terms in the square bracket of
eq.~(\ref{norapdiv}) screen the non-integrable behaviour of the factor $(k^-)^{-1-\ep}$.
Therefore, $\mathcal{F}^{n^2-\text{sing.}}(p,k;n)$ does not produce rapidity divergences 
in the expression of $\mathcal{F}^{(2R)}_{ga,\,\text{corr.}}(p,k;n)$. The $n^2$ dependence of the square bracket term in eq.~(\ref{norapdiv}) leads to a dynamical damping at large values
of $k^-$, and the non-integrable behaviour of the type $(k^-)^{-1-\ep}$ is effective
only inside the region where $(k^-)^2 \ltap (n^-)^2 \,\mathbf{k_T}^2/n^2$. Since 
$\mathbf{k_T}^2 \leq 2 k^+ k^-$ (i.e., $k^2 \geq 0$), this region is contained in the region
with $k^- \ltap 2 (n^-)^2 \,k^+/n^2$, where the collinear approximation of the squared amplitudes is certainly valid. Obviously, the issue of rapidity divergences is not relevant
by considering the differential collinear functions $\mathcal{F}_{ga,\,\text{corr.}}(p,k;n)$
at finite and not large values of~$k^-$.

As discussed at the end of sections~\ref{sec:diffcollfun} and \ref{sec:sldiff}, setting $n^2=0$, $\mb{q_T}=\mb{k_T}$ and $t=2zpk$, the collinear function 
$\mathcal{F}_{ca}(p,k;n)$ is related to the SCET differential beam function 
${\cal B}_{ca}(z,t,\mb{q_T})$ \cite{Jain:2011iu, Mantry:2009qz, Mantry:2010mk}
at the partonic level. At ${\cal O}(\as^2)$ we have
${\cal B}_{ca}^{(2)}(z,t,\mb{q_T})\!=\! {F}^{(1R1V)}_{ca}(p,k;n) + 
{F}^{(2R)}_{ca}(p,k;n)$ ($a=q,\bar q,g$), where ${\cal B}_{ca}^{(2)}$ is the second-order
contribution to ${\cal B}_{ca}$. The azimuthal-independent component 
${\cal B}_{ca, \,\text{az.in.}}^{(2)}$  at ${\cal O}(\as^2)$ was computed in 
refs.~\cite{Gaunt:2014xxa,Gaunt:2020xlc}.
Our expressions in eqs.~(\ref{gg1R1V}), (\ref{gq1R1V}), (\ref{ggcorr2R}) and 
(\ref{gqcorr2R}) give the result
at ${\cal O}(\as^2)$
for the azimuthal-correlation component ${\cal B}_{ga, \,\text{corr.}}^{(2)}$ of the SCET differential beam function.

\subsection{TMD functions}
\label{sec:azcorrtmd}

We use our results at ${\cal O}(\as^2)$ for the differential collinear functions 
$\mathcal{F}_{ga,\, \text{corr.}}(p,k;n)$ to compute the TMD functions
${F}_{ga,\,\text{corr.}}$ at the corresponding perturbative order.
Similarly to our computations of the TMD functions at ${\cal O}(\as)$ (see section~\ref{sec:tmdfun}), we use a time-like auxiliary vector $n^\mu$ to avoid rapidity divergences.
However, we comment on the $n^2$ dependence of the various contributions to 
${F}_{ga,\,\text{corr.}}$.

The ${\cal O}(\as^2)$ contributions to ${F}_{ga,\,\text{corr.}}$ are denoted by
${F}^{(1R1V)}_{ga,\,\text{corr.}}$ and ${F}^{(2R)}_{ga,\,\text{corr.}}$, and they are obtained
according to the definition in eq.~(\ref{sltmdgen})
by integrating $\mathcal{F}^{(1R1V)}_{ga,\,\text{corr.}}(p,k;n)$ and
$\mathcal{F}^{(2R)}_{ga,\,\text{corr.}}(p,k;n)$ over the momentum $k^\mu$.
The integrations over $k^+$ and $\mb{k_T}$ are trivial,
and the integration over $k^-$ extends up to $+\infty$.

The $k^-$ integration of the expressions of $\mathcal{F}^{(1R1V)}_{ga,\,\text{corr.}}(p,k;n)$
($a=g,q,{\bar q}$) in eqs.~(\ref{gg1R1V}) and (\ref{gq1R1V})
is elementary since it simply sets $k^-=\mb{q_T}^2/(2(1-z)p^+)$ by using the delta function
$\delta_+(k^2)$. The ensuing expressions of ${F}^{(1R1V)}_{ga,\,\text{corr.}}$ 
contain double and single poles, $1/\ep^2$ and $1/\ep$, and the dominant contributions in the limit $\mb{q_T}^2 \to 0$ do not depend on $n^2$. These contributions
can be computed by setting $n^2=0$, since the $n^2$ dependence leads to terms of 
${\cal O}(n^2 \mb{q_T}^2/(np)^2)$.
 
The $k^-$ integration of $\mathcal{F}^{(2R)}_{ga,\,\text{corr.}}(p,k;n)$ extends over the region $k^-_{\rm min} \leq k^- < +\infty$. The lower limit 
$k^-_{\rm min}=\mb{q_T}^2/(2(1-z)p^+)$ derives from $k^2 \geq 0$. As noticed in 
section~\ref{sec:azcorrdiff},
the expressions of $\mathcal{F}^{(2R)}_{ga,\,\text{corr.}}(p,k;n)$ contain contributions of the type $(k^2)^{-1-\ep}$ and  $(k^2)^{-1-2\ep}$. The integration of these contributions
over $k^-$ produces single poles $1/\ep$ from the region where 
$k^- \to k^-_{\rm min}$ (i.e., $k^2 \to 0$). The $k^-$ integration in the region
$k^- \to +\infty$ produces rapidity divergences if $n^2=0$. We use an auxiliary vector with
$n^2 > 0$ and the corresponding expression of $\mathcal{F}^{n^2-\text{sing.}}(p,k;n)$
in eq.~(\ref{fnsing}), which eventually leads to contributions of the type 
$\ln(n^2 \mb{q_T}^2/(np)^2)$ to ${F}^{(2R)}_{ga,\,\text{corr.}}$. The
remaining contributions of 
eqs.~(\ref{ggCF})--(\ref{fnreg})
to $\mathcal{F}^{(2R)}_{ga,\,\text{corr.}}(p,k;n)$ produce terms with a smooth $n^2$
dependence in ${F}^{(2R)}_{ga,\,\text{corr.}}$, which can be evaluated by setting 
$n^2=0$ (i.e., neglecting subdominant terms of ${\cal O}(n^2 \mb{q_T}^2/(np)^2)$).
Regarding the dependence of ${F}^{(2R)}_{ga,\,\text{corr.}}$ on the momentum fraction
$z=1-k^+/p^+$, the $k^-$ integration of $\mathcal{F}^{(2R)}_{ga,\,\text{corr.}}(p,k;n)$
can be carried out in terms of rational, logarithmic and di-logarithmic functions of the variable $z$.

Having computed the ${\cal O}(\as^2)$ contributions ${F}^{(1R1V)}_{ga,\,\text{corr.}}$ and
${F}^{(2R)}_{ga,\,\text{corr.}}$ to the TMD function ${F}_{ga,\,\text{corr.}}$
in $\mb{q_T}$ space, we perform their Fourier transformations to obtain the corresponding 
contributions ${\widetilde F}^{(1R1V)}_{ga,\,{\rm corr.}}$ and
${\widetilde F}^{(2R)}_{ga,\,{\rm corr.}}$ of eq.~(\ref{bspacecorr}) to the 
$\mb{b}$ space function ${\widetilde F}_{ga,\,{\rm corr.}}$ in 
eqs.~(\ref{tmdgb}) and (\ref{tmdgbaz}). The explicit Fourier integrals that are required at 
${\cal O}(\as^2)$ are given in eq.~(\ref{ftcor})
with $\delta=\ep, 2\ep$ and $m=0,1$.

The expressions of ${\widetilde F}^{(1R1V)}_{ga,\,{\rm corr.}}$ and
${\widetilde F}^{(2R)}_{ga,\,{\rm corr.}}$ have double and single poles, $1/\ep^2$ and 
$1/\ep$, in the limit $\ep \to 0$. Part of the single-pole terms are cancelled by performing the UV renormalization of the bare coupling: we use eq.~(\ref{eq:alphasren})
up to ${\cal O}(\as)$ and we choose the renormalization scale $\mu_R^2=b_0^2/\mb{b}^2$.
The remaining $\ep$ poles have to be treated by using the
IR factorization formula in eq.~(\ref{IRfactcor}), which we expand up to ${\cal O}(\as^2)$:
\begin{align}
\label{fcorrat2}
{\widetilde F}_{ga,\,{\rm corr.}}(z;\mb{b}^2/b_0^2,{\widetilde \lambda}) =&
\frac{\as(b_0^2/\mb{b}^2)}{\pi} \;
{\widetilde G}^{(1)}_{ga}(z;\ep, {\widetilde \lambda})
+ \left(\frac{\as(b_0^2/\mb{b}^2)}{\pi} \right)^2
\Bigl[ \;Z_g^{(1)}({\widetilde \lambda}) \;{\widetilde G}^{(1)}_{ga}(z;\ep, {\widetilde \lambda}) \Bigr. \nn \\
&- \Bigl. \frac{1}{\ep} \sum_b \int_z^1 \frac{dx}{x} 
\;{\widetilde G}^{(1)}_{gb}(x;\ep, {\widetilde \lambda}) \;P^{(1)}_{ba}(z/x)
+{\widetilde G}^{(2)}_{ga}(z;\ep, {\widetilde \lambda}) \;
\Bigr] + {\cal O}(\as^3),
\end{align}
where
${\widetilde \lambda}= n^2 (b_0^2/\mb{b}^2)/(2zpn)^2$.
The first-order functions $P^{(1)}_{ba}, Z_g^{(1)}$ and ${\widetilde G}^{(1)}_{ga}$ are
given in eqs.~(\ref{AP1}), (\ref{z1tmd}) and (\ref{gtil1}).
Our computation of ${\widetilde F}^{(1R1V)}_{ga,\,{\rm corr.}}$ and
${\widetilde F}^{(2R)}_{ga,\,{\rm corr.}}$ gives an explicit expression up to  
${\cal O}(\ep^0)$ of the entire contribution in the square bracket on the right-hand side of eq.~(\ref{fcorrat2}) and, knowing $P^{(1)}_{ba}, Z_g^{(1)}$ and ${\widetilde G}^{(1)}_{ga}$,
we can explicitly determine ${\widetilde G}^{(2)}_{ga}$ up to ${\cal O}(\ep^0)$.
Our result for ${\widetilde G}^{(2)}_{ga}$ is finite and independent of 
${\widetilde \lambda}$ in the four-dimensional limit $\ep \to 0$, therefore confirming the validity of the IR factorization structure of eq.~(\ref{IRfactcor}).

Setting $\ep = 0$ we have
\begin{equation}
{\widetilde G}^{(2)}_{ga}(z;\ep=0, {\widetilde \lambda}) = {G}^{(2)}_{ga}(z),
\end{equation}
where ${G}^{(2)}_{ga}(z)$ ($a=g,q,{\bar q}$) are the contributions at ${\cal O}(\as^2)$
to the transverse-momentum resummation functions in eq.~(\ref{azcorcolg}).
We find the following results:
\begin{align}
\label{ggcorr2}
G^{(2)}_{gg}(z) =& C_A^2\bigg\{\!\!-\frac{37}{36z}+\frac{31}{18}-\frac{13z}{12}+\frac{11z^2}{36}-\ln(z)\left[\frac{1}{z}+\frac{19}{12}\right]\!+\!\frac{\ln^2(z)}{2}\!+\!\frac{1-z}{z}\left[\text{Li}_2(z)-\frac{\pi^2}{6}\right]\!\!\bigg\} \nn \\
& + C_F N_f\left\{\frac{(1-z)^3}{2z}-\frac{1}{4}\ln^2(z)\right\}+C_A  N_f\left\{-\frac{17}{36z}+\frac{4}{9}+\frac{z}{12}+\frac{z^2}{36}-\frac{1}{6}\ln(z)\right\} \nn \\
&
%- h^{(1)}_g \,G^{(1)}_{gg}(z)
- h^{(1)}_g \,C_A \,\frac{1-z}{z},
\\\nn\\
%\end{align}
%\begin{align}
\label{gqcorr2}
G^{(2)}_{gq}(z) =& C_F^2\left\{-\frac{1-z}{2}+\frac{5}{4}\ln(z)-\frac{1}{4}\ln^2(z)-\frac{1-z}{2z}\bigg[\ln(1-z)+\ln^2(1-z)\bigg]\right\}
\nn\\
&+C_F N_f\left\{-\frac{1-z}{3z}\left[\frac{2}{3}+\ln(1-z)\right] \right\}+ C_A C_F\bigg\{\!-\frac{11}{18z}+\frac{10}{9}-\frac{z}{2}-\ln(z)\left[\frac{1}{z}+\frac{5}{2}\right]
\nn\\
&+\frac{1}{2}\ln^2(z)+\frac{1-z}{z}\left[\frac{5}{6}\ln(1-z)+\frac{1}{2}\ln^2(1-z)+\text{Li}_2(z)-\frac{\pi^2}{6}\right] \!\bigg\}
%- h^{(1)}_g \,G^{(1)}_{gq}(z)
- h^{(1)}_g \,C_F \,\frac{1-z}{z},
\end{align}
and $G^{(2)}_{g{\bar q}}(z)=G^{(2)}_{gq}(z)$.
These results regard a generic resummation scheme, which is specified by the coefficient 
$h^{(1)}_g$ of the IR factorization function $Z_c^{(1)}$ with $c=g$ in eq.~(\ref{z1tmd}). 
We recall that the hard scheme \cite{Catani:2013tia} is defined by setting $h^{(1)}_g=0$,
and the `minimal' scheme has $h^{(1)}_g=  \pi^2 C_A/24$. 
We find full agreement with the results for $G^{(2)}_{ga}(z)$ that were obtained in 
refs.~\cite{Luo:2019bmw, Gutierrez-Reyes:2019rug}
%by using TMD functions defined in a SCET framework
(ref.~\cite{Luo:2019bmw} uses the hard scheme, while 
ref.~\cite{Gutierrez-Reyes:2019rug} uses the minimal scheme)
by using TMD functions defined in a SCET framework.

We note that our results at ${\cal O}(\as^2)$ for the azimuthally-correlated TMD function 
${\widetilde F}_{ga,\,{\rm corr.}}$ are an important and highly non-trivial check of the IR factorization formulae in eqs.~(\ref{IRfactav}) and (\ref{IRfactcor}) and of our framework
to define and compute TMD functions. In particular, we find that the {\it same}
IR singular and $n^2$ dependent contribution $Z_g^{(1)}$ of eq.~(\ref{z1tmd})
is involved in the IR factorization of {\it both} the azimuthal-independent function
${\widetilde F}_{ga,\,{\rm az.in.}}$ at ${\cal O}(\as)$  (see section~\ref{sec:tmdfun})
and the azimuthally-correlated function ${\widetilde F}_{ga,\,{\rm corr.}}$ at 
${\cal O}(\as^2)$. At the purely technical level, we also note that the $n^2$ dependence
of $Z_g^{(1)}$ has an entirely different origin in the calculations of 
${\widetilde F}_{ga,\,{\rm az.in.}}$ at ${\cal O}(\as)$ and 
${\widetilde F}_{ga,\,{\rm corr.}}$ at 
${\cal O}(\as^2)$ (see eq.~(\ref{tmdplus}) vs. eq.~(\ref{fnsing}) and related comments).

The agreement between our results in eqs.~(\ref{ggcorr2}) and 
(\ref{gqcorr2}) and those in refs.~\cite{Luo:2019bmw, Gutierrez-Reyes:2019rug} is also highly non-trivial. Indeed,  this agreement is obtained by using fully independent methods
and, in particular, very different procedures to deal with the issue of rapidity divergences.

\section{TL collinear functions}
\label{sec:tlfunct}

In this section we consider the perturbative computation of the differential collinear functions $\cf^{{\rm TL}}_{ca}$ and of the TMD functions $F^{{\rm TL}}_{ca}$ 
in the TL collinear region. We closely follow the presentation of the analogous 
SL results in sections~\ref{sec:sldiff}, \ref{sec:tmdfun} and \ref{sec:azcorr}, with brief comments to highlight the main
differences between SL and TL functions.

\subsection{TL differential collinear functions at ${\cal O}(\as)$}
%\label{sec:sldiff}
\label{sec:tldiff}

The perturbative expansion of the TL collinear functions in 
eqs.~(\ref{gcolfun}), (\ref{fincor}) and (\ref{qcolfun})
can be written in the following form:
\begin{equation}
%\label{fexp12}
\label{tlfexp12}
\cf^{\rm TL}(p,k;n) = \cf^{{\rm TL} \,(1R)}(p,k;n) + 
\left[ \; \cf^{{\rm TL} \,(2R)}(p,k;n) + \cf^{{\rm TL} \,(1R1V)}(p,k;n) \; \right] 
+ {\cal O}(\as^3),
\end{equation}
where, analogously to eq.~(\ref{fexp12}), we have not explicitly denoted subscripts and superscripts in $\cf^{\rm TL}$.

The first-order term $\cf^{{\rm TL} \,(1R)}_{ca}(p,k;n)$ in eq.~(\ref{tlfexp12}) is obtained from the contribution of the tree-level collinear kernel 
$\cp^{(0)}_{{c}\rightarrow a_1 a}(k,p;n)$ to eqs.~(\ref{gcolfun}) and (\ref{qcolfun}).
Comparing eqs.~(\ref{gcolfun}) and (\ref{qcolfun}) with the analogous expression in 
eq.~(\ref{slcolfun}) for the SL function $\cf^{(1R)}_{ca}(p,k;n)$, we straightforwardly obtain the following relation:
\begin{equation}
\label{1Rcross}
\cf_{ca}^{{\rm TL} \,(1R)}(p,k;n) = \frac{\mathcal{N}_a(\eps)}{\mathcal{N}_c(\eps)}
\;\left[ \cf_{ca}^{(1R)}(p,k;n) \right]_{p \to -p},
\end{equation}
which is valid for both the azimuthal-independent and azimuthal-correlation components of the differential collinear functions.
We note that the SL-TL crossing relation in eq.~(\ref{1Rcross}) is valid since the 
tree-level kernel $\cp^{(0)}_{{c}\rightarrow a_1 a}(k,p;n)$ is process independent and it has a rational dependence on the external momenta, which permits a straightforward replacement $p \leftrightarrow -p$. A similar comment applies to any tree-level kernel
$\cp^{(0)}_{{c}\rightarrow a_1 a_2 \dots a}$, and this leads to ensuing SL-TL crossing relations at higher perturbative orders 
(see, e.g., eq.~(\ref{2Rcross})).

Using the SL results in eqs.~(\ref{f1R}) and (\ref{fgacor}) and the relation in 
eq.~(\ref{1Rcross}), we obtain the following explicit expressions for the 
azimuthal-independent\footnote{Similarly to the case of the SL functions, we introduce the notation $\cf^{{\rm TL}}_{ca}=\cf^{{\rm TL}}_{ca, \,\text{az.in.}}$ and
$F^{{\rm TL}}_{ca}=F^{{\rm TL}}_{ca, \,\text{az.in.}}$ ($c=q,{\bar q}$) for the TL collinear functions in the quark and antiquark partonic channels.}
and azimuthal-correlation components of the TL collinear functions at ${\cal O}(\as)$:
\begin{align}
%\label{f1R}
\label{tlf1R}
\cf^{{\rm TL} \,(1R)}_{ca, \,\text{az.in.}}(p,k;n) &= \frac{\alpha^u_{\rm S} \,\mu_0^{2\eps}
\,S_{\eps}}{\pi} \;\frac{e^{\eps\gamma_E}}{\pi^{1-\eps}} 
\;\frac{\delta_{+}(k^2)}{pk} \;{\widehat P}_{ac}(z_n^{\rm TL}; \ep),
\;\;\; \quad c=g,q,{\bar q}, \\
%\end{equation}
%\begin{equation}
%\label{fgacor}
\label{tlfggcor}
\mathcal{F}^{{\rm TL} \,(1R)}_{gg, \,\text{corr.}}(p,k;n) &= - \;\frac{\alpha^u_{\rm S} \,\mu_0^{2\eps} \,S_{\eps}}{\pi} \;\frac{e^{\eps\gamma_E}}{\pi^{1-\eps}} 
\;\frac{\delta_{+}(k^2)}{pk} \;C_A \;z_n^{\rm TL}(1-z_n^{\rm TL}) 
%\quad\quad a=q,{\bar q}
, \\
%\end{equation}
%\begin{equation}
%\label{fgacor}
\label{tlfgacor}
\mathcal{F}^{{\rm TL} \,(1R)}_{ga, \,\text{corr.}}(p,k;n) &= \;\frac{\alpha^u_{\rm S} \,\mu_0^{2\eps} \,S_{\eps}}{\pi} \;\frac{e^{\eps\gamma_E}}{\pi^{1-\eps}} 
\;\frac{\delta_{+}(k^2)}{pk} \;T_R \;\frac{z_n^{\rm TL}(1-z_n^{\rm TL})}{1-\ep}, 
\quad\quad a=q,{\bar q},
\end{align}
where we have introduced the $n$-dependent variable $z_n^{\rm TL}$,
\begin{equation}
%\label{znvar}
z_n^{\rm TL} = \frac{np}{n(p+k)}.
\end{equation}
The right-hand side of eq.~(\ref{tlf1R}) is proportional to the $d$-dimensional 
Altarelli--Parisi kernel ${\widehat P}_{ac}(x; \ep)$ of eqs.~(\ref{rqq})--(\ref{rqqprime}),
and we note that we have directly used the following well-known (see, e.g., 
ref.~\cite{Catani:1996vz})
crossing symmetry relation:
\begin{equation}
\label{crossap}
\frac{\mathcal{N}_a(\eps)}{\mathcal{N}_c(\eps)}
\; \left[ {\widehat P}_{ca}(x; \ep) \right]_{x=1/z} = 
- \frac{1}{z} \;{\widehat P}_{ac}(z; \ep).
\end{equation}

As discussed at the end of section~\ref{sec:diffcollfun}, the collinear function 
$\cf^{{\rm TL}}(p,k;n)$ is related to the SCET fragmenting jet function 
${\cal G}(z,s,\mathbf{p_\perp})$ of ref.~\cite{Jain:2011iu}. More precisely, setting
$n^2=0$, $s=(p+k)^2$ and $\mathbf{p_\perp}=z\mathbf{k_T}$, the first-order collinear function
$\cf^{{\rm TL} \,(1R)}_{ca}(p,k;n)$ is equal to the first-order contribution
${\cal G}_{ca}^{(1)}(z,s,\mathbf{p_\perp})$ to $\cal G$ at the partonic level.
The azimuthal-independent component of ${\cal G}_{ca}^{(1)}$ was computed in 
ref.~\cite{Jain:2011iu}, and the expression in eq.~(\ref{tlf1R}) agrees with the results presented therein. The azimuthal-correlation component of ${\cal G}_{ga}^{(1)}$ is given by eqs.~(\ref{tlfggcor}) and (\ref{tlfgacor}).

\subsection{TMD functions and IR factorization in the TL region}
%\label{sec:tmdfun}
\label{sec:tltmdfun}

The TL TMD functions $F^{{\rm TL}}_{ca}$ are obtained from the corresponding differential collinear functions $\cf^{{\rm TL}}_{ca}(p,k;n)$ by using eqs.~(\ref{tlgtmd}) and 
(\ref{tmdazin})--(\ref{tlqtmd}).

The contribution of ${\cal O}(\as)$ to $F^{{\rm TL}}_{ca}$ is denoted by 
$F^{{\rm TL} \,(1R)}_{ca}$, and it is computed by using the expressions of 
$\cf^{{\rm TL} \,(1R)}_{ca}(p,k;n)$ in eqs.~(\ref{tlf1R})--(\ref{tlfgacor}).
We define and compute the TL functions $F^{{\rm TL}}_{ca}$ by using a {\it time-like}
auxiliary vector $n^\mu$, since the use of a light-like vector produces rapidity 
divergences. The origin of the rapidity divergences is exactly similar in the TL and SL regions, and we can briefly follow the discussion 
%that we have presented 
in section~\ref{sec:tmdfun}. At ${\cal O}(\as)$ the non-smooth dependence of 
$F^{{\rm TL} \,(1R)}_{ca}$ on $n^2$ is produced by the contribution 
${\widehat P}^{\rm \,sing}_{ac}(z_n^{\rm TL})$
(see eqs.~(\ref{psr})--(\ref{preg})) to the kernel ${\widehat P}_{ac}(z_n^{\rm TL}; \ep)$
in the right-hand side of eq.~(\ref{tlf1R}). Indeed, such contribution is proportional
to the following factor:
\begin{align}
%\label{znsing}
\label{tlznsing}
\frac{1}{1-z_n^{\rm TL}} &= 1+ \frac{p^+}{k^+ + \frac{n^2}{2 np}  \frac{k^-}{n^-} p^+}
= 1+ \frac{z}{1-z +\frac{z^2 n^2 \mathbf{q_T}^2}{(1-z) (2 np)^2}}  \\
\label{tltmdplus}
&= \frac{1}{2} \;\ln\left(\frac{1}{\lambda}\right) \;\delta(1-z) 
+ \left( \frac{1}{1-z} \right)_+ + {\cal O}({\sqrt \lambda}),
\end{align}
where
$\lambda = n^2 \mb{q_T}^2/{(2pn)^2}$.
In the last equality of eq.~(\ref{tlznsing}) we have implemented the kinematics of the TL TMD function at ${\cal O}(\as)$ (i.e., $k^2=0, k^+=p^+ (1-z)/z, 
\mathbf{k_T}= -\mathbf{q_T}$), and in eq.~(\ref{tltmdplus}) we have properly performed the limit $\mathbf{q_T} \to 0$ by using the same procedure as in eq.~(\ref{tmdplus}).
The small-$\mathbf{q_T}$ limit of the remaining dependence of 
$F^{{\rm TL} \,(1R)}_{ca}$ on $z_n^{\rm TL}$ can be evaluated by simply setting 
$z_n^{\rm TL}=z +{\cal O}(\mathbf{q_T}^2)$.
Eventually, we obtain the following final results for the first-order TMD functions in the TL region: 
\begin{align}
%\label{tmdqt}
\label{tmdqttl}
&F^{{\rm TL} \,(1R)}_{ca,\,{\rm az.in.}}\!\left(z;\mb{q_T}^2,\frac{n^2 \mb{q_T}^2}{(2pn/z)^2}\right) 
= \frac{\alpha^u_{\rm S} \,\mu_0^{2\eps}
\,S_{\eps}}{\pi} \;\frac{e^{\eps\gamma_E}}{\pi^{1-\eps} \,\mathbf{q_T}^2}  \\
& \quad \times \left\{{\widehat P}^{\rm \,reg}_{ac}(z; \ep) + \delta_{ca} \;A_c^{(1)}
\left[ \left( \frac{1}{1-z} \right)_+ 
- \frac{1}{2} \;\ln\left( \frac{n^2 \mb{q_T}^2}{(2pn/z)^2} \right) \,\delta(1-z) 
\right] \right\},  \;\;\;\;\;\; c=g,q,{\bar q},\nn \\
%\end{align} 
%\begin{equation}
%\label{tmdgacor}
\label{tmdggcortl}
&F^{{\rm TL} \,(1R)}_{gg,\,{\rm corr.}}\!\left(z;\mb{q_T}^2,\frac{n^2 \mb{q_T}^2}{(2pn/z)^2}\right)
 =- \;\frac{\alpha^u_{\rm S} \,\mu_0^{2\eps} \,S_{\eps}}{\pi} \;\frac{e^{\eps\gamma_E}}{\pi^{1-\eps} \,\mathbf{q_T}^2} 
\; C_A\;z(1-z), \\
%\quad\quad a=q,{\bar q},g \;\;.
%\end{equation}
%\begin{equation}
%\label{tmdgacor}
\label{tmdgqcortl}
&F^{{\rm TL} \,(1R)}_{ga,\,{\rm corr.}}\!\left(z;\mb{q_T}^2,\frac{n^2 \mb{q_T}^2}{(2pn/z)^2}\right)
 = \frac{\alpha^u_{\rm S} \,\mu_0^{2\eps} \,S_{\eps}}{\pi} \;\frac{e^{\eps\gamma_E}}{\pi^{1-\eps} \,\mathbf{q_T}^2} 
\; T_R\;\frac{z(1-z)}{1-\ep}\;\;, 
\quad\quad a=q,{\bar q}.
%\end{equation}
\end{align}

The Fourier transformation ${\widetilde F}^{\rm TL}_{ca}$ of the TL TMD function
${F}^{\rm TL}_{ca}$ is defined analogous to the SL case: we simply use 
eqs.~(\ref{tmdqb})--(\ref{tmdgbaz})
with the replacements ${F}_{ca} \to {F}^{\rm TL}_{ca}$ and ${\widetilde F}_{ca}
\to {\widetilde F}^{\rm TL}_{ca}$. The perturbative expansion of the $\mb{b}$ space
functions ${\widetilde F}^{\rm TL}_{ca}$ can be written as follows:
\begin{align}
%\label{bspaceexp}
\label{bspaceexptl}
{\widetilde F}^{\rm TL}_{ca, \,{\rm az.in.}}\!\left(z;\frac{\mb{b}^2}{b_0^2},\frac{n^2 b_0^2}{(2pn/z)^2 \,\mb{b}^2}\right) \!=& \delta_{ca} \;\delta(1-z) 
+ {\widetilde F}^{{\rm TL} \,(1R)}_{ca, \,{\rm az.in.}}\!\left(z;\frac{\mb{b}^2}{b_0^2},\frac{n^2 b_0^2}{(2pn/z)^2 \,\mb{b}^2}\right)  + {\cal O}(\as^2),
\nn\\ 
& \hspace{5cm}\,\,\,\,\, c=g,q,{\bar q},
\\
\label{bspacecorrtl}
{\widetilde F}^{\rm TL}_{ga,\,{\rm corr.}}\!\left(z;\frac{\mb{b}^2}{b_0^2},
\frac{n^2 b_0^2}{(2pn/z)^2 \,\mb{b}^2}\right) =& 
{\widetilde F}^{{\rm TL} \,(1R)}_{ga,\,{\rm corr.}}\!\left(z;\frac{\mb{b}^2}{b_0^2},
\frac{n^2 b_0^2}{(2pn/z)^2 \,\mb{b}^2}\right)  \\
& \hspace{-4.2cm} + 
\left[ {\widetilde F}^{{\rm TL} \,(2R)}_{ga,\,{\rm corr.}}\!\left(z;\frac{\mb{b}^2}{b_0^2},
\frac{n^2 b_0^2}{(2pn/z)^2 \,\mb{b}^2}\right) + 
{\widetilde F}^{{\rm TL} \,(1R1V)}_{ga,\,{\rm corr.}}\!\left(z;\frac{\mb{b}^2}{b_0^2},
\frac{n^2 b_0^2}{(2pn/z)^2 \,\mb{b}^2}\right) \right]
+ {\cal O}(\as^3),
\nn\\
& \hspace{5cm}\,\,\,\,\, a=g,q,{\bar q}. \nn
%\label{bspacecorr}
%\label{bspacecorrtl}
\end{align}
The first-order contributions ${\widetilde F}^{{\rm TL} \,(1R)}_{ca, \,{\rm az.in.}}$
and ${\widetilde F}^{{\rm TL} \,(1R)}_{ga,\,{\rm corr.}}$ are directly computed from 
eqs.~(\ref{tmdqttl})--(\ref{tmdgqcortl})
by using the Fourier transformations in eqs.~(\ref{ftav}) and (\ref{ftcor}).
We obtain the following results:
\begin{align}
&{\widetilde F}^{{\rm TL} \,(1R)}_{ca,\,{\rm az.in.}}\!\left(z;\frac{\mb{b}^2}{b_0^2},
\frac{n^2 b_0^2}{(2pn/z)^2 \,\mb{b}^2}\right) = \frac{\alpha^u_{\rm S}\,S_{\eps}}{\pi} 
\;\left( \frac{\mu_0^2 \,\mb{b}^2}{b_0^2} \right)^{+\ep}
\;\frac{e^{- \eps\gamma_E} \, \Gamma(1-\ep)}{(-\eps)} 
\Bigg\{ {P}^{(1)}_{ac}(z) + \ep \;{\widehat P}^{\prime}_{ac}(z; \ep)
\nn\\
%\label{azin1b}
\label{azin1btl}
&\hspace{0.5cm}- \delta_{ca} \,\delta(1-z) \left[ 
\frac{A_c^{(1)}}{2} \left( \frac{1}{\ep} + \psi(1-\ep) - \psi(1) 
+ \ln \frac{n^2 b_0^2}{(2pn/z)^2 \,\mb{b}^2}  \right) + \frac{\gamma_c}{2}\right]
\Bigg\}, \quad c=g,q,{\bar q}, 
\\
%\end{align}
%\begin{equation}
%\label{1corb}
\label{1corgtl}
&{\widetilde F}^{{\rm TL} \,(1R)}_{gg,\,{\rm corr.}}\!\left(z;\frac{\mb{b}^2}{b_0^2},
\frac{n^2 b_0^2}{(2pn/z)^2 \,\mb{b}^2}\right) =
\frac{\alpha^u_{\rm S}\,S_{\eps}}{\pi} 
\;\left( \frac{\mu_0^2 \,\mb{b}^2}{b_0^2} \right)^{\!+\ep}
\;e^{- \eps\gamma_E} \, \Gamma(1-\ep) \;C_A \;z(1-z),
\\
%\end{equation}
%\begin{equation}
%\label{1corb}
\label{1corqtl}
&{\widetilde F}^{{\rm TL} \,(1R)}_{ga,\,{\rm corr.}}\!\left(z;\frac{\mb{b}^2}{b_0^2},
\frac{n^2 b_0^2}{(2pn/z)^2 \,\mb{b}^2}\right) = -
\frac{\alpha^u_{\rm S}\,S_{\eps}}{\pi} 
\;\left( \frac{\mu_0^2 \,\mb{b}^2}{b_0^2} \right)^{\!+\ep}
\;e^{- \eps\gamma_E} \, \Gamma(1-\ep) \;T_R \;\frac{z(1-z)}{1-\ep},
\,\, a=q,{\bar q}.
%\end{equation}
\end{align}
 In the right-hand side of eq.~(\ref{azin1btl}) we have used the lowest-order Altarelli--Parisi kernel ${P}^{(1)}_{ac}(x)$ (see eq.~(\ref{AP1}))
and the contribution ${\widehat P}^{\prime}_{ac}$ to the $d$-dimensional real emission kernel ${\widehat P}_{ac}$ (see eqs.~(\ref{psr})--(\ref{preg})).

The perturbative expansion of ${\widetilde F}^{\rm TL}_{ca}$ in powers of the bare QCD coupling $\alpha^u_{\rm S}$ has $\ep$-pole contributions of UV and IR origins. The UV poles
are removed by using eq.~(\ref{eq:alphasren}) and
introducing the renormalized coupling $\as(\mu_R^2)$.
The IR poles can be factorized, similarly to the SL case in eqs.~(\ref{IRfactav}) and 
(\ref{IRfactcor}). The IR factorization formulae for the $\mb{b}$ space TMD functions
${\widetilde F}^{\rm TL}_{ca}$ in the TL region are
\begin{align}
&{\widetilde F}^{\rm TL}_{ca,\,{\rm az.in.}}\!\left(z;\frac{\mb{b}^2}{b_0^2},
\frac{n^2 b_0^2}{(2pn/z)^2 \,\mb{b}^2}\right) = 
Z_c^{\rm TL}\!\left(\as(b_0^2/\mb{b}^2), \frac{n^2 b_0^2}{(2pn/z)^2 \,\mb{b}^2}\right) \nn \\
& \quad \quad \times \sum_b \int_z^1 \frac{dx}{x^{1-2\ep}} \;
{\widetilde C}^{\rm TL}_{cb}\!\left(z/x;\as(b_0^2/\mb{b}^2), \ep, \frac{n^2 b_0^2}{(2pn/z)^2 \,\mb{b}^2}\right) \;\; {\widetilde \Gamma}^{\rm TL}_{ba}(x;b_0^2/\mb{b}^2),
%\label{IRfactav}
\label{IRfactavtl}
\\
&{\widetilde F}^{\rm TL}_{ga,\,{\rm corr.}}\!\left(z;\frac{\mb{b}^2}{b_0^2},
\frac{n^2 b_0^2}{(2pn/z)^2 \,\mb{b}^2}\right) = 
Z_g^{\rm TL}\!\left(\as(b_0^2/\mb{b}^2), \frac{n^2 b_0^2}{(2pn/z)^2 \,\mb{b}^2}\right) \nn \\
& \quad \quad \times \sum_b \int_z^1 \frac{dx}{x^{1-2\ep}} \;
{\widetilde G}^{\rm TL}_{gb}\!\left(z/x;\as(b_0^2/\mb{b}^2), \ep, \frac{n^2 b_0^2}{(2pn/z)^2 \,\mb{b}^2}\right) \;\; {\widetilde \Gamma}^{\rm TL}_{ba}(x;b_0^2/\mb{b}^2).
%\label{IRfactcor}
\label{IRfactcortl}
\end{align}
The factor ${\widetilde \Gamma}^{\rm TL}_{ba}(x;\mu_F^2)$ in the right-hand side of 
eqs.~(\ref{IRfactavtl}) and (\ref{IRfactcortl}) is the customary collinear-divergent function
that defines the scale-dependent PFF $d_b(z;\mu_F^2)$ in the $\overline{{\rm MS}}$
factorization scheme. The relation between $d_b(z;\mu_F^2)$ and the bare PFF 
$d^{(0)}_a(x)$ is
\begin{equation}
%\label{pdfren}
d_b(z;\mu_F^2) = \sum_a \int_z^1 \frac{dx}{x} \;{\widetilde \Gamma}^{\rm TL}_{ba}(z/x;\mu_F^2) \;
d^{(0)}_a(x).
\end{equation}
The perturbative expansion of ${\widetilde \Gamma}^{\rm TL}_{ba}$ is
\begin{equation}
%\label{wgamexp}
\label{wgamexptl}
{\widetilde \Gamma}^{\rm TL}_{ba}(z;\mu_F^2) = \delta_{ba} \;\delta(1-z) 
- \frac{\as(\mu^2_F)}{\pi} \;\frac{{P}^{{\rm TL}\,(1)}_{ba}(z)}{\ep} + {\cal O}(\as^2),
\end{equation}
with 
\begin{equation}
\label{apsltl}
{P}^{{\rm TL}\,(1)}_{ba}(z) = {P}^{(1)}_{ab}(z),
\end{equation}
where ${P}^{(1)}_{ba}(z)$ is the lowest-order Altarelli--Parisi kernel in 
eq.~(\ref{AP1}). We recall that the SL and TL functions 
${\widetilde \Gamma}_{ba}$ and ${\widetilde \Gamma}^{\rm TL}_{ba}$
at ${\cal O}(\as)$ are directly related through the Gribov--Lipatov relation
\cite{Gribov:1972ri, Gribov:1972rt}, namely, through 
the transposition $ba \leftrightarrow ab$
of the flavour indices (see eqs.~(\ref{wgamexp}), (\ref{wgamexptl}) and (\ref{apsltl})).

The structure of the eqs.~(\ref{IRfactavtl}) and (\ref{IRfactcortl}) for the 
TL functions is analogous to that
of the corresponding eqs.~(\ref{IRfactav}) and (\ref{IRfactcor}) for the SL functions.
The IR $\ep$-poles of ${\widetilde F}^{\rm TL}_{ca}$ are factorized in the fragmentation related function ${\widetilde \Gamma}^{\rm TL}_{ba}(x;b_0^2/\mb{b}^2)$ and in the overall
perturbative function $Z_c^{\rm TL}$.
The remaining functions ${\widetilde C}^{\rm TL}_{cb}$ and 
${\widetilde G}^{\rm TL}_{gb}$ are finite and independent of $n^2$ 
(i.e., $n^2(b_0^2/\mb{b}^2)/(2np/z)^2$) in the limit $\ep \to 0$,
order-by-order in the perturbative expansion in powers of $\as(b_0^2/\mb{b}^2)$.
The main difference between eqs.~(\ref{IRfactav}), (\ref{IRfactcor})
and eqs.~(\ref{IRfactavtl}), (\ref{IRfactcortl}) is due to the phase space convolution factor $x^{-1+2\ep}$ in the IR factorization formulae for the TL functions.
The $\ep$ dependence of such convolution factor has a general origin from the $d$-dimensional kinematics of inclusive single-particle production and fragmentation
(see, e.g., ref.~\cite{Catani:1996vz}).

The first-order perturbative results for the TL factors 
$Z_c^{\rm TL}, {\widetilde C}^{\rm TL}_{cb}$ and ${\widetilde G}^{\rm TL}_{gb}$
in eqs.~(\ref{IRfactavtl}) and (\ref{IRfactcortl}) are obtained by using the expressions of 
${\widetilde F}^{{\rm TL} \,(1R)}_{ca}$ in eqs.~(\ref{azin1btl})--(\ref{1corqtl})
and they are reported below. We note that these TL factors depend on $n^2$ through
the variable ${\widetilde \lambda}= n^2 (b_0^2/\mb{b}^2)/(2np/z)^2$.

The IR factor $Z_c^{\rm TL}$ in eqs.~(\ref{IRfactavtl}) and (\ref{IRfactcortl})
has the following perturbative expansion:
\begin{equation}
Z_c^{\rm TL}(\as, {\widetilde \lambda} ) = 1 + \frac{\as}{\pi} \;Z_c^{{\rm TL} \,(1)}({\widetilde \lambda} )+ {\cal O}(\as^2),
\end{equation}
and its first-order contribution is
\begin{equation}
%\label{z1tmd}
\label{z1tmdtl}
Z_c^{{\rm TL} \,(1)}({\widetilde \lambda} ) = \frac{1}{2} \left[ A^{(1)}_c \left( \frac{1}{\ep^2}
+ \frac{1}{\ep} \ln {\widetilde \lambda} \right) + \frac{1}{\ep} \,\gamma_c \right]
- \frac{\pi^2}{24} A^{(1)}_c + h_c^{{\rm TL} \,(1)} + {\widetilde h}_c^{{\rm TL} \,(1)}(\ep,{\widetilde \lambda} ).
\end{equation}
Similarly to eq.~(\ref{z1tmd}) for the SL case, the terms $h_c^{{\rm TL} \,(1)}$ and
${\widetilde h}_c^{{\rm TL} \,(1)}$ in eq.~(\ref{z1tmdtl}) specify the resummation-scheme
dependence. In particular, we have $h_c^{{\rm TL} \,(1)}=0$ in the hard scheme
\cite{Catani:2013tia} and $h_c^{{\rm TL} \,(1)}=\pi^2  A^{(1)}_c/24$ in a minimal scheme
in which $Z_c^{{\rm TL} \,(1)}$ contains only $\ep$ pole contributions in the limit
$\ep \to 0$. The term ${\widetilde h}_c^{{\rm TL} \,(1)}(\ep,{\widetilde \lambda} )$
vanishes in the limit $\ep \to 0$.
We note that the SL contribution $Z_c^{(1)}$ in eq.~(\ref{z1tmd}) and the TL 
contribution $Z_c^{{\rm TL} \,(1)}$ in eq.~(\ref{z1tmdtl}) are equal, modulo
their scheme-dependence arbitrariness. 

The perturbative expansion of the IR finite function ${\widetilde C}_{ca}^{\rm TL}$ 
in eq.~(\ref{IRfactavtl}) is
\begin{align}
{\widetilde C}_{ca}^{\rm TL}(z;\as,\ep,{\widetilde \lambda}) =& \delta_{ca} \;\delta(1-z) +
\frac{\as}{\pi} \;{\widetilde C}_{ca}^{{\rm TL} \,(1)}(z;\ep,{\widetilde \lambda})
+ \left(\frac{\as}{\pi}\right)^2 \;{\widetilde C}_{ca}^{{\rm TL} \,(2)}(z;\ep,{\widetilde \lambda})
+ {\cal O}(\as^3),
\nn\\
&\hspace{8.0cm}c=q,{\bar q},g,
\label{ctildetl}
%\label{ctilde}
\end{align}
and the limit $\ep \to 0$ gives the TL collinear function 
${C}_{ca}^{\rm TL}(z;\as) = {\widetilde C}_{ca}^{\rm TL}(z;\as,\ep=0,{\widetilde \lambda})$
for transverse-momentum resummation (see section~\ref{sec:qcdres}). At ${\cal O}(\as)$ we explicitly obtain
\begin{equation}
%\label{c1coeff}
\label{c1coefftl}
{C}_{ca}^{{\rm TL} \,(1)}(z) = - {\widehat P}^{\prime}_{ac}(z; \ep=0) 
+ 2 {P}^{(1)}_{ac}(z) \,\ln z 
- \delta_{ca}\;\delta(1-z) \;{h}_c^{{\rm TL} \,(1)}, \quad \quad c=q,{\bar q},g,
\end{equation}
which agrees with the known results in the literature
\cite{Nadolsky:1999kb, Echevarria:2016scs, Luo:2019hmp, Luo:2019bmw}.
The complete $\ep$ dependence of ${\widetilde C}_{ca}^{{\rm TL} \,(1)}$ is
\begin{align}
&{\widetilde C}_{ca}^{{\rm TL} \,(1)}(z;\ep,{\widetilde \lambda})= 
- \,e^{- \eps\gamma_E} \, \Gamma(1-\ep) \;{\widehat P}^{\prime}_{ac}(z; \ep)
+ \frac{z^{2\ep} -e^{- \eps\gamma_E} \, \Gamma(1-\ep) }{\ep} {P}^{(1)}_{ac}(z) \nn \\
&
- \delta_{ca}\;\delta(1-z) \; \Bigg\{ 
 \frac{1 -e^{- \eps\gamma_E} \, \Gamma(1-\ep) }{2\ep}
\left[
A_c^{(1)} \left( \frac{1}{\ep} + \ln {\widetilde \lambda}  \right) + \gamma_c \right]
  \\
&
- \frac{e^{- \eps\gamma_E} \, \Gamma(1-\ep) }{2\ep} A_c^{(1)} 
\left( \psi(1-\ep) - \psi(1) \right)- \frac{\pi^2}{24} A^{(1)}_c + h_c^{{\rm TL} \,(1)} + {\widetilde h}_c^{{\rm TL} \,(1)}(\ep,{\widetilde \lambda} )
\Bigg\}, \quad  c=q,{\bar q},g. \nn
%\label{c1all}
\label{c1alltl}
\end{align}
Comparing the SL function in eq.~(\ref{c1coeff}) with the TL function in 
eq.~(\ref{c1coefftl}), we note that they both involve the contribution 
${\widehat P}^{\prime}_{ca}$ of ${\cal O}(\ep)$ to the $d$-dimensional kernel 
${\widehat P}_{ca}(z; \ep)$. In particular, the transposition of flavour indices
(i.e., ${\widehat P}^{\prime}_{ca} \leftrightarrow {\widehat P}^{\prime}_{ac}$)
in eqs.~(\ref{c1coeff}) and (\ref{c1coefftl}) is due to the crossing
relation in eq.~(\ref{crossap}). The TL function ${\widetilde C}_{ca}^{{\rm TL} \,(1)}$
also includes the term $2 {P}^{(1)}_{ac}(z) \,\ln z$, which is due to the collinear factorization of the fragmentation contributions in the $\overline{{\rm MS}}$
factorization scheme (see the convolution factor $x^{-1+2\ep}$ in eq.~(\ref{IRfactavtl})).

The azimuthal-correlation function ${\widetilde G}_{ga}^{\rm TL}$ in eq.~(\ref{IRfactcortl})
has the following perturbative expansion:
\begin{equation}
{\widetilde G}_{ga}^{\rm TL}(z;\as,\ep,{\widetilde \lambda}) = 
\frac{\as}{\pi} \;{\widetilde G}_{ga}^{{\rm TL} \,(1)}(z;\ep,{\widetilde \lambda})
+ \left(\frac{\as}{\pi}\right)^2 \;{\widetilde G}_{ga}^{{\rm TL} \,(2)}(z;\ep,{\widetilde \lambda})
+ {\cal O}(\as^3). 
\label{gtilexptl}
%\label{gtilexp}
\end{equation}
At ${\cal O}(\as)$ the azimuthal-correlation component
${\widetilde F}_{ga,\,{\rm corr.}}^{\rm TL}$ of the $\mb{b}$ space TMD function does not require the factorization of IR divergent contributions, and we directly obtain the following results:
\begin{equation}
%\label{gtil1}
\label{ggtil1tl}
{\widetilde G}_{gg}^{{\rm TL} \,(1)}(z;\ep,{\widetilde \lambda})= 
e^{- \eps\gamma_E} \, \Gamma(1-\ep) \;C_A \;z(1-z),
\end{equation}
\begin{equation}
%\label{gtil1}
\label{gqtil1tl}
{\widetilde G}_{gq}^{{\rm TL} \,(1)}(z;\ep,{\widetilde \lambda})= -
e^{- \eps\gamma_E} \, \Gamma(1-\ep) \;T_R \;\frac{z(1-z)}{1-\ep},
\end{equation}
and ${\widetilde G}_{g{\bar q}}^{{\rm TL} \,(1)}(z;\ep,{\widetilde \lambda}) =
{\widetilde G}_{gq}^{{\rm TL} \,(1)}(z;\ep,{\widetilde \lambda})$.
The limit $\ep \to 0$ of eq.~(\ref{gtilexptl}) gives the transverse-momentum resummation function $G_{ga}^{{\rm TL}}(z;\as)$ of section~\ref{sec:qcdres}, and
we have 
${\widetilde G}_{ga}^{{\rm TL} \,(m)}(z;\ep=0,{\widetilde \lambda}) = 
G_{ga}^{{\rm TL} \,(m)}(z)$.
Setting $\ep =0$ in eqs.~(\ref{ggtil1tl}) and (\ref{gqtil1tl}), we find agreement with the results for $G_{ga}^{{\rm TL} \,(1)}(z)$ in eqs.~(\ref{eq:gg1tl}) and (\ref{eq:gq1tl}).
The computation of the TL azimuthal-correlation functions $G_{ga}^{{\rm TL} \,(2)}(z)$
at ${\cal O}(\as^2)$ is discussed in the following subsection.

\subsection{TL azimuthal correlations at ${\cal O}(\as^2)$}
\label{sec:tlazcorr}

We briefly describe the calculation at ${\cal O}(\as^2)$ of the azimuthal-correlation components $\cf_{ga,\,{\rm corr.}}^{\rm TL}$ and $F_{ga,\,{\rm corr.}}^{\rm TL}$
of the differential and TMD collinear functions in the TL collinear region.

At ${\cal O}(\as^2)$ the differential collinear function 
$\cf_{ga,\,{\rm corr.}}^{\rm TL}(p,k;n)$ receives the two contributions,
$\cf_{ga,\,{\rm corr.}}^{{\rm TL} \,(2R)}$ and 
$\cf_{ga,\,{\rm corr.}}^{{\rm TL} \,(1R1V)}$, in eq.~(\ref{tlfexp12}).

The\,\,\, term\,\,\, $\cf_{ga,\,{\rm corr.}}^{{\rm TL} \,(2R)}$ 
%$\cf_{ga,\,{\rm corr.}}^{{\rm TL} \,(2R)}(p,k;n)$
\,\,is\,\, obtained\,\, by\,\, inserting\,\, 
the\,\, tree-level\,\, collinear\,\, kernel\,\,
$\cp_{{g}\rightarrow a_1 a_2 a}^{(0)}(k_1,k_2,p;n)$
in eq.~(\ref{gcolfun}) and performing the integration over $k_1$ and $k_2$ at fixed momentum
$k=k_1+k_2$. This integration procedure is completely similar to that involved in the SL collinear region and, therefore, by direct comparison of eqs.~(\ref{gcolfun}) and
(\ref{slcolfun}) we obtain
\begin{equation}
%\label{1Rcross}
\label{2Rcross}
\cf_{ga,\,{\rm corr.}}^{{\rm TL} \,(2R)}(p,k;n) = \frac{\mathcal{N}_a(\eps)}{\mathcal{N}_g(\eps)}
\;\left[ \cf_{ga,\,{\rm corr.}}^{(2R)}(p,k;n) \right]_{p \to -p},
\end{equation}
where $\cf_{ga,\,{\rm corr.}}^{(2R)}(p,k;n)$ ($a=g,q,{\bar q}$) are the contributions in eqs.~(\ref{ggcorr2R})--(\ref{fn20})
to the SL collinear functions $\cf_{ga,\,{\rm corr.}}$.
The SL-TL crossing relation in eq.~(\ref{2Rcross}) is analogous to the corresponding relation at ${\cal O}(\as)$ (see eq.~(\ref{1Rcross}) and accompanying comments).

The term $\cf_{ga,\,{\rm corr.}}^{{\rm TL} \,(1R1V)}(p,k;n)$ is directly proportional
(see eqs.~(\ref{gcolfun}) and (\ref{corrker})) to the azimuthal-correlation component 
of the one-loop collinear kernel $\cp_{{g}\rightarrow a_1 a}^{(1) \,\mu \nu}(k,p;n)$. This kernel can be evaluated by squaring the corresponding collinear-factorization results
\cite{Bern:1998sc, Bern:1999ry, Kosower:1999rx, Sborlini:2013jba}
at the amplitude level. We obtain the following explicit expressions:
\begin{align}
%\label{gg1R1V}
\label{gg1R1Vtl}
\mc{F}^{{\rm TL} \,(1R1V)}_{gg,\,{\rm corr.}}(p,k;n)
 =& - \left(\frac{\alpha^u_s\mu_0^{2\eps}S_{\eps}}{\pi}\right)^2\frac{e^{2\eps\gamma_E}{\rm \Gamma}(1+\eps){\rm \Gamma^2(1-\eps)}}{\pi^{1-\eps}{\rm \Gamma(1-2\eps)}}\frac{C_A \,\delta_{+}(k^2)}{(2pk)^{1+\eps}}\bigg\{\!\!-\frac{N_f}{6}+\frac{C_A}{6} 
\\
&\hspace{-3.4cm}+C_A \, z_{\rm TL}(1-z_{\rm TL})\left[ - \frac{1}{\eps^2}+\frac{1}{\eps}
\ln\left(z_{\rm TL}(1-z_{\rm TL})\right) -\frac{\pi^2}{6}
-\frac{1}{2}\ln^2\Big(\frac{1-z_{\rm TL}}{z_{\rm TL}}\Big) \right]\!\!\bigg\} \cos(\pi \ep)
+{\cal O}(\ep),
\nn
\\
%\label{gq1R1V}
\label{gq1R1Vtl}
\mc{F}^{{\rm TL} \,(1R1V)}_{ga,\,{\rm corr.}}(p,k;n) 
=& \left(\frac{\alpha^u_s\mu_0^{2\eps}S_{\eps}}{\pi}\right)^2\frac{e^{2\eps\gamma_E}{\rm \Gamma}(1+\eps){\rm \Gamma^2(1-\eps)}}{\pi^{1-\eps}{\rm \Gamma(1-2\eps)}}\frac{T_R \,\delta_{+}(k^2)}{(2pk)^{1+\eps}}\, \frac{z_{\rm TL}(1-z_{\rm TL})}{1-\ep} \nn\\
&\hspace{-3.4cm}\times \bigg\{ 
 C_A\left[\frac{1}{\eps^2}+\frac{1}{\eps}\left(\frac{11}{3}+ \ln\left(z_{\rm TL}(1-z_{\rm TL})\right)\right)+\frac{76}{9}-\frac{\pi^2}{6}-\frac{1}{2}\ln^2\Big(\frac{1-z_{\rm TL}}{z_{\rm TL}}\Big)\right]
 \nn\\
 &\hspace{-3.4cm}+ C_F\left[-\frac{2}{\eps^2}-\frac{3}{\eps}-8\right]+N_f\left[-\frac{2}{3\eps}-\frac{10}{9}\right]\bigg\} \cos(\pi \ep)+{\cal O}(\ep),  
%\quad \quad \quad \quad \quad \quad
\quad \quad (a=q,{\bar q}),
\end{align}
where the momentum fraction $z_{\rm TL}$ is
\begin{equation}
%\label{zvar}
\label{zvartl}
z_{\rm TL} = \frac{\bar{n} p }{\bar{n}(p+k)},
\end{equation}
and ${\bar n}^\mu$ is the light-like auxiliary vector in eq.~(\ref{nbar}).
Similarly to the SL results in eqs.~(\ref{gg1R1V}) and (\ref{gq1R1V}), we have neglected
contributions to $\mc{F}^{{\rm TL} \,(1R1V)}_{ga,\,{\rm corr.}}(p,k;n)$ with a harmless dependence on $n^2$. We note that the SL expressions in eqs.~(\ref{gg1R1V}) and 
(\ref{gq1R1V}) and the corresponding TL expressions in 
eqs.~(\ref{gg1R1Vtl}) and (\ref{gq1R1Vtl}) are not related by a crossing relation similar to 
eqs.~(\ref{1Rcross}) and (\ref{2Rcross}). This is due to the fact that the corresponding
SL and TL one-loop kernels $\cp_{{g}\rightarrow a_1 a}^{(1)}(k,p;n)$ cannot be directly
related by the replacement $p \leftrightarrow -p$ \cite{Catani:2011st},
as recalled at the beginning of section~\ref{sec:azcorrdiff}.

As discussed at the end of sections~\ref{sec:diffcollfun} and \ref{sec:tldiff}, setting
$n^2=0$, $s=(p+k)^2$ and $\mb{p_\perp} = z \mb{k_T}$, the collinear function 
$\mc{F}^{{\rm TL}}_{ca}(p,k;n)$ is directly related to the SCET fragmenting jet function
${\cal G}_{ca}(z,s,\mb{p_\perp})$ \cite{Jain:2011iu}.
At ${\cal O}(\as^2)$ we have ${\cal G}_{ca}^{(2)}(z,s,\mb{p_\perp})=
\mc{F}^{{\rm TL} \,(1R1V)}_{ca}(p,k;n) + 
\mc{F}^{{\rm TL} \,(2R)}_{ca}(p,k;n)$, where ${\cal G}_{ca}^{(2)}$ is the second-order contribution to ${\cal G}_{ca}$. The expressions in eqs.~(\ref{2Rcross})--(\ref{gq1R1Vtl}) 
give the explicit result for the azimuthal-correlation component of ${\cal G}_{ca}$.

The azimuthal-correlation component of the TL TMD function at ${\cal O}(\as^2)$
can be evaluated\, by\, integrating\, over\, $k$\, (see\, eq.~(\ref{tlgtmd}))
the\, results\, for\, $\mc{F}^{{\rm TL} \,(2R)}_{ga,\,{\rm corr.}}(p,k;n)$ and
$\mc{F}^{{\rm TL} \,(1R1V)}_{ga,\,{\rm corr.}}(p,k;n)$ that we have just presented in 
eqs.~(\ref{2Rcross})--(\ref{gq1R1Vtl}). The integration procedure is completely similar to that of section~\ref{sec:azcorrtmd}
for the case of the SL collinear function. In particular, we recall that we use a time-like auxiliary vector $n^\mu$ to avoid rapidity divergences.
Using eqs.~(\ref{2Rcross})--(\ref{gq1R1Vtl}) we compute the corresponding contributions
${F}^{{\rm TL} \,(2R)}_{ga,\,{\rm corr.}}$ and
${F}^{{\rm TL} \,(1R1V)}_{ga,\,{\rm corr.}}$ to the TMD function in $\mb{q_T}$ space,
and then the $\mb{b}$-space terms ${\widetilde F}^{{\rm TL} \,(2R)}_{ga,\,{\rm corr.}}$
and ${\widetilde F}^{{\rm TL} \,(1R1V)}_{ga,\,{\rm corr.}}$ 
in eq.~(\ref{bspacecorrtl}).

The expression of ${\widetilde F}^{{\rm TL}}_{ga,\,{\rm corr.}}$ at ${\cal O}(\as^2)$
has $\ep$-pole divergences of UV and IR origins. The UV divergences are removed by renormalizing the bare coupling at the scale $\mu_R^2=b_0^2/\mb{b}^2$.
The IR divergences are treated by expanding the IR factorization formula in 
eq.~(\ref{IRfactcortl}) up to ${\cal O}(\as^2)$, and we have
\begin{align}
%\label{fcorrat2}
\label{fcorrat2tl}
{\widetilde F}^{\rm TL}_{ga,\,{\rm corr.}}(z;\mb{b}^2/b_0^2,{\widetilde \lambda}) \!&=\!
\frac{\as(b_0^2/\mb{b}^2)}{\pi}
{\widetilde G}^{{\rm TL} \,(1)}_{ga}(z;\ep, {\widetilde \lambda})
\!+\! \left(\frac{\as(b_0^2/\mb{b}^2)}{\pi} \right)^2
\!\Bigl[Z_g^{{\rm TL} \,(1)}({\widetilde \lambda}) \;{\widetilde G}^{{\rm TL} \,(1)}_{ga}(z;\ep, {\widetilde \lambda}) \Bigr. \nn \\
&- \Bigl. \frac{1}{\ep} \sum_b \int_z^1 \!\frac{dx}{x^{1-2\ep}} 
{\widetilde G}^{{\rm TL} \,(1)}_{gb}(z/x;\ep, {\widetilde \lambda}) P^{{\rm TL} \,(1)}_{ba}(x)
\!+\!{\widetilde G}^{{\rm TL} \,(2)}_{ga}(z;\ep, {\widetilde \lambda})
\Bigr]\! +\! {\cal O}(\as^3),
\end{align}
where
${\widetilde \lambda}= n^2 (b_0^2/\mb{b}^2)/(2np/z)^2$.
The first-order functions $P^{{\rm TL} \,(1)}_{ba},\, Z_g^{{\rm TL} \,(1)}$ and 
${\widetilde G}^{{\rm TL} \,(1)}_{ga}$ are
given in eqs.~(\ref{apsltl}), (\ref{z1tmdtl}), (\ref{ggtil1tl}) and (\ref{gqtil1tl}).
Therefore, using eq.~(\ref{fcorrat2tl}) and our results for 
${\widetilde F}^{{\rm TL} \,(2R)}_{ga,\,{\rm corr.}}$
and ${\widetilde F}^{{\rm TL} \,(1R1V)}_{ga,\,{\rm corr.}}$, we obtain the explicit expressions of the functions 
${\widetilde G}^{{\rm TL} \,(2)}_{ga}(z;\ep, {\widetilde \lambda})$ ($a=g,q,{\bar q}$)
up to ${\cal O}(\ep^0)$.
We find that ${\widetilde G}^{{\rm TL} \,(2)}_{ga}$ are finite and independent of ${\widetilde \lambda}$ in the limit $\ep \to 0$, namely 
${\widetilde G}^{{\rm TL} \,(2)}_{ga}(z;\ep=0, {\widetilde \lambda}) = 
{G}^{{\rm TL} \,(2)}_{ga}(z)$. Our results for the TL azimuthal-correlation functions 
${G}^{{\rm TL} \,(2)}_{ga}(z)$ are
\begin{align}
G^{{\rm TL} \,(2)}_{gg}(z) =& C_F N_f\left\{\frac{1}{18z}+\frac{1}{2}+z-\frac{14z^2}{9}+\ln(z)\left[-\frac{1}{3z}+1+\frac{3z}{2}\right]+\frac{3z}{4}\ln^2(z) \right\}
\nn\\
&+\!C_A N_f\left\{\!-\frac{1}{36z}-\frac{1}{12}-\frac{4z}{9}\!+\!\frac{17z^2}{36}-\frac{z}{6}\ln(z) \!\right\}\!+\! C_A^2\bigg\{\!-\frac{1}{36z}-\frac{5}{12}-\frac{20z}{9}\!+\!\frac{11z^2}{4}
\nn\\
&+\ln(z)\left[\frac{1}{3z}-1-\frac{67z}{12}+z^2 \right]+z(1-z)\bigg[\ln(z)\ln(1-z)-\text{Li}_2(1-z)\bigg]
\nn\\
&-\ln^2(z)\left[3z-\frac{z^2}{2}\right]\bigg\}
-{h}_g^{{\rm TL} \,(1)}\,C_A \,z(1-z),
\label{g2tlgg}
\end{align}
\begin{align}
G^{{\rm TL} \,(2)}_{gq}(z) =& C_F\bigg\{-\frac{1}{8}+\frac{3z}{4}-\frac{5z^2}{8}-\ln(z)\left[\frac{1}{4}+\frac{3z}{8}-\frac{z^2}{4}\right] + \ln^2(z)\left[\frac{3z}{8}-\frac{3z^2}{4}\right]
\nn\\
&+z(1-z)\bigg[\frac{1}{4}\ln(1-z)
-\frac{3}{2}\ln(z)\ln(1-z)+\frac{1}{4}\ln^2(1-z) -\text{Li}_2(z)-\frac{\pi^2}{12}\bigg]\bigg\}
\nn\\
&+N_f\left\{z(1-z)\left[\frac{1}{9}-\frac{1}{6}\ln(z)+\frac{1}{6}\ln(1-z)\right] \right\}+C_A\bigg\{\ln(z)\left[\frac{1}{4}+\frac{13z}{6}-\frac{17z^2}{12}\right] 
\nn\\
&+\ln^2(z)\left[\frac{3z}{4}+\frac{z^2}{2}\right]+z(1-z)\bigg[-\frac{25}{36}-\frac{5}{12}\ln(1-z)-\frac{1}{2}\text{Li}_2(1-z)\nn\\
&-\frac{1}{4}\ln^2(1-z)+\frac{\pi^2}{4}\bigg]\bigg\}
+{h}_g^{{\rm TL} \,(1)}\;\frac{1}{2} \,z(1-z),
\label{g2tlgq}
\end{align}
and $G^{{\rm TL} \,(2)}_{g{\bar q}}(z)=G^{{\rm TL} \,(2)}_{gq}(z)$. The coefficient ${h}_g^{{\rm TL} \,(1)}$ in eqs.~(\ref{g2tlgg}) and (\ref{g2tlgq})
parametrizes the resummation-scheme dependence, and we recall that 
${h}_g^{{\rm TL} \,(1)}=0$ in the hard scheme 
(see eqs.~(\ref{z1tmdtl}) and (\ref{c1coefftl})).

The transverse-momentum resummation functions ${G}^{{\rm TL} \,(2)}_{ga}(z)$ 
were first computed in ref.~\cite{Luo:2019bmw}
(the hard scheme is used therein), by using a SCET framework with properly regularized rapidity divergences. We have presented the first recomputation of ${G}^{{\rm TL} \,(2)}_{ga}(z)$, and we have used a fully independent method. Our results in 
eqs.~(\ref{g2tlgg}) and (\ref{g2tlgq}) agree with those in ref.~\cite{Luo:2019bmw}.

\section{Summary}
\label{sec:sum}

QCD squared amplitudes are singular in the multiparton collinear limit, and the singular behaviour is controlled by the splitting kernels of a factorization formula that has a process-independent structure.

In this paper we have exploited the collinear factorization of QCD amplitudes to introduce collinear functions that contribute to QCD resummation formulae for hard-scattering cross sections. The collinear functions are defined through the integration of the multiparton splitting kernels over a constrained phase space that depends on the hard-scattering observable
of interest. Considering different phase-space constraints, one can define different collinear functions, which embody the logarithmically-enhanced contributions of collinear origin to the corresponding hard-scattering observables. In this paper we have explicitly considered differential collinear functions that can in turn be used to evaluate TMD functions for transverse-momentum resummation and beam functions for $N$-jettiness resummation.

A distinctive and relevant feature of our collinear functions is their dependence on an auxiliary vector $n^\mu$, which directly follows from the corresponding $n$ dependence of the splitting kernels. In applications of collinear factorization of QCD amplitudes, the 
auxiliary vector is usually chosen to be light-like. We use both light-like and time-like auxiliary vectors. In the paper we have discussed how the $n$ dependence controls the behaviour of the splitting kernels in kinematical regions that are far from the collinear region. In particular, in the case of TMD functions we have shown that the use of a 
time-like vector $n^\mu$ avoids the rapidity divergences that are instead present if
$n^\mu$ is light-like.

The collinear functions can be introduced for the cases of both final-state fragmenting partons and initial-state colliding partons. The final-state and initial-state collinear functions use the splitting kernels in the corresponding TL and SL collinear regions, respectively. The TL splitting kernels and, consequently, the final-state collinear functions are process independent. In contrast, the initial-state collinear functions are,
in general, process dependent beyond ${\cal O}(\as^2)$ (although they have a 
process-independent structure). Such process dependence is a consequence of the violation of strict collinear factorization of the QCD squared amplitudes in the SL collinear regions.

As discussed in the paper, our TMD and beam functions with a light-like auxiliary vector
$n^\mu$ can be related to the analogous SCET functions in the literature. In the TL region our collinear functions are equivalent to the parton level SCET functions. In the SL region
the same equivalence is limited up to ${\cal O}(\as^2)$, since at higher perturbative orders our collinear functions are process dependent.

We have discussed the perturbative computation of the collinear functions. Such computation
leads to UV and IR divergences, which we have regularized by the customary procedure of analytic continuation in $d=4-2\ep$ space-time dimensions. The $\ep$-pole divergences 
of UV origin are removed by the renormalization of the QCD coupling $\as$. The IR divergences can instead be factorized from IR finite collinear terms that directly contribute to QCD resummation formulae of hard-scattering observables.
At the cross section level the IR divergences of the collinear functions 
are partly removed by the `renormalization' of the bare parton densities and fragmentation functions, and the remaining part is then cancelled
by the IR terms due to the soft and purely-virtual contributions to the hard-scattering observable.

We have illustrated the perturbative features of the collinear functions by performing their calculation at ${\cal O}(\as)$ and discussing the explicit dependence on the auxiliary vector
$n^\mu$.

In the case of TMD observables, the collinear functions have both azimuthal-independent and
azimuthal-correlation components. The azimuthal-correlation component is specific of the gluon partonic channels and is also known as the contribution of linearly-polarized gluons.
We have presented the calculation at ${\cal O}(\as^2)$ of the azimuthal-correlation component
of the differential and TMD collinear functions. Performing UV renormalization and  factorization of the IR divergences, we have computed the ${\cal O}(\as^2)$ contribution
of linearly-polarized gluons  to transverse-momentum resummation. Our result for both the SL and TL regions agree with the results obtained by other authors using SCET functions
and related theoretical methods.  

The computation at ${\cal O}(\as^2)$ of the azimuthal-independent component of the collinear functions will be presented in future work, where we shall also discuss the related computation of the $n$-dependent soft factor for transverse-momentum resummation. We also plan to study collinear functions for other hard-scattering observables.

Certainly, an important future step can be the explicit extension of our theoretical framework to ${\cal O}(\as^3)$ and higher perturbative orders. In particular, the explicit computation of the process dependence of the collinear functions  
in the SL region is very relevant. This computation requires the preliminary calculation at ${\cal O}(\as^3)$
of the splitting kernels for the SL collinear limit of the QCD scattering amplitudes. 
At present, such SL splitting kernels are not known at the required order in the $\ep$ expansion. 

\section*{Acknowledgement}
We would like to thank Roberto Bonciani, Leandro Cieri, Giancarlo Ferrera, Massimiliano Grazzini, and Germ\'{a}n Rodrigo for useful discussions.

\bibliographystyle{JHEP}
\bibliography{refcoll}
\end{document}